\documentclass[12pt]{article}
\usepackage{amsmath,amssymb,epsfig,amsfonts}
\usepackage{graphicx,subfigure}
\usepackage[usenames, dvipsnames]{color}
\usepackage[backref]{hyperref}
\usepackage{cite}
\usepackage{lscape}
\usepackage{verbatim} 
\usepackage{float}
\restylefloat{table}
\usepackage[all]{xy}
\usepackage{tikz}

\makeatletter

\newcommand{\be}{\begin{equation}}
\newcommand{\ee}{\end{equation}}
\newcommand{\ba}{\begin{aligned}}
\newcommand{\ea}{\end{aligned}}

\def\unit{{1\kern-.65ex {\rm l}}}
\def\1{{1\kern-.65ex {\rm l}}}

\def\bbP{\mathbb{P}}

\newcommand{\nocontentsline}[3]{}
\newcommand{\tocless}[2]{\bgroup\let\addcontentsline=\nocontentsline#1{#2}\egroup}

\def\CC{{\cal C}}

\def\CL{{\cal L}}

\def\CO{{\cal O}}
\def\CP{{\cal P}}
\def\CQ{{\cal Q}}

\def\bbP{{\mathbb{P}}}

\def\bbZ{{\mathbb{Z}}}

\def\fkb{{\mathfrak{b}}}
\def\fkc{{\mathfrak{c}}}

\newcount\hour \newcount\minute
\hour=\time \divide \hour by 60
\minute=\time
\def\now{%
\ifnum \hour<13
  \ifnum \hour=0 \advance \hour by 12 \number\hour:\else \number\hour:\fi%
     \ifnum \minute<10 0\fi%
     \number\minute%
\ A.M.%
\else \advance \hour by -12 \number\hour:%
  \ifnum \minute<10 0\fi%
  \number\minute%
  \ P.M.%
\fi%
}

\makeatother

\begin{document}

\baselineskip=18pt  
\numberwithin{equation}{section}  
\allowdisplaybreaks  



%
%


\thispagestyle{empty}

\vspace*{-2cm} 
\begin{flushright}
{\tt KCL-MTH-14-10} \\
\end{flushright}

\vspace*{1cm} 
\begin{center}
{\LARGE   Tate Trees for Elliptic Fibrations with\\
 \smallskip
  Rank one Mordell-Weil group}\\

 \vspace*{1.5cm}
{Moritz K\"untzler and Sakura Sch\"afer-Nameki}\\

 \vspace*{1.0cm} 
{\it Department of Mathematics, King's College, London \\
 The Strand, London WC2R 2LS, England }\\
 {\tt moritz.kuentzler@kcl.ac.uk, {gmail:$\,$ sakura.schafer.nameki}}\\

\vspace*{0.8cm}
\end{center}
\vspace*{.1cm}

\noindent
$U(1)$ symmetries  play a central role in constructing phenomenologically viable
F-theory compactifications that realize Grand Unified Theories (GUTs). 
In F-theory, gauge symmetries with abelian gauge factors are modeled 
by singular elliptic fibrations with additional rational sections, i.e.  a non-trivial Mordell-Weil rank.
To determine the full scope of possible low energy theories with abelian gauge factors, which allow for  
 an F-theory realization, it is central to obtain a comprehensive list of all 
singular elliptic fibrations with extra sections. We answer this question for the case of one abelian factor 
by applying Tate's algorithm to the elliptic fiber realized as a quartic  in the weighted projective space $\mathbb{P}^{(1,1,2)}$, which guarantees, in addition to the zero section, the existence of an additional rational section. 
The algorithm gives rise to a tree-like enhancement structure, where each fiber is  characterized by 
a Kodaira fiber type, that governs the non-abelian gauge factor, and the separation of the two sections. 
We determine Tate-like forms for elliptic fibrations with one extra section for all Kodaira fiber types. 
In addition to standard Tate forms that are determined by the  vanishing order of the coefficient sections in the quartic (so-called canonical models), 
the algorithm also gives rise to fibrations that require non-trivial relations among the coefficient sections. 
Such non-canonical models have phenomenologically interesting properties, as they allow for a richer charged matter content, and thus codimension two fiber structure, than the canonical 
models that have been considered thus far in the literature. 
As an application we determine the complete set of codimension one fibers types, matter spectra, both canonical and non-canonical, for 
$SU(5) \times U(1)$ models.

\newpage


\tableofcontents


\section{Introduction}

The Tate forms of singular elliptic fibrations with a section are the starting point for modeling non-abelian gauge symmetries in F-theory  \cite{Tate,Bershadsky:1996nh, Katz:2011qp}. The goal of this paper is to determine Tate-like forms for singular elliptic fibrations with an extra section, corresponding to an additional abelian gauge factor. The associated F-theory compactifications give rise to four-dimensional gauge groups of the type $G\times U(1)$, with $G$ a simple Lie group. 

An elliptic fibration with a section has an associated Weierstrass model realized in  the weighted projective space $\mathbb{P}^{(1,2,3)}$ with homogenous coordinates $[w, x,y]$ by the hypersurface equations
\be
y^2 = x^3 + fx w^4  +g w^6 \,,
\ee 
where $f,g$ are sections of  $K_B^{-4}$ and $K_B^{-6}$, respectively, and $K_B$ is the canonical bundle of the base $B$ of the fibration. 
The elliptic fiber becomes singular whenever the discriminant $\Delta= 4 f^3 + 27 g^2$ vanishes. 
Let $z$ be a local coordinate in the base and $z=0$ a component of the vanishing locus of the discriminant. 
The possible singular fibers above a codimension one locus $z=0$ have been classified by Kodaira and N\'eron
 \cite{Kodaira, Neron}\footnote{Kodaira's proof applies to elliptic surfaces, but is equally applicable in codimension one in the base for higher dimensional fibrations.} and are characterized in terms of the vanishing orders of $(f, g,\Delta)$ in $z$. For instance, for an $I_n$ type fiber, which realizes $SU(n)$ gauge groups in F-theory, the vanishing orders are ord$(f, g, \Delta)= (0,0,n)$. This requires a suitable tuning of the expansion coefficients of $f$ and $g$, to give rise to a cancellation in the discriminant up  to order $n$. 

Tate's algorithm \cite{Tate} determines an alternative representation of the elliptic fibration
\be
y^2 + b_1 x y + b_3 y = x^3 + b_2 x^2 + b_4 x + b_6 \,,
\ee
which makes the Kodaira singular fiber type apparent, in terms of the vanishing order in $z$ of the coefficient sections $b_i$ \cite{Bershadsky:1996nh, Katz:2011qp}.  These Tate forms come with the caveat that coordinate changes applied throughout the algorithm  may be locally not well-defined, as one cannot perform certain divisions over the local ring of functions on the base of the fibration. As shown in \cite{Katz:2011qp}, this occurs in particular for $I_{2m+1}$, $m>5$, fibers with monodromy, and furthermore for the outlier cases $I_n$, $n=6,7,8,9$. In these cases, the coefficient sections $b_i$ satisfy non-trivial relations and the locally attainable form of the fibration is not only characterized in terms of vanishing orders. We shall refer to such models as {\it non-canonical}\footnote{These are not to be confused with non-canonical singularities.} forms, as opposed to the standard Tate forms, which are in this sense {\it canonical},  i.e. are specified entirely by the vanishing orders of the coefficients $b_i$.

In applications to particle physics, F-theory compactifications on singular elliptic Calabi-Yau fourfolds with section 
provide a rich framework for constructions of supersymmetric GUT models. The singular fiber in codimension one determines the non-abelian gauge symmetry, the codimension 2 and 3 fibers realize matter and Yukawa couplings, respectively. 
However, this structure alone does not yield a fully realistic framework, as for instance it does not provide additional symmetries
that are pivotal to constrain the generation of dangerous proton decay operators. Abelian gauge symmetries have 
been instrumental to this effect. 

Abelian gauge symmetries are realized in F-theory by elliptic fibrations with extra sections, or equivalently, a non-trivial Mordell-Weil rank, i.e. a non-torsion components of the Mordell-Weil group \cite{Morrison:1996na, Morrison:1996pp}. The Mordell-Weil group is the set of rational sections of an elliptic fibration, which form a group with the standard group law on the elliptic curve. 
Each additional section defines a $(1,1)$ form along which the M-theory $C_3$ can be reduced to give rise to an abelian gauge field in four dimensions. It was shown in \cite{Morrison:2012ei} that for an elliptic fibration with one extra section, the fiber can be embedded into the weighted projective space $\mathbb{P}^{(1,1,2)}$ in terms of  a quartic equation. It is then natural to ask, how non-abelian gauge symmetries, i.e. Kodaira fibers, are realized in terms of these quartic hypersurface equations, and whether Tate-like forms exist that make the additional section apparent. 

This is the question we set out to answer in this paper. The fibers are characterized in terms of the codimension one Kodaira fiber, as well as the position of the two sections, $\sigma_0$ and $\sigma_1$, i.e. the intersection of these with the  fiber components of the singular fiber. In applications to F-theory, the section separation  governs the $U(1)$ charges of matter realized in codimension two. 
By application of a Tate type algorithm, we determine realizations of these, which are  characterized either by the vanishing orders of the coefficient sections of the quartic, or for non-canonical cases, specific relations among these coefficients. The presence of the extra section results in a tree-like enhancement structure, that we will refer to as the {\it Tate tree}: For instance in the $I_n$ branch, there are multiple ways to enhance from $I_n$ to $I_{n+1}$, which differ by the separation of the sections $\sigma_0$ and $\sigma_1$. An excerpt of the tree starting from $I_1$ up to $I_5$ is shown in figure \ref{fig:TheTree}. 
The complete set of  canonical forms  for all Kodaira fibers with additional section can be found in tables \ref{tab:TateTable} and \ref{tab:FiberTypeTable}. 

In the last few years, several constructions of models with abelian gauge groups have appeared in the literature, in particular, for phenomenological reasons, focusing on the case of $SU(5)\times U(1)$ gauge groups. The first examples were constructed in fact starting with the standard Tate form in $\mathbb{P}^{(1,2,3)}$   \cite{Grimm:2010ez, Mayrhofer:2012zy}, which give lifts of spectral cover models with $U(1)$ symmetries studied in \cite{Marsano:2009gv, Marsano:2009wr, Dolan:2011iu}. More recently, applying the toric top construction \cite{Candelas:2012uu} several $SU(5)\times U(1)$ models were obtained in
\cite{Braun:2013yti, Borchmann:2013jwa, Borchmann:2013hta, Braun:2013nqa}. 
Subsequently, methods for multiple $U(1)$ factors were developed, where the fiber was shown to have a realization in terms of a cubic in $\mathbb{P}^2$ and several models with $SU(5)$ non-abelian gauge factor were constructed \cite{Borchmann:2013jwa, Borchmann:2013hta, Cvetic:2013nia, Cvetic:2013uta, Cvetic:2013jta, Cvetic:2013qsa, Krippendorf:2014xba}. All these models are determined by a set of vanishing orders of the coefficients in the hypersurface equation that realizes the singular fiber, and are thus of canonical type\footnote{The model in \cite{Mayrhofer:2012zy} based on the split spectral cover with $3+2$ factorization is the only exception, which has two differently charged ${\bf 10}$ matter loci. }.

The focus of the present paper is the case of one extra section, which realizes gauge groups $G \times U(1)$ in four dimensions. In the  companion paper \cite{LSN} the case of two extra sections with gauge groups $G\times U(1)\times U(1)$ is considered.  
{The main motivation is to determine all possible $U(1)$ charge assignments in an F-theory model, which forms the basis of carrying out a survey of the phenomenological properties of F-theory compactification. These questions can be reformulated in terms of two mathematical goals:} to provide  Tate-like forms for elliptic fibrations with an extra section, i.e. specifying a set of vanishing orders for a quartic in $\mathbb{P}^{(1,1,2)}$ for each  Kodaira fiber type (and the separation of the two sections on the fiber). Secondly, to study the validity of these forms, i.e. determining whether they can be reached by locally well defined changes of coordinates, without division, over the local ring of functions on the base of the fibration. As discussed above, the latter issue was addressed for the standard Tate forms in \cite{Katz:2011qp}. In the present context we find that non-canonical fibrations are much more common, and in fact occur also in the phenomenologically interesting case of $I_5$ fibers, which implies that there is a larger class of models than previously obtained with gauge group $SU(5)\times U(1)$. 

Non-canonical models can arise, whenever the discriminant has a {non-linear} polynomial factor. To enhance the vanishing order, one has to solve for vanishing of this polynomial over the local ring of functions on the base of the elliptic fibration. Using the  unique factorization property of this ring results in non-trivial relations among the coefficients of the quartic equation. 
Whenever there is no coordinate change that does not require division by a section  that brings this form back to a canonical form, the resulting models are non-canonical. 
In $\mathbb{P}^{(1,1,2)}$, most shifts are locally obstructed, leading to a large class of non-canonical forms.  

The main difference between canonical and non-canonical models is the codimension two fiber structure, namely, non-canonical models generically correspond to models with multiple codimension two fibers of the same Kodaira type, with however different separation between the zero section and the extra section. In terms of the phenomenological models realizing $SU(5)$, for instance, this results in models with multiple, differently $U(1)$ charged ${\bf 10}$ matter curves, which open up further possibilities for model building in F-theory.

The plan of this paper is as follows: in section \ref{sec:MWG} we summarize the setup and fix notation for the realization of elliptic fibrations with rank one Mordell-Weil group. The general structure of Tate's algorithm, and the associated Tate tree, is discussed in section \ref{sec:TT}, including the analysis of the starting points of Tate's algorithm, i.e. the realizations of $I_1$ and $I_2$ fibers, and a discussion of the  symmetries that allow identifications of models. Resolutions of the singularities throughout the algorithm allows us to determine the Kodaira types (which of course one could also determine by mapping back to Weierstrass) and more importantly, the intersection with the sections. The resolution method is discussed in section \ref{sec:Fibration}, and applied to the $I_n$ and $I_n^*$ infinite series in appendix \ref{sec:ResolvedGeometries}.
Our main results are summarized in section \ref{sec:Summary}, providing the canonical forms for all fiber types, including a summary table for the canonical forms of  the infinite $I_n$ and $I_n^*$ series.  
Tate's algorithm is then discussed in detail up to $I_5$ in section \ref{sec:Tate}. The exceptional cases  $II^*, III^*, IV^*$ as well as canonical forms for  $I_n$ and $I_n^*$ for general $n$ are derived in section \ref{sec:TreeTopsInfty}. 
The non-canonical models for $I_n$ for $n=3, 4, 5$ are discussed in section \ref{sec:P112NonCanonical}, including the matter spectrum and $U(1)$ charges.   
We close in section \ref{sec:P123Reloaded} by reconsidering the outlier cases in $\mathbb{P}^{(1,2,3)}$ first noted in \cite{Katz:2011qp}, and derive the non-canonical forms for them. The appendices collect several technical results normal forms for elliptic curves with extra sections, on the solutions of polynomial equations over UFDs  used to derive the non-canonical forms, an alternative representation for the $I_5$ models as well as a discussion of the relation between the tops  and split spectral cover models, that have appeared in the literature and the $I_5$ models obtained from Tate's algorithm.

\section{Elliptic Fibrations with one extra section and Tate Trees}
\label{sec:TT}

The purpose of this section is to collect general structural properties of Tate's algorithm for elliptic fibrations with rank one Mordell-Weil group. Such fibrations can be realized as quartics in $\mathbb{P}^{(1,1,2)}$, which we will review in section \ref{sec:MWG}. 
The singular fibers are characterized by their Kodaira type as well as the separation of the two rational sections in the singular fiber.  The resulting enhancement structure is tree-like and we collect general properties of this Tate tree in sections \ref{sec:TateSetup} and \ref{sec:Constr}. In the remaining sections \ref{sec:Order0}, \ref{sec:Sym} and \ref{sec:Lop} we determine the starting points of the algorithm, which are $I_1$ and $I_2$ fibers with different section separation, and discuss symmetries (lops) which map quartics, which describe the same fiber type, but have different vanishing orders, into each other. We give a summary of the results of each section at the start, allowing the reader to skip the rather technical proofs. Section \ref{sec:Fibration} summarizes the resolution of the singular fibrations, which are used throughout the algorithm in order to determine the fiber types.


\subsection{Rank one Mordell-Weil group}
\label{sec:MWG}

As is shown in the appendix \ref{app:Deligne}, an elliptic curve with rank one Mordell-Weil group can be realized in terms of homogeneous quartic polynomials in the weighted projective space $\mathbb{P}^{(1,1,2)}$ 
\begin{equation}  
	\mathcal{C}: \qquad \fkc_0 w^4 + \fkc_1 w^3x + \fkc_2 w^2  x^2 + \fkc_3 w x^3+ \fkc_4 x^4  =  \mathfrak{a} y^2 + \fkb_0 x^2 y + \fkb_1 y w  x + \fkb_2 w^2 y \,.
\end{equation}
where the condition of rank one Mordell-Weil group is shown to imply\footnote{Compared to the form obtained in \cite{Park:2011ji}, the analysis in the appendix shows that there is an additional constraint that $\fkb_0\not=0$. This will be of particular importance for the Tate's algorithm, where this condition translates into $\mathfrak{b}_0 \not=0$ in codimension one.}
\be
\mathfrak{a} = 1 \,,\qquad \mathfrak{c}_4=0\,,\qquad  \fkb_0 \not=0 \,.
\ee
With these restrictions, one can also consider  $\mathrm{Bl}_{[0,1,0]}\bbP^{(1,1,2)}$, which is  the blowup at $[0,1,0]$ of $\mathbb{P}^{(1,1,2)}$. 
Let $[\tilde w{\,:\,}x{\,:\,}\tilde y]$ be the coordinates of the weighted projective space $\mathbb{P}^{(1,1,2)}$. 
Blowing this up at $\tilde w=\tilde y=0$, yields an exceptional divisor $s=0$ and new coordinates $\tilde w = s w$, $\tilde y = s y$. The projective relation from this new divisor is $[w{\,:\,} y]$, as well as the relation $[s w{\,:\,}x{\,:\,}s y]$. Consider the homogeneous  polynomial of degree four in $\mathrm{Bl}_{[0,1,0]}\bbP^{(1,1,2)}[4]$ 
\begin{equation}  \label{eq:Qpoly}
\CQ: \qquad \fkc_0 w^4s^3 + \fkc_1 w^3s^2 x + \fkc_2 w^2s  x^2 + \fkc_3 w x^3 = y^2 s + \fkb_0 x^2 y + \fkb_1 y w s x + \fkb_2 w^2 s^2 y \,.
\end{equation}
We consider singular elliptic fibrations, where the fiber is realized in terms of the quartic (\ref{eq:Qpoly}). In this case, 
the $\fkb_i$, $\fkc_i$ are sections of suitable line bundles on the base $B$ of the fibration, described in more detail in section \ref{sec:Fibration}.  The goal of this paper is to determine conditions on the coefficients $\fkb_i$ and $\fkc_i$ for them to realize Kodaira singular fibers above a codimension one locus $z=0$.

By shifting and scaling the $y$ coordinate, the quartic can be put into the form
\begin{equation}\label{QuarticSimple}
 \fkc_0 w^4 s^3 + \fkc_1 w^3 s^2 x + \fkc_2 w^2 s x^2 + \fkc_3 w x^3 = s y^2 + \fkb_0 x^2 y \,,
\end{equation}
possibly with new $\fkb_i$ and $\fkc_i$. The rational points $\sigma_0= [0:1:0]$ and $\sigma_1=[0:1: -\mathfrak{b}_0]$ of this elliptic curve in the blow-up are 
\be
\ba
\sigma_0: & \qquad s=0 ,\  x=1 ,\, y= \mathfrak{c}_3,\,  w=\mathfrak{b}_0 \cr
\sigma_1 : &\qquad   w=0,\  y=1,\,  s=-\fkb_0 x^2\,.
\ea
\ee
The elliptic curve (\ref{QuarticSimple}) has a representation in terms of a Weierstrass model, for instance with respect to the zero-section $\sigma_0$ the Weierstrass form is given by \cite{Morrison:2012ei}
\begin{equation}\label{eq:WstrMap}
  \hat{y}^2 = \hat{x}^3 + \left( \fkc_1 \fkc_3 - \fkb_0^2 \fkc_0 - \frac{\fkc_2^2}{3} \right) \hat{x} \hat{w}^4 + \left( \fkc_0 \fkc_3^2 - \frac{1}{3}\fkc_1 \fkc_2 \fkc_3 + \frac{2}{27} \fkc_2^3 - \frac{2}{3} \fkb_0^2 \fkc_0 \fkc_2 + \frac{\fkb_0^2 \fkc_1^2}{4} \right) \hat{w}^6 \,.
\end{equation}
The singular loci of the elliptic curve (\ref{QuarticSimple}) are characterized by the vanishing of the discriminant 
\begin{equation}
  \begin{aligned}
    \Delta = & \, 256 \bigl( 64 \fkb_0^6 \fkc_0^3 - 16 \fkc_0^2 \left( 8 \fkb_0^4 \fkc_2^2 + 12 \fkb_0^4 \fkc_1 \fkc_3 - 36 \fkb_0^2 \fkc_2 \fkc_3^2 + 27 \fkc_3^4 \right) + \\
    & + 8 \fkc_0 \left( 8 \fkc_2^3 \left(\fkb_0^2 \fkc_2 - \fkc_3^2 \right) + 4 \fkc_1 \fkc_2 \fkc_3 \left(9 \fkc_3^2 - 10 \fkb_0^2 \fkc_2 \right) + 3 \fkc_1^2 \left( 6 \fkb_0^4 \fkc_2 - \fkb_0^2 \fkc_3^2 \right) \right) + \\
    & + \fkc_1^2 \left( -27 \fkb_0^4 \fkc_1^2 + 16 \fkc_2^2 \left( \fkc_3^2 - \fkb_0^2 \fkc_2 \right) + 8 \fkc_1 \left( 9 \fkb_0^2 \fkc_2 \fkc_3 - 8 \fkc_3^3 \right) \right) \bigr) \,.
    \label{eq:Delta}
  \end{aligned}
\end{equation}
The Tate forms that we determine for models with extra section have $\mathfrak{b}_1$ and $\mathfrak{b}_2$ coefficients, 
and so in order to map back to Weierstrass by (\ref{eq:WstrMap}), one has to shift those away first to reach the form (\ref{QuarticSimple}). This is useful when determining the simple Kodaira type of the fiber, without for instance resolving the singularity first.  The shift that mapes (\ref{eq:Qpoly}) back to (\ref{QuarticSimple}) is
\be
y \quad \rightarrow \quad y - {1\over 2}\mathfrak{b}_{1} w x   - {1\over 2}  \mathfrak{b}_{2}  s  w^2   \,,
\ee
which leads to the new coefficients
\be\label{ShiftedCs}
\ba
\fkc_{0} &\quad\rightarrow\quad \fkc_{0}+ {1\over 4} \fkb_2^2 \cr
\fkc_{1}  &\quad \rightarrow \quad  \fkc_{1} + {1\over 2} \fkb_{1} \fkb_{2} \cr
\fkc_{2} &\quad \rightarrow \quad  \fkc_{2} + {1\over 4} \fkb_{1}^2 + {1\over 2} \fkb_0 \fkb_2 \cr
\fkc_{3} &\quad\rightarrow \quad \fkc_{3} + {1\over 2} \fkb_{0} \fkb_{1} \,.
\ea
\ee
The coefficients $f$ and $g$ in the Weierstrass form $y^2 = x^3 + fx +g$ after this shift
have the following leading order when expanded as a power series in $z$
\be
\ba
f=& -b_{0,0}^2 c_{0,0}-\frac{1}{48} (b_{1,0}^2+2 b_{0,0} b_{2,0}+4 c_{2,0})^2+{1\over 4}({ b_{1,0} b_{2,0}}+ 2c_{1,0})
  ({b_{0,0} b_{1,0} }+2 c_{3,0})- {b_{2,0}^2 b_{0,0}^2\over 4}
+ O(z) \cr
 g= &  \frac{1}{864} \left(b_{1,0}^2+2 b_{0,0} b_{2,0}+4 c_{2,0}\right){}^3-\frac{1}{24} b_{0,0}^2 \left(b_{2,0}^2+4 c_{0,0}\right) \left(b_{1,0}^2+2
   b_{0,0} b_{2,0}+4 c_{2,0}\right)\cr
&   -\frac{1}{48} \left(b_{1,0} b_{2,0}+2 c_{1,0}\right) \left(b_{0,0} b_{1,0}+2 c_{3,0}\right) \left(b_{1,0}^2+2
   b_{0,0} b_{2,0}+4 c_{2,0}\right)\cr
&   +\frac{1}{16} b_{0,0}^2 \left(b_{1,0} b_{2,0}+2 c_{1,0}\right){}^2+{1\over 16}\left({b_{2,0}^2}+4c_{0,0}\right)
   \left(b_{0,0} b_{1,0}+2 c_{3,0}\right)^2+ O(z) \,.
\ea
\ee
The lowest order term that does not vanish will always be the leading coefficient of $\mathfrak{b}_{1}$, which thereby determines the vanishing order of $f$ and $g$\, \footnote{As will be explained in the next section in codimension 1 any vanishing order in $\mathfrak{b}_0$ can be shifted or `lopped' away, so that this always has a zeroeth order term. }. From the Kodaira classification this implies that for instance that $I_n$ fibers, which have $f$ and $g$ of vanishing orders $0$, it is necessary that  $b_{1,0} \not=0$, whereas for $I_n^*$, which have ord$(f)=2$ and ord$(g)=3$, $b_{1,0}=0$ and $b_{1,1}\not=0$. These conditions will appear naturally in Tate tree.


\subsection{Tate's algorithm, Trees and Canonicality}
\label{sec:TateSetup}

A singular elliptic fibration with a section can be realized in terms of a Weierstrass model  
\be
y^2 = x^3 + fx w^4 + g w^6  \,,
\ee
where $[w, x, y]$ are  homogenous coordinates  in  $\bbP^{(1,2,3)}$.
Let $z$ be a local coordinate on the base of the fibration and let $z=0$ be a component of the discriminant of the Weierstrass model $\Delta= 4 f^3 + 27 g^2$. We will assume throughout that the divisor $z=0$ in the base is smooth. 
The possible singular fibers in codimension one in the base were classified by Kodaira and N\'eron \cite{Kodaira, Neron}\footnote{This is under the assumption that the classification for surfaces obtained by Kodaira and N\'eron carries over to codimension one in a higher-dimensional elliptic fibration.}. For a given singular  Weierstrass model  in $\bbP^{(1,2,3)}$
Tate's algorithm \cite{Tate, Bershadsky:1996nh,Katz:2011qp} allows a systematic determination of the singular fibers in codimension one in the base of the fibration. The algorithm is based upon successively determining the conditions for the vanishing of the discriminant in the coordinate $z$ in the base. The coordinate ring in a sufficiently small neighborhood on the divisor $z=0$ in the base is a unique factorization domain (UFD) \cite{MR0103906}. Tate's algorithm proceeds then by solving the conditions $\Delta=0$ order by order in an expansion in $z$ over a UFD. 
In the process the Weierstrass form can be brought into the so-called Tate form
\be
  y^2  + b_1 x y + b_3 y = x^3 + b_2 x^2 + b_4 x + b_6 \,,
\ee
where the coefficients $b_i$ are sections of suitable line bundles, and have an expansion in powers of $z$  that characterize the singular fibers. We will refer to a Tate form as  {\it canonical}, if it is characterized solely by the vanishing orders of the  coefficients $b_i$. 
As was shown in \cite{Katz:2011qp}, most Weierstrass forms in $\mathbb{P}^{(1,2,3)}$ can be  locally put into (canonical) Tate forms, albeit there exist outliner cases, which cannot be reached without allowing for divisions, in which case only generalized Tate forms can be achieved locally. These {\it non-canononical} forms are not specified solely by a vanishing order of the coefficients, but require non-trivially relation among the coeffficients $b_i$, which cannot be removed by well-defined coordinate changes like shifts. 

The goal of this paper is to apply Tate's algorithm in the context of elliptic fibrations with a rank one Mordell-Weil group, and determine Tate-like forms for these models realized in terms of a quartic equation (\ref{QuarticSimple}) in  $\bbP^{(1,1,2)}$. 

We will find, that unlike in $\mathbb{P}^{(1,2,3)}$, non-canonicality of the models is quite generic, i.e., the vanishing orders alone will not determine the complete set of forms of a given fiber type. In addition to the codimension one fiber type we also analyze the possible enhancements in codimension two and three, which will depend on the position of the two sections on the fiber in codimension one. 
The fibers will be characterized by the following data:
\begin{itemize}
  \item Kodaira fiber type in codimension one
  \item Location of sections $\sigma_0$ and $\sigma_1$ on the fiber
  \item Singular fiber type in codimension two
\end{itemize}
For each Kodaira fiber type there is an additional choice of position of the sections $\sigma_i$. This leads to a tree-like structure of the algorithm even when truncating it to one type of  Kodaira fiber e.g. $I_n$. We will refer to these as {\it Tate trees}, and the first few branches  for $\mathbb{P}^{(1,1,2)}$ are shown in figure \ref{fig:TheTree}.

To keep track the sections, it is useful to characterize the fibers by their Kodaira type with an additional superscript that encodes the separation of the two sections $\sigma_i$ : 
$I_n$ fibers (i.e. $\mathbb{P}^1$s intersecting in an affine $A$-type Dynkin diagram) will be labeled by $I_n^{(0||\cdots | | 1)}$with $k$ separations $|$ between $0$ and $1$ corresponding to $\sigma_0$ and $\sigma_1$ intersecting $\mathbb{P}^1$s which are separated by $k-1$ $\mathbb{P}^1$s, e.g. $I_n^{(0|1)}$ if the sections intersect nearest neighbor $\mathbb{P}^1$s or  $I_n^{(0||1)}$ for next to nearest neighbors. Subscripts ${nc}$ denote non-canonical forms. 


We will show in section \ref{sec:Constr}, that the sections $\sigma_i$ can only intersect components of Kodaira fibers in codimension one, with multiplicity one. The location of the sections for $I_n^*$ fiber, which has the structure of a $D$ type affine Dynkin diagram,  the sections can only be on the four end-nodes (which are the only fiber components with multiplicity one), and  modulo symmetries of the diagram, there are  three distinct types of fibers $I_n^{*(01)}$, $I_n^{*(0|1)}$, and $I_n^{*(0||1)}$, shown in figure \ref{fig:In*}. Similar restrictions apply for the type $II^*, III^*, IV^*$ fibers. 

Note that the codimension two fibers also have an interpretation in terms of representations of the associated Lie algebra of the codimension one fiber, and the distribution of the sections corresponds in this context to different $U(1)$ charge assigements to the representation. {This will play a key role in the application to F-theory model building, where the codimension two fiber type determines the matter, and the intersection with the (Shioda mapped) sections corresponds to the $U(1)$ charges of the matter.} The existence of additional sections also plays a key role in the possible topologically inequivalent resolutions of the singular fibers in higher codimension as discussed in \cite{Hayashi:2014kca}. 

Tate's algorithm applied to the quartic in $\mathbb{P}^{(1,1,2)}$ will result in multiple Tate-type forms for each fiber type. For canonical models, i.e. those characterized in terms of simple vanishing orders of the sections $\mathfrak{c}_i$ and $\mathfrak{b}_j$ in the local coordinate $z$ in the base, which characterizes a component of the discriminant, it is useful to write the power series expansion  
\begin{equation}
  \fkc_i = \sum_{j} c_{i,j} z^j \,,
\end{equation}
which for canonical models with certain higher vanishing orders in $z$ will be specialized to
\begin{equation}
\fkc_{i,j} z^j = \sum_{k=j}^\infty c_{i,k} z^{k} = c_{i, j} z^j + c_{i, j+1} z^{j+1} + \cdots\,,
\end{equation}
i.e., a truncated power series, starting with the terms $z^j$, and $\fkc_{i,j}= c_{i, j} + z c_{i, j+1} + \cdots$. For models that are characterized in terms of vanishing orders alone, i.e. models that we refer to as {\it canonical models}, we will use the shorthand notation
\begin{equation}
  \begin{aligned}
    \CQ(i_1,i_2,i_3,i_4,i_5,i_6,i_7): \qquad & \fkc_{0,i_1} z^{i_1} w^4 s^3 + \fkc_{1,i_2} z^{i_2} w^3 s^2 x + \fkc_{2,i_3} z^{i_3} w^2 s x^2 + \fkc_{3,i_4} z^{i_4} w x^3 \\
    &= y^2 s + \fkb_{0,i_5} z^{i_5} y x^2 +  \fkb_{1,i_6} z^{i_6} y w s x + \fkb_{2,i_7} z^{i_7} y w^2 s^2 \,.
    \label{eq:CanonicalDefinition}
  \end{aligned}
\end{equation}
In many instances, Tate's algorithm will run into local obstruction in reaching a canonical form\footnote{There are potentially global obstructions as pointed out in \cite{Katz:2011qp}, which will depend on the base of the fibration. Local obstructions refer to changes of coordinates that would require divisions by sections that can vanish along $z=0$. The type of coordinate changes that we will allow should be locally well-defined in this sense. },
so-called {\it non-canonical models}, in which case there are relations among the leading-order coefficients. 
Such relations between coefficients are typically described by the vanishing of a polynomial $P$ in these coefficients, and we will therefore denote the corresponding non-canonical form as $\CQ(i_1,i_2,i_3,i_4,i_5,i_6,i_7)|_P$. A vanishing order of $\infty$ indicates that the term is completely absent from the fibration.


\subsection{Constraints on Sections}
\label{sec:Constr}

The  sections $\sigma_i$ can only intersect the multiplicity one components of the  Kodaira fibers. To see this, first note that 
in the quartic in $\mathbb{P}^{(1,1,2)}$, 
the sections are on equal footing, and only by mapping to a Weierstrass model, do we single out one of the sections as the origin of the elliptic curve, e.g. $\sigma_0$  in (\ref{eq:WstrMap}). There is a symmetry that exchanges the two sections, and we can construct a Weierstrass model, with origin given by $\sigma_1$:  blow-down  $s=0$, 
after which there is a holomorphic coordinate shift that exchanges the sections 
\be
\sigma_0 \leftrightarrow \sigma_1   \qquad\Longleftrightarrow \qquad  y \rightarrow y \pm \fkb_0 x^2 \,.
\ee
Then $\sigma_1$ now has projective coordinates $[0{\,:\,}1{\,:\,}0]$, and will be mapped to the zero section under (\ref{eq:WstrMap}). 
If the fibration under consideration is singular, one can find a birational map between its desingularization and a smooth Weierstrass model with either $\sigma_0$ or $\sigma_1$ as origin by passing to the singular model, mapping to the Weierstrass model with either $\sigma_0$ or $\sigma_1$ chosen as the origin, and resolving the singular Weierstrass model.

As the intersection of the section with every fiber equals one, a section can only meet a component of the fiber that 
 has multiplicity one \cite{Miranda}. 
In terms of the intersections of $\sigma_i$ with the fiber components, this means that they can only meet the multiplicity one components of the resolved Kodaira fibers in codimension 1. 
These are exactly the nodes  that are in the orbit of the affine node under an outer automorphism of the affine Dynkin diagram, which is the dual graph to the Kodaira fiber.

This considerably restricts the position of the sections for Kodaira fibers  $I_n^*$, $IV^*$, $III^*$ and $II^*$, which have higher multiplicity fiber components, while posing no constraint on the $I_n$ fibers. All distributions of sections on the fibers consistent with this restriction arise in Tate's algorithm.  
The $I_n^*$ fibers consistent with this restriction are shown in figure \ref{fig:In*}, where  the sections can be distributed over the four multiplicity one fibers of the affine $D_n$ Dynkin diagram, and figures \ref{fig:IV*}, \ref{fig:III*} and \ref{fig:II*} for 
$IV^*$, $III^*$ and $II^*$, respectively. 

\subsection{Starting points for Tate's algorithm}
\label{sec:Order0}

Instead of directly solving the rather complicated leading order term in the discriminant (\ref{eq:Delta}), we will determine where the fiber is singular, by considering the loci where the tangent space becomes degenerate to leading order in the $z$ expansion. 
This will be done in local affine coordinates by covering $\mathrm{Bl}_{[0,1,0]}\bbP^{(1,1,2)}$ with open patches. 

This analysis is rather technical, so we first summarize the result: 
There are three distinct starting point fibrations
 \be\label{StartSum}
 \ba
   I_1^{(01)}: &\qquad \CQ(1,1,0,0,0,0,1)\cr
  \big[ I_2^{(01)}: &\qquad \CQ(0,0,1,1,1,0,0)\big]\cr 
   I_2^{(0|1)}: &\qquad \CQ(1,1,1,1,0,0,0) \,.
\ea
\ee 
Of these $I_1^{(01)}$ and $I_2^{(01)}$ are contained within a single affine patch, whereas $I_2^{(0|1)}$ is not. 
In the next section we will show that the two fibers with zero separation between the sections $I_1^{(01)}$ and $I_2^{(01)}$ are in fact related, so that effectively there are only two starting points $I_1^{(01)}$, which gives rise to the Tate tree for $I_n$ and $I_n^*$, and $I_2^{(0|1)}$, which generates the $I_{2m}^{ns (0|1)}$ part of the tree. 

We will now derive these results. 
A complete set of patches for $\mathrm{Bl}_{[0,1,0]}\bbP^{(1,1,2)}$ is given by \footnote{The patches are characterized by the non-vanishing of certain coordinates, which we  then locally set to one.}:
\begin{equation}
  \begin{tabular}{c|c}
    Coordinate patch & Affine coordinates \\ \hline
    $w=s=1$ & $x$, $y$ \\
    $w=x=1$ & $s$, $y$ \\
    $y=s=1$ & $w$, $x$ \\
    $y=x=1$ & $s$, $w$
  \end{tabular}
  \label{eq:Patches}
\end{equation}
First consider the patch $w=s=1$. 
Assume the elliptic fiber over $z=0$ admits a singularity in this patch at a point $(x_0,y_0)$. Then the equations describing the quartic and its derivatives with respect to $x$ and $y$ have to vanish. Explicitly,
\begin{equation}
  \begin{aligned}
    0 &= \,\,\,\,\,\, \CQ|_{z=0} = - c_{0,0}+y_0^2 + x_0^2 y_0 b_{0,0} + x_0 y_0 b_{1,0} + y_0 b_{2,0}  - x_0 c_{1,0} - x_0^2 c_{2,0} - x_0^3 c_{3,0} \\
    0 &= \partial_x \CQ|_{z=0} =- c_{1,0}+  2 x_0 y_0 b_{0,0} + y_0 b_{1,0}  - 2 x_0 c_{2,0} - 3 x_0^2 c_{3,0} \\
    0 &= \partial_y \CQ|_{z=0} =  b_{2,0} +2 y_0 + x_0^2 b_{0,0} + x_0 b_{1,0} \,.
    \label{eq:DervsInSWpatch}
  \end{aligned}
\end{equation}
Solving these equations for $c_{0,0}$, $c_{1,0}$ and $b_{2,0}$ indeed yields a discriminant vanishing to $O(z)$. Furthermore, one can perform a coordinate shift
\begin{equation}
  \left( \begin{array}{c} x \\ y \end{array} \right) \rightarrow \left( \begin{array}{c} x - x_0 s w \\ y  -y_0 s w^2 \end{array} \right) \,,
\end{equation}
to put the singularity at the origin of the $w=s=1$ patch. There, the quartic and its derivatives read
\begin{equation}
  \begin{aligned}
    \CQ|_{x=y=z=0} &= -c_{0,0} \\
    \partial_x \CQ|_{x=y=z=0} &= -c_{1,0} \\
    \partial_y \CQ|_{x=y=z=0} &= b_{2,0} \,.
  \end{aligned}
\end{equation}
Thus, having a singularity in the fiber over $z=0$ in $w=s=1$ is, after a coordinate shift, equivalent to having a fiber with $c_{0,0}=c_{1,0}=b_{2,0}=0$ that is otherwise generic. These conditions also solve the zeroth-order term of the discriminant. The canonical form for such an $I_1$ fiber is
\begin{equation}
    \CQ_{I_1}: \quad \fkc_{0,1} z w^4 s^3 + \fkc_{1,1} z  w^3 s^2 x + \fkc_2 w^2 s x^2 + \fkc_3 w x^3 = s y^2 + \fkb_0 x^2 y + \fkb_1 s w x y + \fkb_{2,1} s^2 w^2 y \,,
\end{equation}
or equivalently
\begin{equation}
 I_1^{(01)}:\qquad  \CQ(1,1,0,0,0,0,1) \,.
  \label{eq:QI1}
\end{equation}
Its discriminant at leading order reads
\begin{equation}
  \Delta_{I_1} = c_{0,1} \left( b_{1,0}^2 + 4 c_{2,0} \right)^3 \left( b_{0,0}^2 c_{2,0} - b_{0,0} b_{1,0} c_{3,0} - c_{3,0}^2 \right) z + O(z^2) \,.
\end{equation}
Next consider the patch $w=x=1$, and assume a singularity at $(s_0,y_0)$. The equations for the quartic and its $s$- and $y$-derivatives are
\begin{equation}
  \begin{aligned}
    0 = \,\,\,\,\,\, \CQ|_{z=0} &= s_0 y_0^2 + y_0 b_{0,0} + s_0 y_0 b_{1,0} + s_0^2 y_0 b_{2,0} - s_0^3 c_{0,0} - s_0^2 c_{1,0} - s_0 c_{2,0} - c_{3,0} \\
    0 = \partial_s \CQ|_{z=0} &= y_0^2 + y_0 b_{1,0} + 2 s_0 y_0 b_{2,0} - 3 s_0^2 c_{0,0} - 2 s_0 c_{1,0} - c_{2,0} \\
    0 = \partial_y \CQ|_{z=0} &= 2 s_0 y_0 + b_{0,0} + s_0 b_{1,0} + s_0^2 b_{2,0} \,.
  \end{aligned}
\end{equation}
Solving for $b_{0,0}$, $c_{2,0}$ and $c_{3,0}$ and inserting into the discriminant, one finds a vanishing at leading order. Note that any singularity in this patch will also be in the patch $w=s=1$, unless it also is on $s=0$, i.e., it has inhomogeneous coordinates $(s,y)=(0,y_0)$. By the coordinate shift $y\rightarrow y - y_0 w x$ any such singularity is moved to the origin of the $w=x=1$ patch. There, the derivative conditions read
\begin{equation}
  \begin{aligned}
    \CQ|_{s=y=z=0} &= -c_{3,0} \\
    \partial_s \CQ|_{s=y=z=0} &= -c_{2,0} \\
    \partial_y \CQ|_{s=y=z=0} &= b_{0,0} \,.
    \label{eq:DervsInXWpatch}
  \end{aligned}
\end{equation}
Any singular fibration in this patch that is not also in the $s=w=1$ patch can therefore be brought into the form
\begin{equation}
 I_2^{(01)}:\qquad  \CQ(0,0,1,1,1,0,0) \,.
\end{equation}
Note furthermore that over the locus $z=0$, this fiber splits into two components:
\begin{equation}
  \CQ(0,0,1,1,1,0,0)|_{z=0} = s \left( y^2 + b_{1,0} w x y + b_{2,0} s w^2 y - c_{0,0} s^2 w^4 - c_{1,0} s w^3 x \right) \,. \\
\end{equation}
Since they intersect in the two points $s=y=0$ and $s=y+b_{1,0}w x=0$, the Kodaira fiber type of this model is $I_2$. After a blow-up of the form $(s,z;\zeta_1)$ in the notation of \cite{Lawrie:2012gg}, see also  section \ref{sec:Fibration},
one finds that the two components are given by $z=0$ and $\zeta_1=0$ in the proper transform of $\CQ(0,0,1,1,1,0,0)$, respectively. Both sections $\sigma_0$ and $\sigma_1$ intersect the same fiber component $\zeta_1=0$, so that the fiber type is $I_2^{(01)}$. The leading-order discriminant of this model is order $z^2$
\begin{equation}
  \Delta_{I_2^{(01)}} = b_{1,0}^4 c_{3,1} \left( c_{3,1} + b_{1,0} b_{0,1} \right) \left( b_{1,0}^2 c_{0,0} - b_{1,0} b_{2,0} c_{1,0} - c_{1,0}^2 \right) z^2 + O(z^3)\,.
  \label{eq:DeltaI2x}
\end{equation}
In the third coordinate patch $y=s=1$, the conditions on the quartic and its derivatives for an assumed singularity at $(w_0,x_0)$ are
\begin{equation}
  \begin{aligned}
    0 = \,\,\,\,\,\, \CQ|_{z=0} &= 1 + x_0^2 b_{0,0} + w_0 x_0 b_{1,0} + w_0^2 b_{2,0} - w_0^4 c_{0,0} - w_0^3 x_0 c_{1,0} - w_0^2 x_0^2 c_{2,0} - w_0 x_0^3 c_{3,0} \\
    0 = \partial_w \CQ|_{z=0} &= x_0 b_{1,0} + 2 w_0 b_{2,0} - 4 w_0^3 c_{0,0} - 3 w_0^2 x_0 c_{1,0} - 2 w_0 x_0^2 c_{2,0} - x_0^3 c_{3,0} \\
    0 = \partial_x \CQ|_{z=0} &= 2 x_0 b_{0,0} + w_0 b_{1,0} - w_0^3 c_{1,0} - 2 w_0^2 x_0 c_{2,0} - 3 w_0 x_0^2 c_{3,0} \,.
  \end{aligned}
\end{equation}
There are no solutions of these equations in the coefficients $b_{i,0}$, $c_{i,0}$ that hold for any point $(w_0,x_0)$ in this patch. However, the only locus in the third coordinate patch that is not in $w=s=1$ is the $w=0$ locus. Here, the $x$-derivative of the quartic equation is given by
\begin{equation}
  \partial_x \CQ|_{w=z=0} = 2 b_{0,0} x_0 \,,
\end{equation}
and a singularity at this locus hence requires $b_{0,0} x_0=0$. Then, however,
\begin{equation}
  \CQ|_{w=z=0} = s y^2 + b_{0,0} x_0^2 = s y^2 = 1 \,,
\end{equation}
and thus $\CQ$ can never vanish there, no matter how the $\fkb_i$ and $\fkc_i$ are chosen. Therefore, any singularity of the fiber in the $y=s=1$ patch is also contained in either $w=s=1$ or $w=x=1$, and can therefore be described by the standard forms found above.

Lastly, the only remaining locus in the $x=y=1$ patch that is not contained in either of the patches is $(s,w)=(0,0)$. Here, the $s$-derivative of $\CQ$ cannot vanish, since 
\begin{equation}
  \partial_s  \CQ|_{s=w=z=0} = y^2 = 1 \,,
\end{equation}
and the fiber will always be regular over this point.

There is a third starting fibration not covered by the analysis performed until here, because the singularity of this fibration is not contained within a single patch of the ambient $\bbP^{(1,1,2)}$ over the entire codimension one locus $z=0$ in the base $B$:
Consider again the derivatives of the elliptic fibration (\ref{eq:DervsInSWpatch}) in the patch $w=s=1$. If $c_{0,0}=c_{1,0}=c_{2,0}=c_{3,0}=0$ and for a generic point $b$ on the base with non-vanishing values of $b_{0,0}$ or $b_{1,0}$, there exists an $x_0(b)$ such that all derivatives vanish at $x=x_0(b)$, $y=0$. This is not the case on any point $b$ on $B$ where $b_{0,0}(b)=b_{1,0}(b)=0$. However, on such a point, the fibration is singular in the $w=x=1$ patch, on the locus $y=s=0$. This can be seen explicitly from the derivatives (\ref{eq:DervsInXWpatch}) in this patch. This  results in the third starting point fiber, which is an $I_2^{(0|1)}$ 
\be\label{ExtraStart}
I_2^{(0|1)}:\qquad \CQ(1,1,1,1,0,0,0) \,,
\ee
which is singular on the entire locus $z=0$, although its singularity is not contained in a single patch over $z=0$.
This starting point will generate the infinite series of $I_{2m}^{ns(0|1)}$ fibers, as is shown in section \ref{sec:In}.


\subsection{Symmetries and Pruning of the Tree}
\label{sec:Sym}

In the last section we have seen that there are  {three} starting points for Tate's algorithm in $\bbP^{(1,1,2)}${, two of which are contained in a single patch of $\bbP^{(1,1,2)}$}. We will show that two of these are related
\be\label{SymPrun}
I_1^{(01)}: \quad  \CQ(1,1,0,0,0,0,1)  \qquad \leftrightarrow \qquad I_2^{(01)} :\quad \CQ(0,0,1,1,1,0,0) \,.
\ee
This implies that there is a single $I_1^{(01)}$ starting point for the algorithm, giving rise to the $I_n$ and $I_n^*$ part of the Tate tree, and a second starting point (\ref{ExtraStart}) with fiber type $I_2^{(0|1)}$, which enhances to the $I_{2m}^{ns(0|1)}$ part of the tree.

{To show how the fibers in (\ref{SymPrun}) are related}, we will use the fact that there is an exchange of the two sections $\sigma_0$ and $\sigma_1$, which maps these two fibrations into each other. For this symmetry to be manifest, we blow down
 the divisor $s=0$, whereby the coordinate shift $y\rightarrow y - \frac{1}{2} \fkb_0 x^2 - \frac{1}{2} \fkb_2 w^2$ becomes holomorphic. After applying this shift the quartic takes the form
\begin{equation}
  \fkc_0 w^4 + \fkc_1 w^3 x + \fkc_2 w^2 x^2 + \fkc_3 w x^3 + \fkb_0^2 x^4 = y^2 + \fkb_1 w x y \,.
  \label{eq:shiftedfib}
\end{equation}
The sections are now at $[0{\,:\,}1{\,:}\pm \fkb_0]$. One can again analyze whether this fibration is singular in the three affine coordinate patches of $\bbP^{(1,1,2)}$, given by $w=1$, $x=1$ and $y=1$. Indeed one recovers the $I_1$ singularity in $w=1$ with conditions $c_{0,0}=c_{1,0}=0$. It has discriminant
\begin{equation}
  \Delta_{I_1} = c_{0,1} \left( b_{1,0}^2 + 4 c_{2,0} \right)^3 \left( b_{0,0}^2 c_{2,0} - b_{0,0} b_{1,0} c_{3,0} - c_{3,0}^2 \right) z + O(z^2) \,,
\end{equation}
and setting $c_{0,1}=0$ enhances it into an $I_2^{(01)}$ fibration, given by
\begin{equation}
  \fkc_{0,2} z^2 w^4 + \fkc_{1,1} z w^3 x + \fkc_2 w^2 x^2 + \fkc_3 w x^3 + \fkb_0^2 x^4 = y^2 + \fkb_1 w x y \,.
  \label{eq:shifteddegeni1}
\end{equation}
In the $x=1$ patch, one again finds the $I_2^{(01)}$ singularity with $c_{3,0}=b_{0,0}=0$. Since $\fkb_0$ appears as a square in (\ref{eq:shiftedfib}), $b_{0,0}=0$ implies that the coefficient of the $x^4$-term vanishes to second order in $z$, and one has
\begin{equation}
  \fkc_0 w^4 + \fkc_1 w^3 x + \fkc_2 w^2 x^2 + \fkc_{3,1} z w x^3 + \fkb_{0,1}^2 z^2 x^4 = y^2 + \fkb_1 w x y \,.
  \label{eq:shiftedi2}
\end{equation}
Now, if one interchanges $x \leftrightarrow w$ in (\ref{eq:shiftedi2}), one recovers (\ref{eq:shifteddegeni1}). This symmetry thus relates the two singular fibrations. It also explains why there is no $I_1$ underlying (\ref{eq:shiftedi2}): $\fkb_{0,1}^2$ simply cannot vanish to linear order in $z$.

The quartic (\ref{eq:shifteddegeni1}) has (full, not leading order) discriminant
\begin{equation}
  \begin{aligned}
    \Delta_{(\ref{eq:shifteddegeni1})} &= 256 z^2 \bigg( P_0^2 \left( P_0 \fkc_{0,2} - \fkc_{1,1}^2 \right) \left( P_0 \fkb_0^2 - \fkc_3^2 \right) \\ 
    &- 8 \fkc_{1,1} \fkc_3 z \left( 8 \fkc_{1,1}^2 \fkc_3^2 - 9 P_0 \left( \fkc_{0,2} \fkc_3^2 + \fkb_0^2 \fkc_{1,1}^2 \right) + 10 P_0 \fkc_{0,2} \fkb_0^2 \right) \\
    &- 16 z^2 \left( 27 \left( \fkc_{0,2}^2 \fkc_3^4 + \fkb_0^4 \fkc_{1,1}^4 \right) - 36 \fkc_{0,2} \fkb_0^2 P_0 \left( \fkc_{0,2} \fkc_3^2 - \fkc_{1,1}^2 \fkb_0^2 \right) + 2 \fkc_{0,2} \fkb_0^2 \left( 3 \fkc_{1,1}^2 \fkc_3^2 + 4 P_0 \fkc_{0,2} \fkb_0^2 \right) \right) \\
    &- 3072 z^3 \fkc_{0,2}^2 \fkc_{1,1} \fkc_3 \fkb_0^4 + 4096 z^4 \fkc_{0,2}^3 \fkb_0^6 \bigg) \,.
  \end{aligned}
\end{equation}
with $P_0 = \fkb_1^2 + 4 \fkc_2$. 
One can easily check that it is invariant under the exchange of $w \leftrightarrow x$, which amounts to the interchanges $\fkc_{0,2} \leftrightarrow \fkb_0^2$ and $\fkc_{1,1} \leftrightarrow \fkc_3$. On the other hand, the discriminant of (\ref{eq:shiftedi2}) is identical to the discriminant of (\ref{eq:shifteddegeni1}) if one replaces $\fkc_{0,2}$ by $\fkc_0$, $\fkc_{1,1}$ by $\fkc_1$, $\fkc_3$ by $\fkc_{3,1}$ and $\fkb_0$ by $\fkb_{0,1}$. The discriminants of two quartics are structurally identical, so that they have the same enhancements. It suffices therefore to consider Tate's algorithm only for enhancements of either (\ref{eq:shifteddegeni1}) or (\ref{eq:shiftedi2}). 


\subsection{Lops}
\label{sec:Lop}

With the arguments in the last section, we can concentrate on the branch of the Tate tree, that starts from the $I_1^{(01)}$ fiber in 
(\ref{StartSum})\footnote{The additional $I_2^{(0|1)}$ has very simple enhancements and we discuss it separately in section \ref{sec:In}.}, realized in terms of $\mathcal{Q}(1,1,0,0,0,0,1)$. 
In this section, we will show that there is an additional symmetries, which we call {\it lops or lopping transformations}\footnote{Lops are arboricultural operations on trees. Lopping refers to the removal of large side branches (the making of vertical cuts).[Arboricultural Association]} (in analogy to flops)  that identify different branches of the tree.  In summary we show that the following two $I_2$ models are equivalent
\be\label{WeirdMap0}
\mathcal{Q}(2,1,1,0,0,0,1)  \equiv 
\mathcal{Q}(0,0,1,1,1,0,0) \,,
\ee
i.e. they correspond to the same fiber type, 
and more generally, for non-negative vanishing orders $n_i$ and $m_i$
\begin{equation}
  \label{WeirdMap}
  \mathcal{Q}(n_0+2,  n_1+1,  n_2, n_3 , m_0, m_1, m_2+1)  \equiv 
  \mathcal{Q}(n_0 ,  n_1 ,  n_2, n_3+1 , m_0+ 1, m_1, m_2) \,,
\end{equation}
Here equivalence here means isomorphism of the fiber, which implies that the fiber types of the two models are identical in all codimension. 

We can considerably trim the Tate tree that starts at $I_1^{(01)}$ by successive application of the lops. One important implication is that without loss of generality the vanishing order of the  coefficient $\fkb_0$ can always be set to zero, i.e. 
\be
\fkb_0 = b_{0,0} + b_{0,1} z + \cdots \,, \qquad b_{0,0} \not=0\,.
\ee
This is similar to the specialization in fibrations realized in $\mathbb{P}^{(1,2,3)}$, where the coefficients of $y^2$ and $x^3$ have been set to be one. 
Note that the lopping operation does not restrict the branch growing out of the  second starting point, the $I_2^{(0|1)}$ model
$\CQ(1,1,1,1,0,0,1)$.\footnote{Applying the same type of arguments as in the following for the $I_1^{(01)}$ branch,  after the proper transform a term $y^2 \zeta_1$ in introduced, and thus is 
already reduced.}

\smallskip

We will now prove that these lops are equivalences of the fibers.
First consider the two $I_2^{(01)}$ models (\ref{WeirdMap0}).
Resolving the model on the left with $(x, y, z; \zeta_1)$ results, after the proper transform, in 
\be\label{I2res1}
z^2 \fkc_{0,2} w^4 + z \fkc_{1,1} w^3 x +  z\zeta_1  \fkc_{2, 1} w^2  x^2 + \zeta_1  \fkc_3 w x^3 
=  y^2  + \zeta_1  \fkb_0 x^2 y + \fkb_1 y w  x + z  \fkb_{2, 1} w^2  y  \,.
\ee
where in $\fkc$ and $\fkb$ the exansions are now in terms of $z\zeta_1$. 
Likewise, the resolution of the model on the right of (\ref{WeirdMap0}) with $(w, y, \tilde{z}; \tilde\zeta_1)$ results in 
\be\label{I2res2}
\tilde{\zeta}_1^2 \fkc_{0} w^4 + \tilde{\zeta}_1 \fkc_{1} w^3 x +   \tilde{\zeta}_1 \tilde{z} \fkc_{2, 1} w^2 x^2
+ \tilde{z} \fkc_{3, 1} w x^3 
=  y^2 + \tilde{z}\fkb_{0, 1} x^2 y + \fkb_1 y w  x +  \tilde{\zeta}_1\fkb_{2} w^2  y  \,.
\ee
Again each of the coefficient sections are now series in $\tilde{z}\tilde{\zeta}_1$. 
Comparing the two resolved equations, we see that indeed, swapping 
\be\label{zzetaEx}
\tilde{z} \quad \leftrightarrow \quad \zeta_1 \qquad\hbox{and}\qquad  \tilde{\zeta}_1 \quad \leftrightarrow \quad z 
\ee
maps (\ref{I2res1}) and (\ref{I2res2}) into each other. 
Furthermore, from the projective relations of the blow-up $\mathcal{Q}(2,1,1,0,0,0,1)$, we see that in (\ref{I2res1}) the two sections sit on $z=0$, 
and in (\ref{I2res2}) on $\tilde{\zeta}_1=0$, which exactly are mapped into each other. The birational map between these
two forms is thus, to first resolve as in (\ref{I2res1}), and then blow-down $z=0$,\footnote{More detailed studies of when such blow-downs exist are discussed in \cite{BSSN}.} which is precisely realized in terms of the singular model $\mathcal{Q}(0,0,1,1,1,0,0)$. 

More generally, consider the quartic, without the blow-up with respect to $s$. We will now show that there is a symmetry between models, whose vanishing orders differ by the vector $(2,1,0,-1,-1,0,1)$, i.e. the lopping transformation (\ref{WeirdMap}). To prove this, consider the left hand side 
\be
\ba
&\mathcal{Q}(n_0+2,  n_1+1,  n_2, n_3 , m_0, m_1, m_2+1) :\cr
&\qquad \qquad \fkc_{0, n_0+2}{z}^{n_0+2} w^4 + \fkc_{1, n_1+1} {z}^{n_1+1} w^3  x 
+ \fkc_{2, n_2} {z}^{n_2} w^2  x^2 + \fkc_{3, n_3} {z}^{n_3} w x^3 \cr
&\qquad \qquad = y^2  + \fkb_{0, m_0} {z}^{m_0}  x^2 y + \fkb_{1, m_1} {z}^{m_1} y w  x + \fkb_{2, m_2+1}{z}^{m_2+1} w^2  y \,.
\ea
\ee
Then applying one big resolution
\be
(x, y, z ; \zeta_1) 
\ee
results, after the proper transform, in 
\be
\ba
& 
z^2\,  \fkc_{0, n_0+2}(z \zeta_1)^{n_0} w^4 
+ z\,  \fkc_{1, n_1+1} ({z}\zeta_1)^{n_1} w^3  x 
+ \fkc_{2, n_2} ({z}\zeta_1)^{n_2} w^2  x^2 
+  \zeta_1 \, \fkc_{3, n_3} ({z}\zeta_1)^{n_3} w x^3 \cr
&\qquad  
= y^2  
+ \zeta_1\ \fkb_{0, m_0} ({z}\zeta_1)^{m_0}  x^2 y 
+ \fkb_{1, m_1} ({z}\zeta_1)^{m_1} y w  x 
+  z\, \fkb_{2, m_2+1}({z}\zeta_1)^{m_2} w^2  y \,.
\ea
\ee
On the other hand, resolving the right hand side of (\ref{WeirdMap}), denoting the component of the discriminant by $\tilde{z}$ with
\be
(w, y, \tilde{z}; \tilde\zeta_1) 
\ee
yields after the proper transform 
\be
\ba
& \tilde\zeta_1^2 \, \fkc_{0, n_0}(\tilde{z}\tilde\zeta_1)^{n_0} w^4 
+ \tilde\zeta_1\, \fkc_{1, n_1} (\tilde{z}\tilde\zeta_1)^{n_1} w^3  x 
+ \fkc_{2, n_2} (\tilde{z}\tilde\zeta_1)^{n_2} w^2  x^2 
+   \tilde{z} \, \fkc_{3, n_3+1} (\tilde{z}\tilde\zeta_1)^{n_3} w x^3 \cr
&\qquad  
= y^2  
+ \tilde{z}\,  \fkb_{0, m_0+1} (\tilde{z}\tilde\zeta_1)^{m_0}  x^2 y 
+ \fkb_{1, m_1} (\tilde{z}\tilde\zeta_1)^{m_1} y w  x 
+  \tilde{\zeta}_1 \, \fkb_{2, m_2}(\tilde{z}\tilde\zeta_1)^{m_2} w^2  y \,.
\ea
\ee
Again, the lop transformation (\ref{zzetaEx}) applied to these partially resolved 
elliptic fibrations, is a symmetry, and maps the fiber component that intersects both sections,
into each other.


\subsection{Resolutions of singular elliptic fibrations}
\label{sec:Fibration}

To determine each fiber type, including the separation of the two section, in the Tate tree, we need to resolve the fiber and compute intersections. In practice the computations in this paper were done using {\tt Smooth} \cite{Smooth}, where the algebraic resolution procedure and intersections are implemented for singular (elliptic) fibrations. Algebraic resolutions of the singularities of elliptic fibrations, including, the higher codmension structure of the fibers, realized in $\mathbb{P}^{(1,2,3)}$ have been discussed in \cite{MS, EY, Lawrie:2012gg, Hayashi:2013lra, BSSN}. 
We consider crepant resolutions, and allow for up to codimension 3 fibers, i.e. the base of the fibration can be up to three-dimensional. The geometric setting thereby allows not only the analysis of the codimension one fibers, but also the higher codimension structure, which has an intricate pattern depending on the location of the sections in codimension one.
This is mostly motivated by model building in F-theory, where the relevant geometries are elliptic Calabi-Yau fourfolds with extra section.  
We now summarize the data determining the fibration, in terms of sections of line bundles of the base. 
The elliptic fibration is realized  in the ambient five-fold $X_5=\mathrm{Bl}_{[0,1,0]}\bbP^{(1,1,2)}(\CO \oplus \CO(\alpha) \oplus \CO(\beta))$ as a hypersurface
\begin{equation}
\CQ: \qquad y^2 s + \fkb_0 x^2 y + \fkb_1 y w s x + \fkb_2 y w^2 s^2 = \fkc_0 w^4s^3 + \fkc_1 w^3s^2 x + \fkc_2 w^2s  x^2 + \fkc_3 w x^3
\end{equation}
with
\begin{equation}
  \begin{tabular}{c|c}
    Section & Bundle \\ \hline
    $w$ & $\CO(\sigma - F)$ \\
    $x$ & $\CO(\sigma + \alpha)$ \\
    $y$ & $\CO(2 \sigma + \beta - F)$ \\
    $s$ & $\CO(F)$ \\
    $z$ & $\CO(S)$
  \end{tabular}
\end{equation}
Here, $\sigma$ is the section of the hyperplane class of $\bbP^{(1,1,2)}$ before blowing up at the point $[0{\,:\,}1{\,:\,}0]$, and $F$ is the section of the new exceptional $\bbP^1$ introduced by the blow-up. $\alpha$ and $\beta$ are two sections of line bundles on the base manifold $B$, which are related by $\beta = \alpha + c_1$ as shown below, where $c_1=c_1(B)$. $S$ is the divisor class of the singular surface $z=0$ in $B$. From the equation of $\CQ$ one infers the class of the four-fold to be
\begin{equation}
  [Y_4] = 4 \sigma + 2 \beta - F \,,
\end{equation}
and the $\fkb_i$ and $\fkc_i$ are sections of the following bundles
\begin{equation}
  \begin{tabular}{c|c}
    Section & Bundle \\ \hline
    $\fkb_i$ & $\CO(c_1 + (i-1) \alpha)$ \\
    $\fkc_i$ & $\CO(2 c_1 + (2-i) \alpha)$ \\
    $\fkb_{i,j}$ & $\CO(c_1 + (i-1) \alpha - j S)$ \\
    $b_{i,j}$ & $\CO(c_1 + (i-1) \alpha - j S)$ \\
    $\fkc_{i,j}$ & $\CO(2 c_1 + (2-i) \alpha - j S)$ \\
    $c_{i,j}$ & $\CO(2 c_1 + (2-i) \alpha - j S)$
  \end{tabular}
  \label{eq:bcSections}
\end{equation}
One then finds for the Chern class of $X_5$ that
\begin{equation}
  \begin{aligned}
    c(X_5) &= c(B) \cdot (1+[w]) \cdot (1+[x]) \cdot (1+[y]) \cdot (1+[s]) \big|_{X_5} \\
	   &= 1 + c_1 + 4 \sigma + \alpha + \beta - F + \cdots\,,
  \end{aligned}
\end{equation}
with the dots indicating higher-rank forms. By adjunction, the Chern class of $Y_4$ is given by
\begin{equation}
  c(Y_4) = \frac{c(X_5)}{1+[Y_4]}\bigg|_{Y_4}=1+c_1+\alpha-\beta+\cdots\,.
\end{equation}
The Calabi-Yau condition, which we shall impose in most practical applications to F-theory, $c_1(Y_4)=0$ thus restricts the possible choices of sections of line bundles $\alpha$ and $\beta$ by imposing the condition
\begin{equation}
  \beta = \alpha + c_1 \,.
\end{equation}
Furthermore, the second Chern class of $Y_4$ is
\begin{equation}
  c_2(Y_4) = c_2 + c_1^2 + \alpha^2 + 6 \alpha \sigma + 7 \sigma^2 - 2 F \left( \alpha + 2 \sigma \right) + c_1 \left( 3 \alpha + 7 \sigma - 2 F \right) \,.
\end{equation}
The projective relations 
\begin{equation}
  [s w{\,:\,}x{\,:\,}s y] \qquad \text{and} \qquad [w{\,:\,}y]
\end{equation}
imply the following relations in the intersection ring of $X_5$
\begin{equation}
  \begin{aligned}
    \sigma \cdot \left( \sigma + \alpha \right) \cdot \left( 2 \sigma + \alpha + c_1 \right) &= 0 \\
    \left( \sigma - F \right) \cdot \left( 2 \sigma + \alpha + c_1 - F \right) &= 0 \,.
  \end{aligned}
\end{equation}
Repeated applications of these -- and similar ones for exceptional divisors introduced by blowing up singularities -- allow us to compute intersections in $X_5$, similar to the computations for the standard Tate models in \cite{MS,Lawrie:2012gg}.
Furthermore, we will use the following notation for resolutions: a big resolution along $x=y=\zeta_0=0$ with the new exceptional section $\zeta_1$ will be denoted in the notation of \cite{Lawrie:2012gg} by 
\be
(x, y, \zeta_0; \zeta_1) \,.
\ee
Likewise a small resolution $y= x=0$ with $\delta$ is denoted by $(x, y; \delta)$. 

Finally, we should discuss the Mordell-Weil group, and how we compute the actual $U(1)$ charges of matter representations that are engineered in codimension 2. Recall that the Mordell-Weil group, since it is a finitely generated abelian group, can be written as
\begin{equation}
  \bbZ \oplus \dots \oplus \bbZ \oplus T \,,
\end{equation}
with the torsion subgroup $T$. Let $\{\sigma_1, \dots, \sigma_n\}$ be a set of rational sections generating the non-torsion part of the Mordell-Weil group. In \cite{Morrison:2012ei}, it was shown that the abelian vector fields $A_i$ of an F-theory vacuum are dual to the images $s(\sigma_i)$ of the rational sections $\sigma_i$ under the so-called Shioda map. The Shioda map is a map from the Mordell-Weil group to the homology group $H^{(1,1)}(Y_4)$ of the fourfold, and has been given and discussed e.g. in \cite{Park:2011ji}.
The  $U(1)$ charge, associated to the abelian gauge field $A_i$ with section $\sigma_i$, of any matter coming from a rational curve $C$ in the fiber is given by $C \cdot s(\sigma_i)$.

The Shioda map has the property that the intersection of $s(\sigma_i)$ with any Cartan divisor $D_{-\alpha_i}$ vanishes, i.e.,
\begin{equation}
  s(\sigma_i) \cdot D_{-\alpha_i} = 0 \,.
  \label{eq:ShiodaCartanDiv}
\end{equation}
Therefore, as one would expect, no vector multiplets are charged under $A_i$. Further, its intersection with any horizontal divisor $\pi^*(D_H)$ pulled back from a base divisor $D_H$ also vanishes
\begin{equation}
  s(\sigma_i) \cdot \pi^* D_H = 0 \,.
  \label{eq:ShiodaBaseDiv}
\end{equation}
To construct $s(\sigma)$ explicitly, we use (\ref{eq:ShiodaCartanDiv}) and (\ref{eq:ShiodaBaseDiv}) as a set of constraints on the Shioda map. This set is sufficient to fully specify $s(\sigma)$ up to redefinitons of the abelian fields $A_i$ that preserve charge minimality.


\section{Summary of Results}
\label{sec:Summary}

Tate's algorithm for Weierstrass forms in $\bbP^{(1,2,3)}$ with a 
given Kodaira singular fiber above $z=0$, derives alternative forms of the fibration, where the vanishing order of the coefficients around $z=0$ completely determines the fiber type.
We find that for  elliptic fibrations with additional rational section, similar Tate forms exist  however there are additional form, which are non-canonical, and are not determined fully  by the vanishing orders of the coefficients, but require non-trivial correlations among them. In applications to F-theory these open up interesting model building options.

The starting point of our analysis is the quartic (\ref{eq:Qpoly}) in $\mathbb{P}^{(1,1,2)}$. The codimension one fibers are characterized in terms of their Kodaira type and the separation of the two sections $\sigma_0$ and $\sigma_1$.  There is additional data, that distinguishes models in codimension two. Canonical and non-canonical models can have the same codimension one fiber, however they differ in the codimesion two fibers, and thus in terms of applications in F-theory, have different, charged matter content. E.g. canonical $I_n$ models have a single matter curve in the antisymmetric representation, non-canonical models will have several, with different $U(1)$ charges. 

\begin{landscape}
\begin{figure}
    \centering
    \includegraphics[width=20cm]{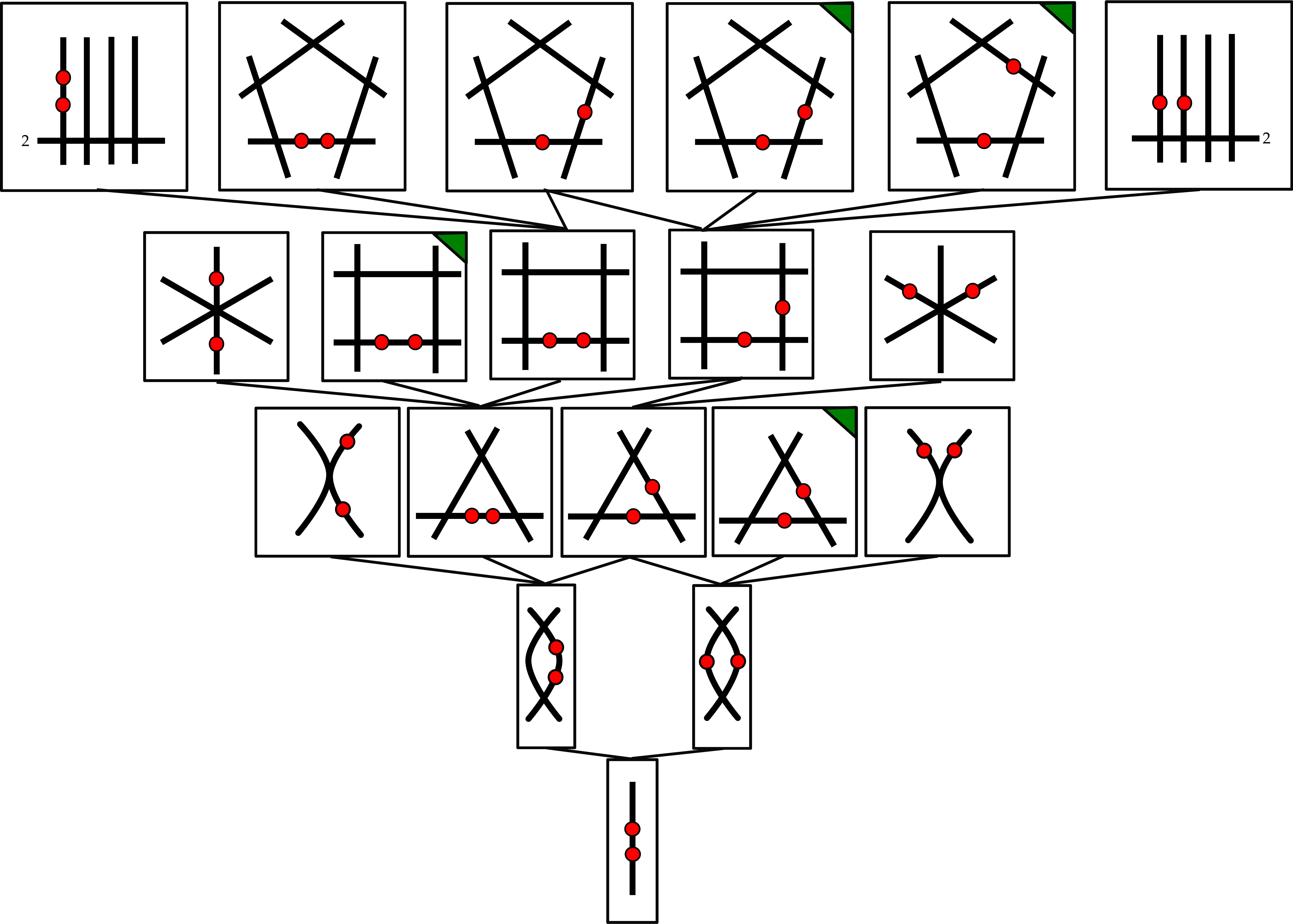}
    \caption{The Tate tree  for $\mathbb{P}^{(1,1,2)}$, up until and including the $O(z^5)$ discriminant fibers. Black lines are irreducible fiber components of the singular fibers, red nodes show where the sections intersect the irreducible components. Fiber types with green corners are non-canonical models. From each of these, there is another branch of the tree sprouting off, with multiply non-canonical fiber types. Enhancements from type $II, III, IV$ are not shown, but discussed in the text.   \label{fig:TheTree}}

\end{figure}
\end{landscape}

Due to the additional data specifying the separation of the two sections, the enhancement structure becomes tree-like. 
The first few enhancements of this Tate tree are shown in figure \ref{fig:TheTree}, based on the Tate's algorithm in section \ref{sec:Tate}. 

{The main results can be summarized as follows: 
\begin{itemize}
\item[(1.)] {\it Canonical Tate-like forms:}

The canonical forms for the low rank are summarized in 
table \ref{tab:TateTable} and for the infinite series $I_n$ and $I_n^*$ with any section separation in codimension one, can be found in table \ref{tab:FiberTypeTable}. 
These form the closest analog to the known Tate forms in $\mathbb{P}^{(1,2,3)}$.

\item[(2.)] {\it Non-canonical forms:}

Non-canonical models are discussed in section \ref{sec:P112NonCanonical}, focusing on the low rank cases\footnote{We have not studied the full structure of non-canonical enhancements of all $I_n$ models, however most non-canonical models will have only non-minimal codimension 2 loci, as in $\mathbb{P}^{(1,2,3)}$, which usually implies that those sections can be set to one, thus allowing shifts to canonical forms.}. Each of the low rank non-canonical models gives rise to a new branch of the Tate tree, with multiply non-canonical enhancements. 

\item[(3.)] {\it Applications for F-theory GUT model builing:}

For $I_5$, which is of phenomenological interest in F-theory model building with GUT group $SU(5)$, there are three fiber types:  $I_5^{(01)}$, $I_5^{(0|1)}$ and $I_5^{(0||1)}$. All of these have canonical (section  \ref{sec:I5s}, in particular table \ref{tab:I5Canw=1}) and non-canonical models (tables \ref{tab:I5NonCanCharges} and \ref{tab:I5ncncCharges}). There are non-canonical models for $I_5^{(0|1)}$ and $I_5^{(0||1)}$ which we analyze in \ref{sec:NonCanI5}.  As shown in section \ref{sec:I5ncnc}, the $I_5^{(01)}$ fiber only arises as a doubly non-canonical form. 
Finally, one can explicitly check where the models that are already present in the literature are located within the Tate tree. For the toric models arising from tops and for the split spectral cover models this is done in appendix \ref{sec:tops}. It turns out that tops 2 and 3 are special cases of the non-canonical $I_5^{(0|1)}$ model, and top 4 is a special case of the non-canonical $I_5^{(0||1)}$ model. Top 1 is precisely the canonical $I_5^{(01)}$ model. The $2+3$-factorized Tate model is  the non-canonical $I_5^{(0|1)}$ model, the $4+1$-factorized Tate model is a special case of the non-canonical $I_5^{(0||1)}$ model.

\end{itemize}

}
    
\begin{table}
  \centering
    \begin{tabular}{|c|c|c|c|c|c|c|c|c|c|}
    \hline
      Fiber & $ \hbox{ord}(\Delta)$ & Group  & $\fkc_0$ & $\fkc_1$ & $\fkc_2$ & $\fkc_3$ & $\fkb_0$ & $\fkb_1$ & $\fkb_2$ \\ \hline\hline
      
      $I_0$              & 0 & ---     & 0 & 0 & 0 & 0 & 0 & 0 & 0 \\ \hline 
      $I_1$              & 1 & ---     & 1 & 1 & 1 & 0 & 0 & 0 & 1 \\ \hline
      $I_2^{(01)}$       & 2 & $SU(2)$ & 2 & 1 & 1 & 0 & 0 & 0 & 1 \\ \hline
      $I_2^{(0|1)}$      & 2 & $SU(2)$ & 1 & 1 & 1 & 1 & 0 & 0 & 1 \\ \hline
      $I_3^{(01)}$       & 3 & $SU(3)$ & 3 & 2 & 1 & 0 & 0 & 0 & 1 \\ \hline
      $I_3^{(0|1)}$      & 3 & $SU(3)$ & 2 & 1 & 1 & 1 & 0 & 0 & 1 \\ \hline
      $I_4^{(01)}$       & 4 & $SU(4)$ & 4 & 2 & 1 & 0 & 0 & 0 & 2 \\ \hline
      $I_4^{(0|1)}$      & 4 & $SU(4)$ & 3 & 2 & 1 & 1 & 0 & 0 & 1 \\ \hline
      $I_4^{(0||1)}$     & 4 & $SU(4)$ & 2 & 2 & 2 & 2 & 0 & 0 & 1 \\ \hline
      $I_5^{(01)}$       & 5 & $SU(5)$ & 5 & 3 & 1 & 0 & 0 & 0 & 2 \\ \hline
      $I_5^{(0|1)}$      & 5 & $SU(5)$ & 4 & 2 & 1 & 1 & 0 & 0 & 2 \\ \hline
      $I_5^{(0||1)}$     & 5 & $SU(5)$ & 3 & 2 & 2 & 2 & 0 & 0 & 1 \\ \hline \hline
      
      $II$               & 2 & ---     & 1 & 1 & 1 & 0 & 0 & 1 & 1 \\ \hline
      $III^{(01)}$       & 3 & $SU(2)$ & 2 & 1 & 1 & 0 & 0 & 1 & 1 \\ \hline
      $III^{(0|1)}$      & 3 & $SU(2)$ & 1 & 1 & 1 & 1 & 0 & 1 & 1 \\ \hline
      $IV^{(01)}$        & 4 & $SU(3)$ & 3 & 2 & 1 & 0 & 0 & 1 & 1 \\ \hline
      $IV^{(0|1)}$       & 5 & $SU(3)$ & 2 & 1 & 1 & 1 & 0 & 1 & 1 \\ \hline \hline
      
      $I_0^{*ns(01)}$    & 6 & $G_2$   & 4 & 2 & 0 & 0 & 0 & 0 & 2 \\ \hline
      $I_0^{*ss(01)}$    & 6 & $SO(7)$ & 4 & 2 & 1 & 0 & 0 & 1 & 2 \\ \hline
      $I_0^{*ss(0|1)}$   & 6 & $SO(7)$ & 2 & 2 & 1 & 1 & 0 & 1 & 1 \\ \hline
      $I_0^{*s(01)}$     & 6 & $SO(8)$ & 4 & 2 & 1 & 0 & 0 & 1 & 2 \\ \hline
      $I_0^{*(0|1)}$     & 6 & $SO(8)$ & 3 & 2 & 1 & 1 & 0 & 1 & 1 \\ \hline
      $I_1^{*(01)}$      & 7 & $SO(10)$& 5 & 3 & 1 & 0 & 0 & 1 & 2 \\ \hline
      $I_1^{*(0|1)}$     & 7 & $SO(10)$& 4 & 2 & 1 & 1 & 0 & 1 & 2 \\ \hline
      $I_1^{*(0||1)}$    & 7 & $SO(10)$& 3 & 2 & 2 & 1 & 0 & 1 & 1 \\ \hline \hline
      
      $IV^{*ns(01)}$     & 8 & $F_4$   & 4 & 3 & 2 & 0 & 0 & 1 & 2 \\ \hline
      $IV^{*(01)}$       & 8 & $E_6$   & 5 & 3 & 2 & 0 & 0 & 1 & 2 \\ \hline
      $IV^{*(0|1)}$      & 8 & $E_6$   & 3 & 2 & 2 & 1 & 0 & 1 & 2 \\ \hline
      $III^{*(01)}$      & 9 & $E_7$   & 5 & 3 & 2 & 0 & 0 & 1 & 3 \\ \hline
      $III^{*(0|1)}$     & 9 & $E_7$   & 3 & 3 & 2 & 1 & 0 & 1 & 2 \\ \hline
      $II^{*(01)}$       &10 & $E_8$   & 5 & 4 & 2 & 0 & 0 & 1 & 3 \\ \hline \hline
      
      non-min            &12 & ---     & 6 & 4 & 2 & 0 & 0 & 1 & 3 \\ \hline
      non-min            &12 & ---     & 4 & 3 & 2 & 1 & 0 & 1 & 2 \\ \hline
      
    \end{tabular}
  \caption{Fiber types and vanishing orders for low-rank canonical fibrations with rank-1 Mordell-Weil group, from Tate's algorithm for quartics in $\bbP^{(1,1,2)}$. $\Delta$ specifies the vanishing order of the discriminant. If not explicitly stated otherwise, models are of split-type. The monodromy condition that differentiates between the $I_0^{*ss(01)}$ fiber from  $I_0^{*s(01)}$  is given in equation (\ref{eq:I0sMonodromy}){, and the additional monodromy condition for $I_0^{*ns(01)}$ in (\ref{eq:I0nsMonodromy})}.}
  \label{tab:TateTable}
\end{table}
\newpage


\begin{landscape}
\begin{table}
  \centering
  \begin{tabular}{|c|c|c|c|c|c|c|c|c|c|c|}\hline
    Fiber & Vanishing & Gauge & \multicolumn{7}{c|}{Vanishing orders of coefficient sections} & \\
    type & order of $\Delta$ & group  & $\fkc_0$ & $\fkc_1$ & $\fkc_2$ & $\fkc_3$ & $\fkb_0$ & $\fkb_1$ & $\fkb_2$  &\\ \hline\hline
      $I_{2m+k}^{(01)}$    & $2m$   & $SU(2m)$ & $2m$ & $m$ & $1$ & $0$ & $0$ & $0$ & $m$ & \\
      $I_{2m+k}^{(0|^k 1)}$& $2m+k$ & $SU(2m+k)$ & $2m$ & $m$ & $k$ & $k$ & $0$ & $0$ & $m$ & $1\leq k \leq m$ \\
      $I_{2m}^{(0|^m 1)}$  & $2m$   & $SU(2m)$ & $m$ & $m$ & $m$ & $m$ & $0$ & $0$ & $1$ & \\
      $I_{2(m+k)}^{(0|^m 1)}$&$2(m+k)$& $SU(2(m+k))$ & $m+2k$ & $m$ & $m$ & $m$ & $0$ & $0$ & $2k$ & $1 \leq k \leq \left\lfloor \frac m 2 \right\rfloor$ \\ \hline
      
      $I_{2m+1}^{(01)}$    & $2m+1$   & $SU(2m+1)$ & $2m+1$ & $m+1$ & $1$ & $0$ & $0$ & $0$ & $m$ & \\
      $I_{2m+k+1}^{(0|^k1)}$& $2m+k+1$   & $SU(2m+k+1)$ & $2m+1$ & $m+1$ & $k$ & $k$ & $0$ & $0$ & $m$ & $1\leq k \leq m$ \\
      $I_{2m+1}^{(0|^m 1)}$& $2m+1$ & $SU(2m+1)$ & $m+1$ & $m$ & $m$ & $m$ & $0$ & $0$ & $1$ & \\
      $I_{2(m+k)+1}^{(0|^m 1)}$ & $2(m+k)+1$ & $SU(2(m+k)+1)$ & $m+2k+1$ & $m$ & $m$ & $m$ & $0$ & $0$ & $2k+1$ & $1 \leq k \leq \left\lfloor \frac m 2 \right\rfloor$ \\ \hline
      
      $I_{2m}^{ns(01)}$    & $2m$ & $Sp(m)$ & $2m$ & $m$ & $0$ & $0$ & $0$ & $0$ & $m$ & \\
      $I_{2m}^{ns(0|1)}$ & $2m$ & $Sp(m)$ & $m$ & $m$ & $m$ & $m$ & $0$ & $0$ & $0$ & \\
      $I_{2m+1}^{ns(01)}$  & $2m+1$ & --- & $1+2m$ & $1+m$ & $0$ & $0$ & $0$ & $0$ & $m+1$ & \\ \hline
      
      $I_{2m}^{*(01)}$     & $2m+6$ & $SO(4m+8)$ & $4+2m$ & $2+m$ & $1$ & $0$ & $0$ & $1$ & $2+m$ & \\
      $I_{2m+1}^{*(01)}$   & $2m+7$ & $SO(4m+10)$& $5+2m$ & $3+m$ & $1$ & $0$ & $0$ & $1$ & $2+m$ & \\
      $I_{2m}^{*(0|1)}$    & $2m+6$ & $SO(4m+8)$ & $3+2m$ & $2+m$ & $1$ & $1$ & $0$ & $1$ & $1+m$ & \\
      $I_{2m+1}^{*(0|1)}$  & $2m+7$ & $SO(4m+10)$& $4+2m$ & $2+m$ & $1$ & $1$ & $0$ & $1$ & $2+m$ & \\
      $I_{2m}^{*(0||1)}$   & $2m+6$ & $SO(4m+8)$ & $2+m$  & $2+m$ &$1+m$&$1+m$& $0$ & $1$ & $1$   & \\
      $I_{2m+1}^{*(0||1)}$ & $2m+7$ & $SO(4m+10)$& $3+m$  & $2+m$ &$2+m$&$1+m$& $0$ & $1$ & $1$   & \\ \hline
      
      $I_{2m}^{*ns(01)}$   & $2m+6$ & $SO(4m+7)$ & $3+2m$ & $2+m$ & $1$ & $0$ & $0$ & $1$ & $2+m$ & \\
      $I_{2m+1}^{*ns(01)}$ & $2m+7$ & $SO(4m+9)$ & $4+2m$ & $3+m$ & $1$ & $0$ & $0$ & $1$ & $2+m$ & \\
      $I_{2m}^{*ns(0|1)}$  & $2m+6$ & $SO(4m+7)$ & $2+2m$ & $2+m$ & $1$ & $1$ & $0$ & $1$ & $1+m$ & \\
      $I_{2m+1}^{*ns(0|1)}$& $2m+7$ & $SO(4m+9)$ & $3+2m$ & $2+m$ & $1$ & $1$ & $0$ & $1$ & $2+m$ & \\ \hline
      
  \end{tabular}     
  \caption{Vanishing orders for fibrations realizing $I_n$, $I_n^*$ fiber types, both split and non-split (ns) for $n\geq 1$,  with two sections in canonical form, i.e., characterized entirely in terms of vanishing        order of the coefficients $\fkb_i$ and $\fkc_j$.
  There are three distinct distributions of the two sections on the fibers for the split $I_n^*$, modulo symmetry of the affine $D$ type Dynkin diagram, and two for the non-split case. 
  The $I_n$ fibers are grouped into whether $n$ is even or odd, and whether the separation between the two sections is small or large. For $I_n$ the forms can be put into a single, slightly more complicated form given in (\ref{Ininfty}).
   \label{tab:FiberTypeTable} }
\end{table}
\end{landscape}


\section{Tate Tree: Canonical Forms}
\label{sec:Tate}

In this section we will determine all the canonical models, i.e. those determined solely by vanishing orders with generic coefficients $c_i$ and $b_i$. The non-canonical enhancements will be discussed separately in section \ref{sec:P112NonCanonical}. 
In this sense the present section results in the analog of the standard Tate models in $\mathbb{P}^{(1,2,3)}$ in 
\cite{Bershadsky:1996nh}, whereas the section on non-canonical forms also encondes local obstructions such as those studied for $\mathbb{P}^{(1,2,3)}$ in \cite{Katz:2011qp}. The main difference to $\mathbb{P}^{(1,2,3)}$ is that non-canonical models are much more generic in $\mathbb{P}^{(1,1,2)}$ and also arise prominently in the $I_n$ branch. 
We run the algorithm in detail up until and  including $O(z^5)$, i.e. in the $I_5$, and derive $I_0^*$ $IV^*$, $III^*$ and $II^*$ fibers in the next section. Finally, we give canonical forms for all $I_n$ and $I_n^*$ fibers, however the (multiply) non-canonical progression for the infinite series in the algorithm is left for future work. 


\subsection{Monodromy}
\label{sec:Mono}

In section \ref{sec:Order0}, it was found that there is a single $I_1$ fibration with canonical form
\begin{equation}
I_1^{(01)}:\qquad   \CQ(1,1,0,0,0,0,1)
\end{equation}
and discriminant at leading order
\begin{equation}
  \Delta_{I_1} = c_{0,1} \left( b_{1,0}^2 + 4 c_{2,0} \right)^3 \left( b_{0,0}^2 c_{2,0} - b_{0,0} b_{1,0} c_{3,0} - c_{3,0}^2 \right) z + O(z^2) \,.
\end{equation}
Upon performing the coordinate shift $y\rightarrow y-\frac{1}{2}b_{1,0} w x$, one obtains the quartic
\begin{equation}
  \begin{aligned}
 &y^2 s + \fkb_{0,0} x^2 y + \fkb_{1,1} z y w s x + \fkb_{2,1} z w^2 s^2 y \cr
&=   \fkc_{0,1} z w^4s^3 +  \left(\fkc_{1,1} + \frac{1}{2} b_{1,0} b_{2,1}  z  \right) w^3s^2 x + \left( \fkc_{2,0} + \frac{1}{4} b_{1,0}^2  \right) w^2 s  x^2 
    + \left( \fkc_{3,0} + \frac{1}{2} b_{0,0} b_{1,0} \right) w x^3 
  \,.
  \end{aligned}
\end{equation}
Defining shifted leading coefficients of the series $\mathfrak{c}_{ij}$ as $\hat c_{1,1} = c_{1,1}+\frac{1}{2}b_{1,0}b_{2,1}$, $\hat c_{2,0} = c_{2,0} + \frac{1}{4} b_{1,0}^2$, $\hat c_{3,0}=c_{3,0} + \frac{1}{2}b_{0,0}b_{1,0}$ and dropping the hats, the fibration above is described by the canonical form
\begin{equation}
  \CQ(1,1,0,0,0,1,1) \,,
\end{equation}
with discriminant
\begin{equation}
  \Delta = c_{0,1} c_{2,0}^3 \left( b_{0,0}^2 c_{2,0} - c_{3,0}^2 \right) z + O(z^2) \,.
\end{equation}
The monodromy condition for $I_n$, determining whether the local gauge group is given by $SU(n)$ or $Sp\left(\lfloor\frac n 2\rfloor\right)$, is checked by testing whether $c_{2,0}$ of this fiber has a square root: One can always write $c_{2,0}=\mu \tilde c_{2,0}^2$, and choose $\mu$ such that $\mu=1$ if $\mu$ has no zeros. Then, the local gauge group will be $SU(n)$ if $\mu=1$, and $Sp\left(\left\lfloor \frac n 2 \right\rfloor \right)$ if $\mu$ has zeros. Note that $\mu=1$ is a necessary condition for the term in brackets in the discriminant above to vanish.
Further, if $\mu=1$, one can perform the coordinate shift
$y \rightarrow y - \tilde c_{2,0} w x$. This coordinate shift yields the canonical form 
\begin{equation}
I_1^{(01)}:\qquad   \CQ(1,1,1,0,0,0,1) \,.
    \label{eq:Q1110001}
\end{equation}
As we are interested in $I_n^{split}$ fibers, we proceed assuming the starting $I_1$ singularity to be of the form (\ref{eq:Q1110001}).
In the following we proceed to enhance the type of this singularity, which in the case of the quartic in $\mathbb{P}^{(1,1,2)}$ has a tree-like structure, which is characterized by the Kodaira fiber type as well as the location of the sections.  


\subsection{Discriminant at $O(z^2)$}
\label{sec:I1Order2}

The $I_1$ singularity (\ref{eq:Q1110001}) has leading-order discriminant
\begin{equation}
  \Delta_{I_1} = c_{0,1} b_{1,0}^6 c_{3,0} \left( b_{0,0} b_{1,0} + c_{3,0} \right) z + O(z^2)\,.
  \label{eq:DeltaI1}
\end{equation}
The possible fiber enhancements are given by setting factors of this expression to zero.
\begin{itemize}
  \item $c_{0,1}=0$: $I_2^{(01)}$ \\
    This fiber trivially has canonical form
    \begin{equation}
I_2^{(01)}:\qquad       \CQ(2,1,1,0,0,0,1) \,.
      \label{eq:Q2110001}
    \end{equation}
Resolving the singular fiber for instance with the big resolution $(x,y,z;\zeta_1)$, using the notation of section \ref{sec:Fibration}, we see that the two sections intersect the same fiber component, and are thus of type $I_2^{(01)}$, and is shown on the left hand side in figure \ref{I2Fibs}. 

\begin{figure} 
  \centering
  \includegraphics[height=3cm]{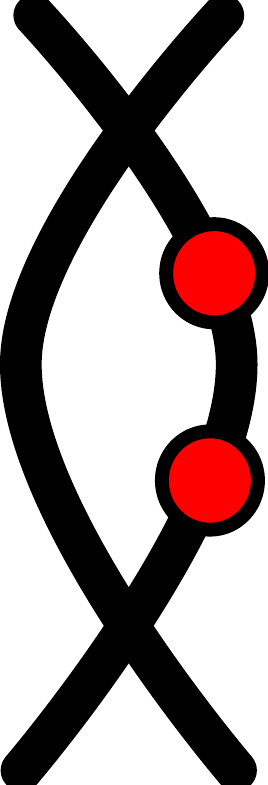} \qquad\qquad 
  \includegraphics[height=3cm]{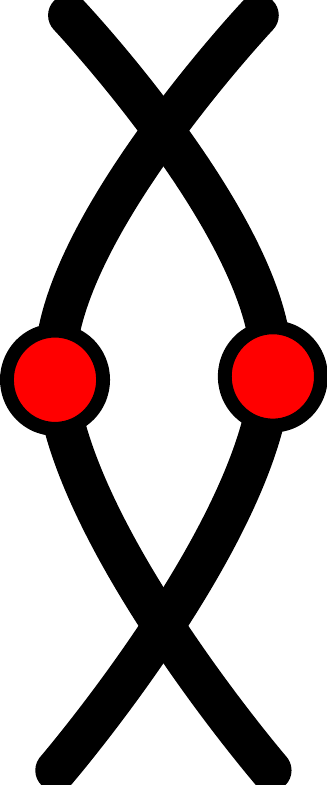}
  \caption{$I_2^{(01)}$ and $I_2^{(0|1)}$ fibers, with black lines corresponding to the two $\mathbb{P}^1$ fiber components, and the red nodes to the two sections, $\sigma_0$ and $\sigma_1$. \label{I2Fibs}}
\end{figure}

  \item $c_{3,0}=0$: $I_2^{(0|1)}$ \\
    The canonical form for this fiber is
    \begin{equation}
   I_2^{(0|1)}:\qquad    \CQ(1,1,1,1,0,0,1) \,.
      \label{eq:Q1111001}
    \end{equation}
    Here, the two sections intersect neighbouring components of the resolved fiber, i.e. of type $I_2^{(0|1)}$, shown on the right hand side in figure \ref{I2Fibs}. 
  
  \item $b_{0,0}b_{1,0}+c_{3,0}=0$: $I_2^{(0|1)}$ \\
    This enhancement is equivalent to setting $c_{3,0}=0$, as can be seen as follows. Applying the coordinate shift $y \rightarrow y - b_{1,0} w x$ turns $\CQ_{I_1}$ into a form in which the section $\tilde c_{3,0}$ in the new coordinates is given by $\tilde c_{3,0}= b_{0,0} b_{1,0} + c_{3,0}$, and all other sections are still generic. Hence, $b_{0,0} b_{1,0} + c_{3,0}=0$ is equivalent to $\tilde c_{3,0}=0$ in the new coordinates.
    
  During later stages of the algorithm, one encounters a few more discriminants with factors of the form $c_{3,j} \left( b_{0,j} b_{1,0} + c_{3,j} \right)$. Let us note here that all enhancnements arising from the bracketed part of this expression are always equal to enhancements arising from $c_{3,j}$, and that there is always a coordinate shift of the form discussed here linking the two. We therefore do not treat $b_{0,j} b_{1,0} + c_{3,j}$ explicitly in the following.
  \item $b_{1,0}=0$: $II$ \\
    Setting $b_{1,0}=0$ enhances the singularity in a way that leaves the $I_n$ branch: $\CQ_{I_1}|_{z=0}$ has a double root at $x=y=0$, and a Taylor expansion around this double root yields
    \begin{equation}
      \CQ_{I_1}|_{z=0,w=s=1}: y^2 + b_{1,0} x y + O(x^3, y^3)
    \end{equation}
    whose discriminant is given by $\left( \partial_{xy} \CQ_{I_1} \right)^2 - \partial_{xx} \CQ_{I_1} \partial_{yy} \CQ_{I_1} = b_{1,0}^2$. Vanishing of this discriminant indicates a cusp singularity with Kodaira type $II$.
\end{itemize}


\subsection{Discriminant  at $O(z^3)$}

Each distinct $I_2$ fiber type opens a new branch of the algorithm, or Tate tree, yielding different enhancements. There are two $I_2$ fibers, where  the two rational sections intersect either  the same or  distinct fiber components of the resolved singular fiber. Following the discriminant we now determine all the fiber types for each branch. 

\begin{figure}
  \centering
  \includegraphics[width=3cm]{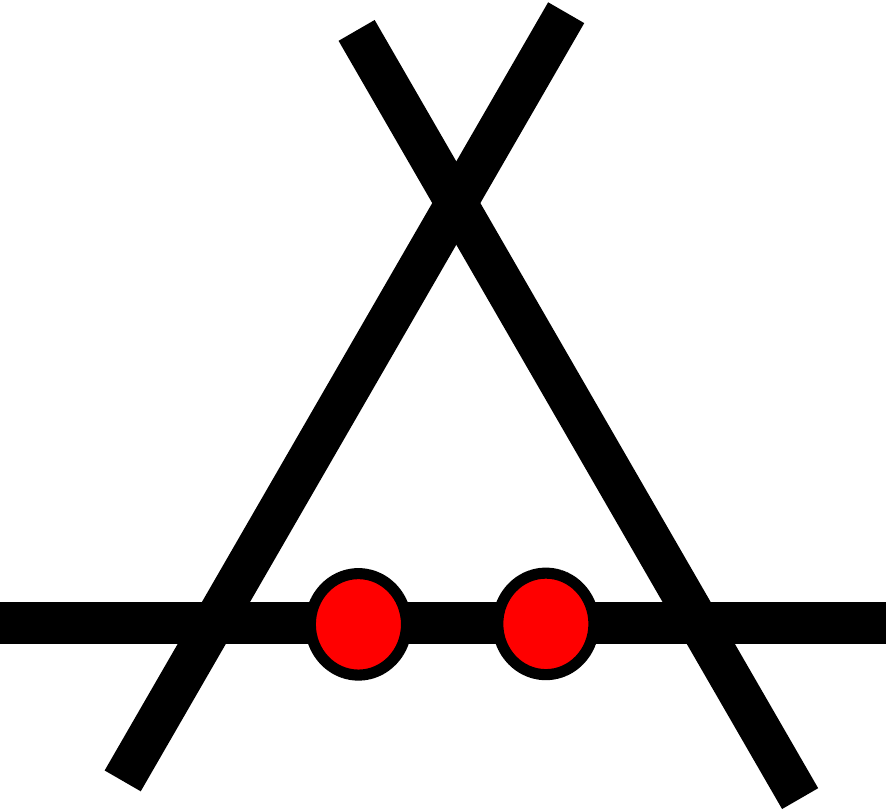} \qquad
  \includegraphics[width=3cm]{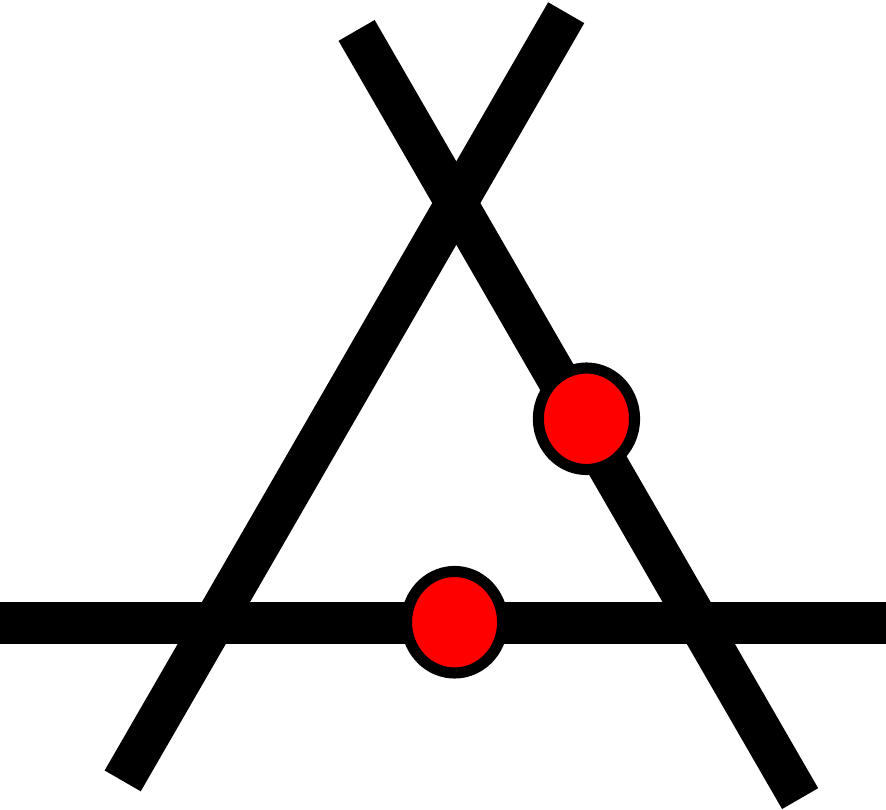}
  \caption{$I_3^{(01)}$ and $I_3^{(0|1)}$ fibers. The black lines correspond to the 
  $\mathbb{P}^1$ fiber components, and the red nodes to the two sections, $\sigma_0$ and $\sigma_1$. Due to the symmetry of the diagram, there are only two distinct distributions of the two sections. \label{fig:I3Fibs}}
\end{figure}

\subsubsection{$I_2^{(01)}$ Branch}

Consider first the branch starting from the  $I_2^{(01)}$ fiber, where both sections lie on one fiber component, realized in terms of (\ref{eq:Q2110001}). The discriminant for this fibration is
\begin{equation}
  \Delta_{I_2^{(01)}} = b_{1,0}^4 c_{3,0} \left( b_{1,0} b_{0,0} + c_{3,0} \right) P_0 z^2 + O(z^3)
\end{equation}
with
\begin{equation}\label{P0z2}
  P_0 = b_{1,0}^2 c_{0,2} - b_{1,0} b_{2,1} c_{1,1} - c_{1,1}^2 \,.
\end{equation}
Each factor corresponds to an enhancement type, which we will consider in turn. The polynomials appearing in the discriminant generically give rise to non-canonical models and will be discussed later in detail. Here we will focus on the canonical branch. 

\begin{figure}
  \centering
  \includegraphics[height=3cm]{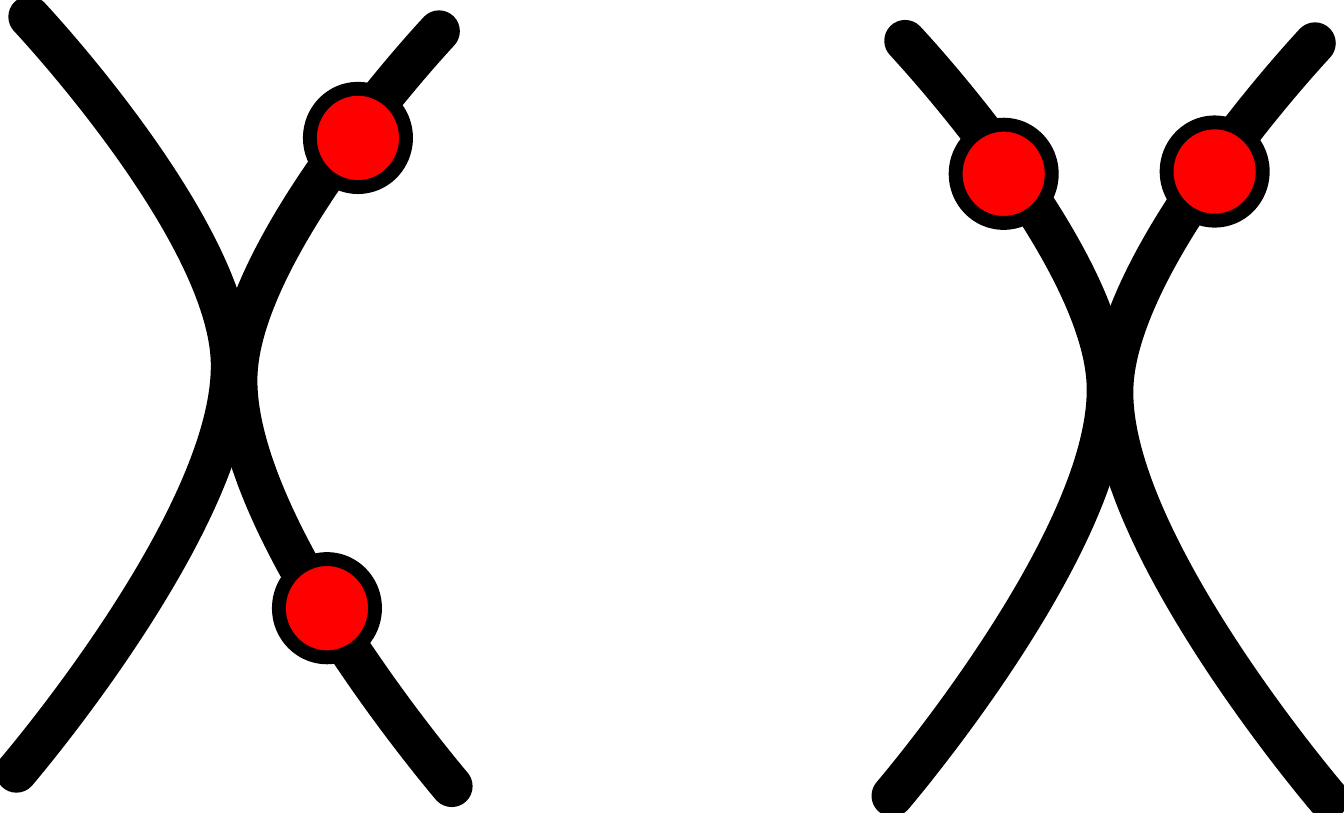}
  \caption{$III^{(01)}$ and $III^{(0|1)}$ fibers, again with black lines corresponding to the fiber components, and the red dots to the extra sections. \label{fig:IIIFibs}}
\end{figure}

\begin{itemize}

  \item $P_0=0$: $I_3^{(01)}$ \\\
    The general solution to the vanishing of the  polynomial (\ref{P0z2}) over a UFD is determined in appendix \ref{sec:ThreeTermPoly} as
    \begin{equation}\label{P0Sol}
      c_{1,1} = b_{1,0} \tilde c_{1,1} \,, \qquad
      c_{0,2} = b_{2,1} \tilde c_{1,1} + \tilde c_{1,1}^2 \,.
    \end{equation}
    The corresponding quartic significantly simplifies upon application of the coordinate shift $y \rightarrow y+\tilde c_{1,1} z s w^2$, where it takes the canonical form
    \begin{equation}      \label{eq:Q3210001}
I_3^{(01)}:\qquad       \CQ(3,2,1,0,0,0,1) \,.
    \end{equation}
    Slight variations of the polynomial $P_0$ will reappear at later stages of the algorithm. After applying the solution from appendix \ref{sec:ThreeTermPoly}, one can always find a coordinate shift that brings these into canonical form by enhancing the vanishing order of $\fkc_0$ and $\fkc_1$.  The fiber type is determined by computing the intersections as described in section  \ref{sec:Fibration} and the fiber is depicted in figure \ref{fig:I3Fibs}.

  \item $c_{3,0}=0$: $I_3^{(0|1)}$ \\
    The canonical form for this fiber is 
    \be \label{eq:Q2111001}
  I_3^{(0|1)}:\qquad     \CQ(2,1,1,1,0,0,1) \,.
        \ee
Here the sections are located on distinct fiber components, as shown in figure \ref{fig:I3Fibs}. 

  \item $b_{1,0}=0$: $III^{(01)}$ \\
    This branch corresponds to a type $III$ fiber,  is shown in figure \ref{fig:IIIFibs}, and has canonical form
    \begin{equation}
   III^{(01)}:\qquad    \CQ(2,1,1,0,0,1,1) \,.
    \end{equation}

\end{itemize}

\subsubsection{$I_2^{(0|1)}$ Branch}
\label{sec:Q1111001}

The second $I_2$ branch starts with $I_2^{(0|1)}$, which is  realized in terms of (\ref{eq:Q1111001}). In this case the sections are on separate fiber components with the discriminant given by
\begin{equation}
  \Delta_{I_2^{(0|1)}} = b_{1,0}^4 b_{0,0} c_{0,1} P_0 z^2 + O(z^3) \,,
\end{equation}
with
\begin{equation}
  P_0 = b_{0,0}^3 c_{0,1} - b_{0,0}^2 b_{1,0} c_{1,1} + b_{0,0} b_{1,0}^2 c_{2,1} - b_{1,0}^3 c_{3,1} \,.
  \label{eq:D1111001P0}
  \end{equation}
The component $c_{0,1}=0$ of the discriminant gives the model $I_3^{(0|1)}$ realized in terms of (\ref{eq:Q2111001}), i.e. it joins back with the branch starting from $I_2^{(01)}$. 

Furthermore, the branch $b_{0,0}=0$ has been removed by the lopping, as it is equivalent to other models, that we considered already.\footnote{More precisely setting $b_{0,0}=0$ yields a (non-extremal) $I_3^{(01)}$  model, realized by $\CQ(1,1,1,1,1,0,1)$, which is related by a lop transition to  (\ref{eq:Q3210001}).}

The other discriminant components result in the following fibers:
  
  \begin{itemize}  
  \item $P_0=0$: ${I_{3, nc}^{(0|1)}}$ \\
    $P_0$ is an example of the four-term polynomial discussed in appendix \ref{sec:FourTermPoly}, and we can directly substitute the general solution found there into $I_2^{(0|1)}$. We denote the resulting non-canonical ($nc$) form as
    \begin{equation}
   {I_{3, nc}^{(0|1)}}:\qquad   \CQ(1,1,1,1,0,0,1)|_{(\ref{eq:D1111001P0})} \,.
      \label{eq:Q1111001P0}
    \end{equation}
Note that this gives the same fiber type as (\ref{eq:Q2111001}). However, due to the non-canonical nature of the enhancement, solving $P_0=0$,  as in appendix \ref{sec:FourTermPoly}, results in the section $b_{1,0} = \sigma_1 \sigma_2$ to factor. This implies that compared to the model  
(\ref{eq:Q2111001}), where $b_{1,0}$ is generically irreducible, the structure of the codimension 2 fibers will be different. This effect yields multiple, differently charged matter curves. We will study these models in the next section.
    
  \item $b_{1,0}=0$: $III^{(0|1)}$ \\
This yields a type $III$ fiber with canonical form
    \begin{equation}
  III^{(0|1)}:\qquad     \CQ(1,1,1,1,0,1,1) \,.
    \end{equation}
    

\end{itemize}


\subsection{Discriminant  at $O(z^4)$}

We have seen in the last section, that at order  $z^3$ the following fiber types occur: $I_3^{(01)}$, $I_3^{(0|1)}$, $I_{3, nc}^{(0|1)}$, as well as the type $III^{(01)}$ and $III^{(0|1)}$ fibers, each of these occured once in the algorithm. We continue here with the canonical tree growing out from $I_3^{(01)}$ and $I_3^{(0|1)}$. The enhancements of the non-canonical models will be discussed in section \ref{sec:P112NonCanonical}.

\subsubsection{$I_3^{(01)}$ Branch}

The $I_3^{(01)}$ fiber realized by  $\CQ(3,2,1,0,0,0,1)$ in (\ref{eq:Q3210001}) has discriminant
\begin{equation}
  \Delta_{I_3^{(01)}} = b_{1,0}^3 c_{3,0} \left( b_{0,0} b_{1,0} + c_{3,0} \right) P_0 z^3 + O(z^4) \,,
\end{equation}
where the polynomial term is 
\begin{equation}
  P_0 = b_{1,0}^3 c_{0,3} - b_{1,0}^2 b_{2,1} c_{1,2} + b_{1,0} b_{2,1}^2 c_{2,1} - b_{2,1}^3 c_{3,0} \,.
  \label{eq:D3210001P0}
\end{equation}

\begin{figure}
  \centering
  \includegraphics[width=3cm]{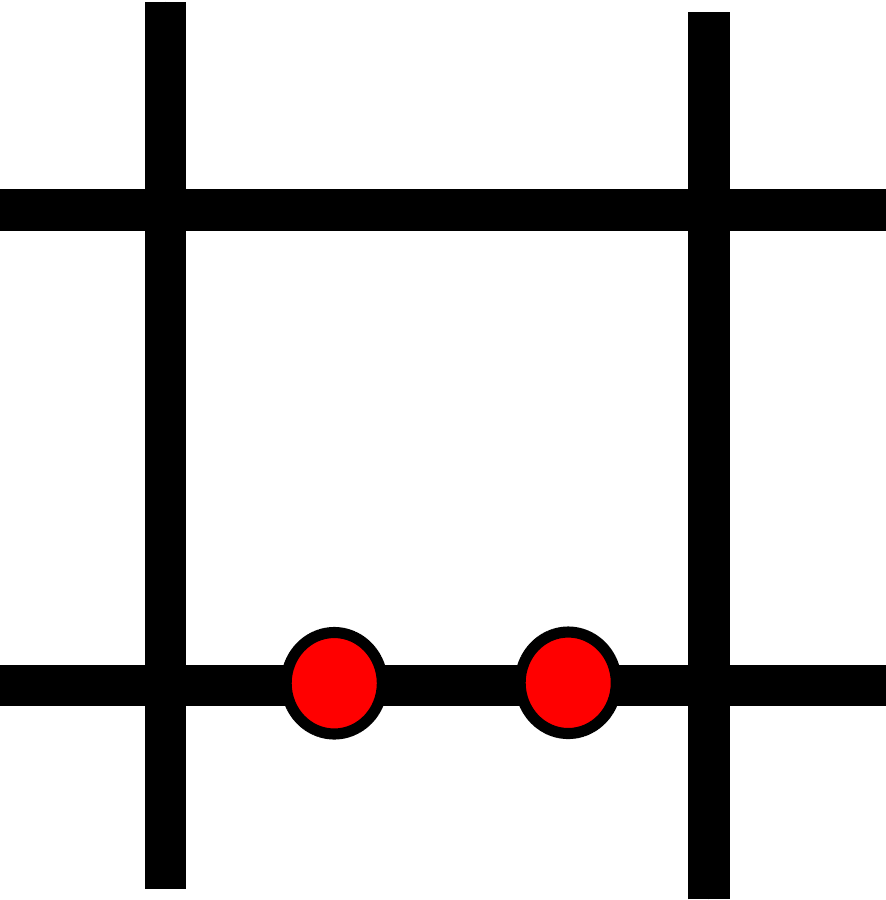} \qquad
  \includegraphics[width=3cm]{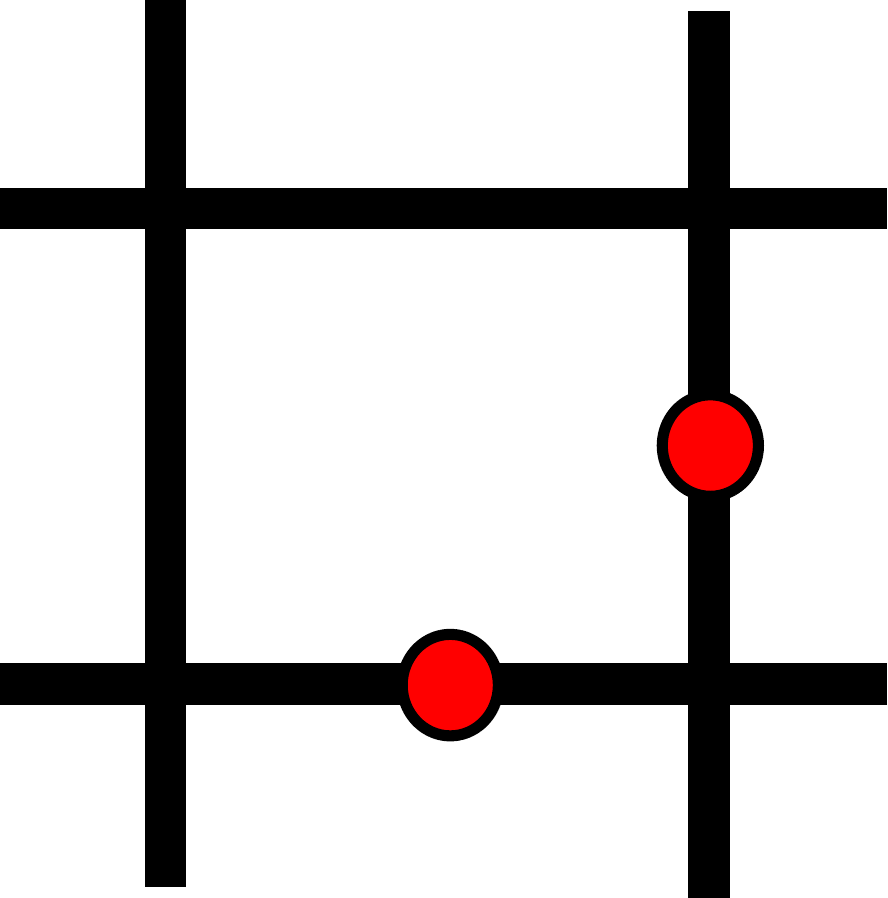} \qquad
  \includegraphics[width=3cm]{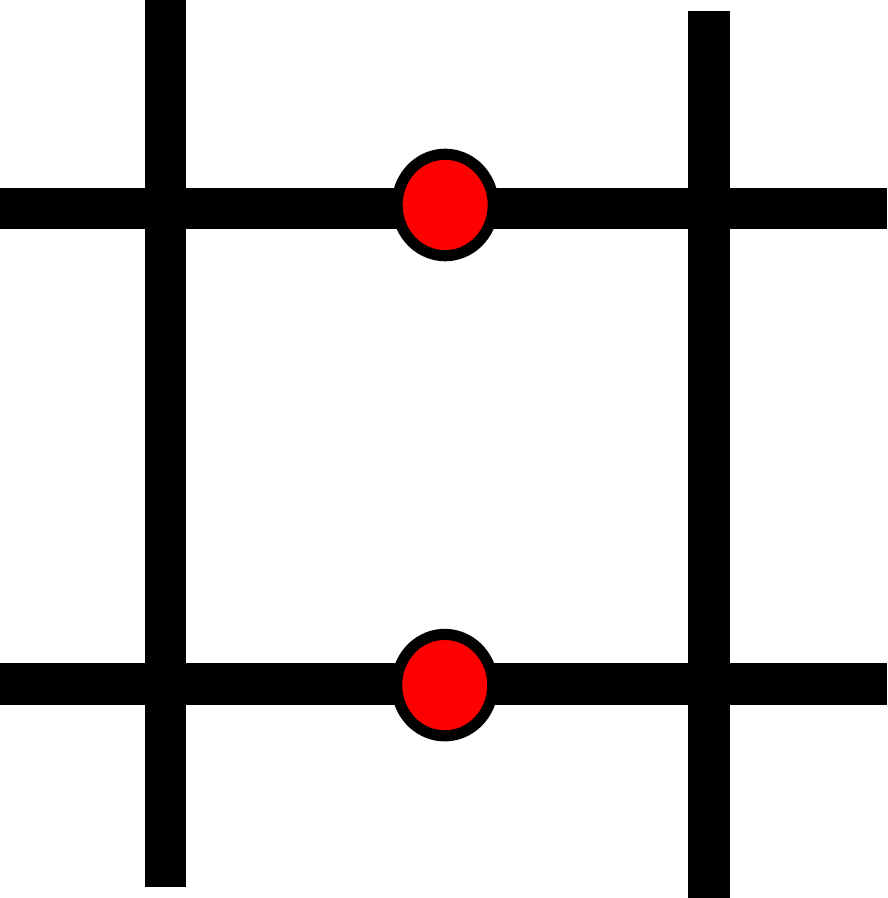}
  \caption{From left to right, showing the $I_4^{(01)}$, $I_4^{(0|1)}$ and $I_4^{(0||1)}$ fibers, respectively, with sections indicated by the red nodes. \label{fig:I4Fibs}}
\end{figure}

\begin{itemize}
  \item $c_{3,0}=0$: $I_4^{(0|1)}$ \\
    This enhancement splits the two sections to lie on separate, neighboring, fiber components, as shown in figure \ref{fig:I4Fibs}, with canonical form
    \begin{equation}      \label{eq:3211001}
I_4^{(0|1)}:\qquad     \CQ(3,2,1,1,0,0,1) \,.
    \end{equation}
  \item $P_0=0$: $I_{4}^{(01)}$  and $I_{4, nc}^{(01)}$ \\
    Applying appendix \ref{sec:FourTermPoly} to solve $P_0=0$, has two solutions:
    $b_{2,0}=c_{0,3}=0$ or the solution given in (\ref{FourTermSol}). 
   The former gives a canonical model, the latter a non-canonical one, with the same distribution of sections, however due to the non-canonicality the second one has multiple matter curves
    \begin{align}
     I_{4 }^{(01)}:\qquad & 		\CQ(4,2,1,0,0,0,2) \label{Q4210002}\\
     I_{4, nc}^{(01)}:\qquad &     \CQ(3,2,1,0,0,0,1)|_{(\ref{eq:D3210001P0})}  \label{eq:Q3210001P0}\,.
    \end{align}
The canonical model $I_4^{(01)}$ has one, whereas the non-canonical has two  codimension 2 curve over $b_{1,0}=0$. In the non-canonical fiber, there are two such loci, as $b_{1,0}$ factors into the product $\sigma_1 \sigma_2$ in order to solve $P_0=0$.
  \item $b_{1,0}=0$: $IV^{(01)}$ \\
    As before, for the low-rank cases, the $b_{1,0}=0$ enhancement moves us out of the $I_n$ branch, in this case to a type $IV$ fiber
        \begin{equation}
IV^{(01)}:\qquad       \CQ(3,2,1,0,0,1,1) \,.
    \end{equation}
\end{itemize}


\subsubsection{$I_3^{(0|1)}$ Branch}

The second branch continues from $\CQ(2,1,1,1,0,0,1)$ as in (\ref{eq:Q2111001}), which has leading order discriminant 
\begin{equation}
  \Delta_{I_3^{(0|1)}} = b_{1,0}^3 b_{0,0} P_0 P_1 z^3 + O(z^4)\,,
\end{equation}
where the polynomial terms are now
\begin{align}
  P_0 &= b_{1,0}^2 c_{0,2} - b_{1,0} b_{2,1} c_{1,1} - c_{1,1}^2 \label{eq:D2111001P0}\,, \\
  P_1 &= b_{0,0}^2 c_{1,1} - b_{0,0} b_{1,0} c_{2,1} + b_{1,0}^2 c_{3,1} \label{eq:D2111001P1}\,.
\end{align}
The $P_0=0$ enhancement, which in fact after a shift is again canonical, is precisely the model that we discussed already following the other branch of the algorithm in (\ref{eq:3211001}), i.e. this is another instance when the branches join back together. 
Furthermore, $b_{0,0}=0$ is removed by the lopping operation  explained in section \ref{sec:Lop},\footnote{Setting $b_{0,0}=0$ here would give rise to an $I_4^{(01)}$ model, which one can check explicitly, but which moreover is expected by the lopping.} so that we are left with the following branches:

\begin{figure}
  \centering
  \includegraphics[width=3cm]{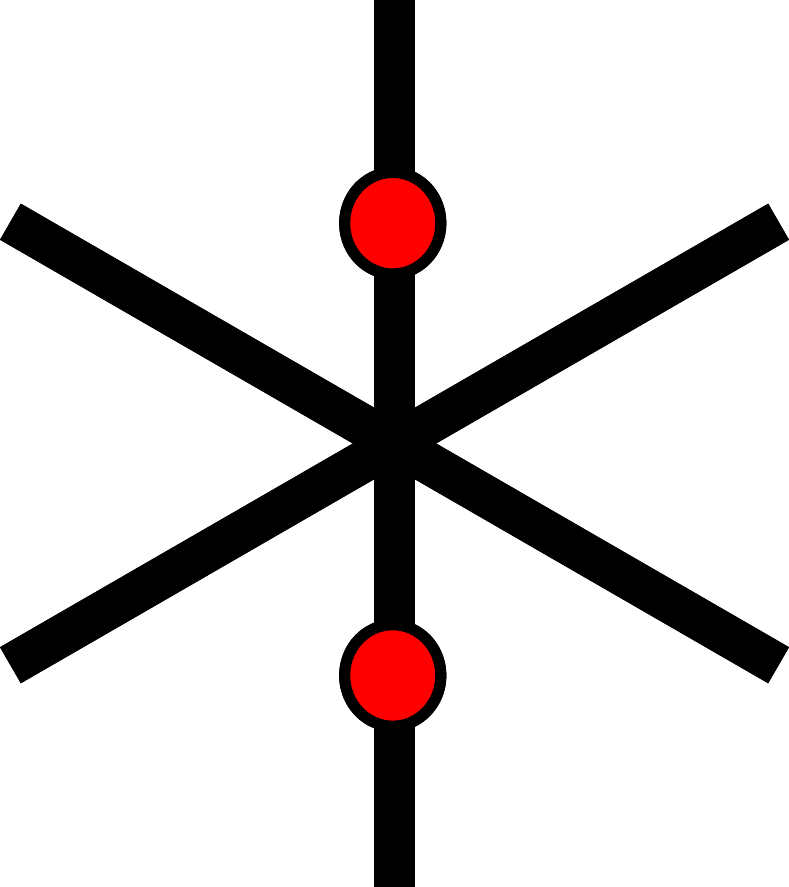} \qquad
  \includegraphics[width=3cm]{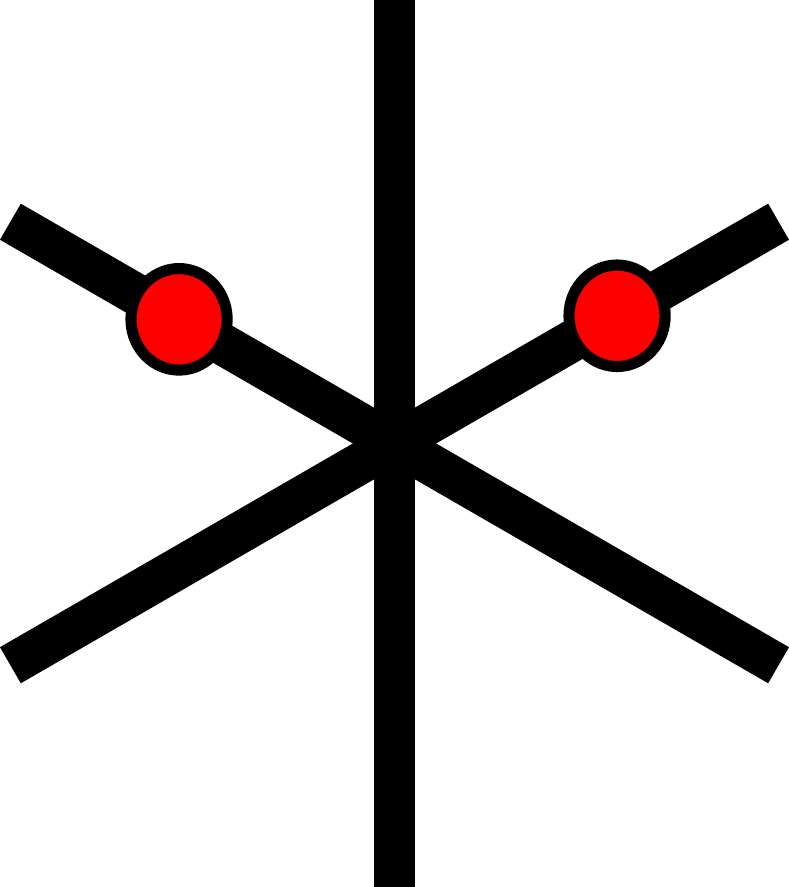}
  \caption{$IV^{(01)}$ and $IV^{(0|1)}$ fibers.}
\end{figure}

\begin{itemize} 
  \item $P_1=0$: $I_{4, nc}^{(0||1)}$ \\
   $P_1$ can be solved along the lines of appendix \ref{sec:ThreeTermPoly}, yielding the non-canonical form, that we will discuss later, in section \ref{sec:P112NonCanonical}
   \begin{equation}
     I_{4, nc}^{(0||1)}:\qquad      \CQ(2,1,1,1,0,0,1)|_{(\ref{eq:D2111001P1})} \,.
     \label{eq:Q2111001P1}
   \end{equation}
   This fiber is also depicted in figure \ref{fig:I4Fibs}, as the codimension one fiber structure does not depend on canonical versus non-canonical realization. However the codimension 2 structure will be different. 
   \item $b_{1,0}=0$: $IV^{(0|1)}$ \\
    Finally, $b_{1,0}=0$ moves us out of the $I_n$ branch again to give another $IV$ fiber, with the sections located on separate fiber components
    \begin{equation}
      IV^{(0|1)}:\qquad      \CQ(2,1,1,1,0,1,1) \,.
      \label{eq:IV0s1}
    \end{equation}
\end{itemize}


\subsection{Discriminant at $O(z^5)$}

In the last subsection we have seen that at order $z^4$ in the discriminant the $I_4$ fiber types are $I_4^{(01)}$, $I_4^{(0|1)}$ and the non-canonical $I_{4, nc}^{(01)}$ and $I_{4, nc}^{(0||1)}$. The non-canonical models will be discussed in detail in section \ref{sec:P112NonCanonical}. 
Continuining with the canoncial branches in this section, we now study the enhancements starting from $I_4^{(01)}$ and $I_4^{(0|1)}$ realized by  $\CQ(4,2,1,0,0,0,2)$ and $\CQ(3,2,1,1,0,0,1)$, respectively.

\subsubsection{$I_4^{(01)}$ Branch}

The discriminant of $I_4^{(01)}$ realized by $\CQ(4,2,1,0,0,0,2)$ in (\ref{Q4210002}) at leading order is
\begin{equation}
  \Delta_{I_4^{(01)}} = b_{1,0}^4 c_{3,0} \left( b_{0,0} b_{1,0} + c_{3,0} \right) P_0 z^4 + O(z^5)\,,
\end{equation}
with
\begin{equation}
  P_0 = b_{1,0}^2 c_{0,4} - b_{1,0} b_{2,2} c_{1,2} - c_{1,2}^2 \,.
\end{equation}

\begin{figure}
  \centering
  \includegraphics[width=3.5cm]{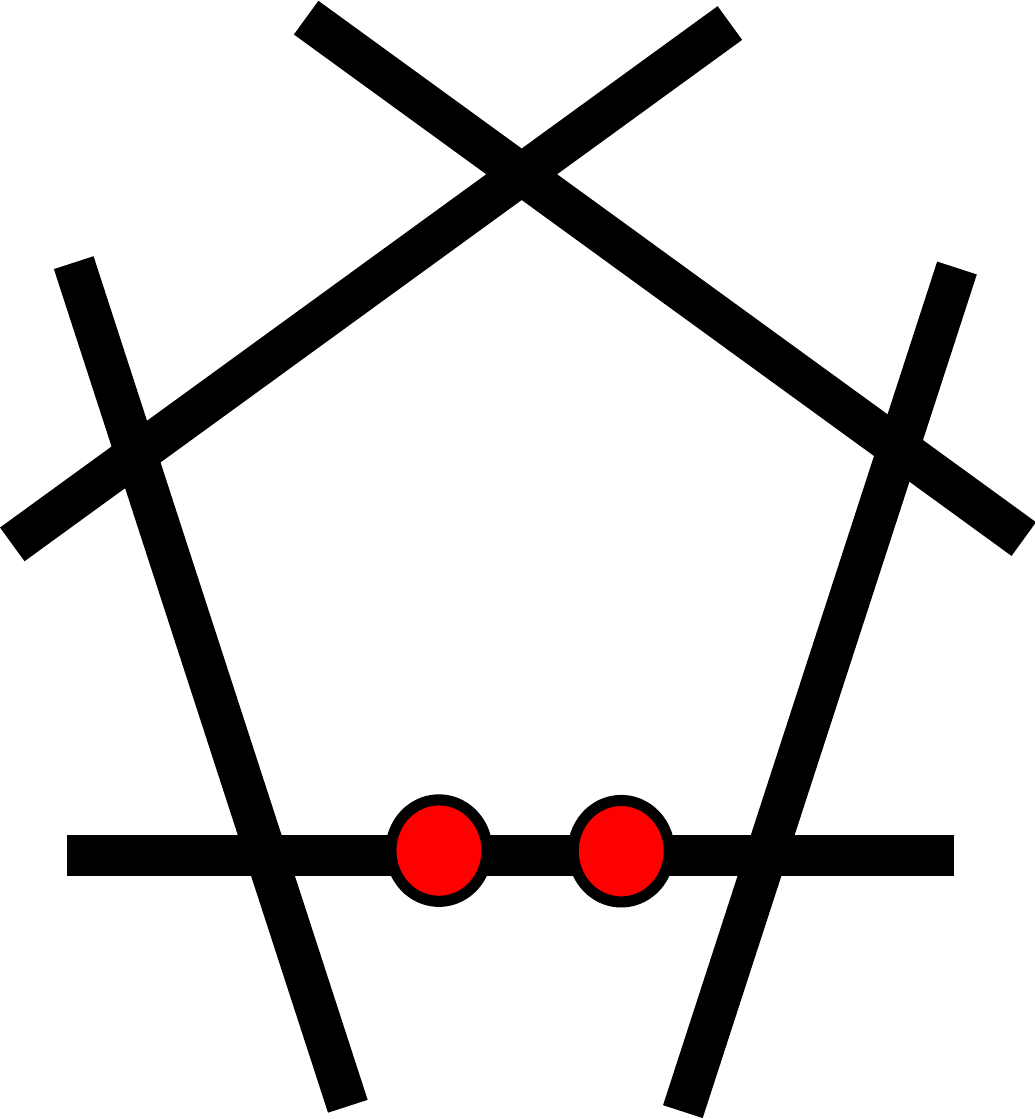} \qquad
  \includegraphics[width=3.5cm]{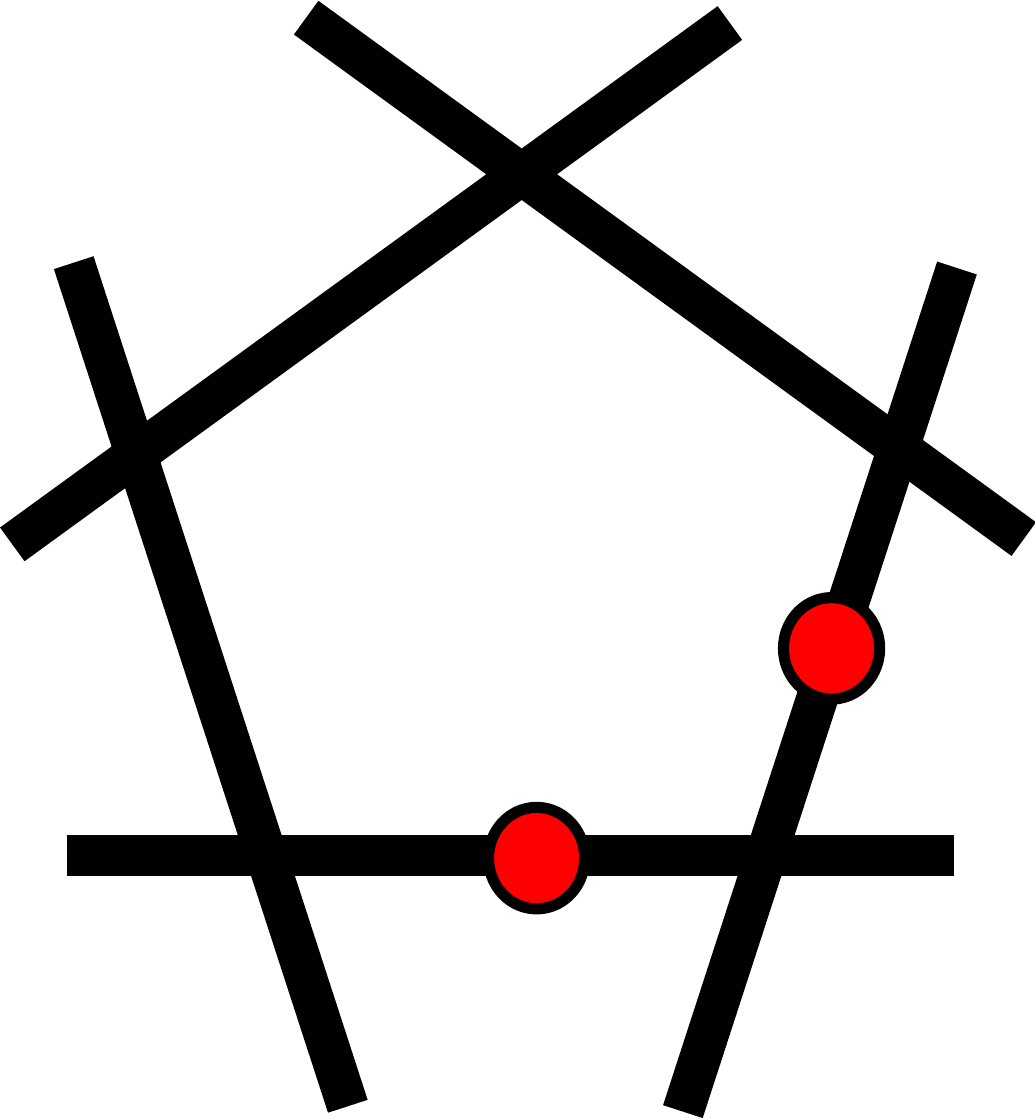} \qquad
  \includegraphics[width=3.5cm]{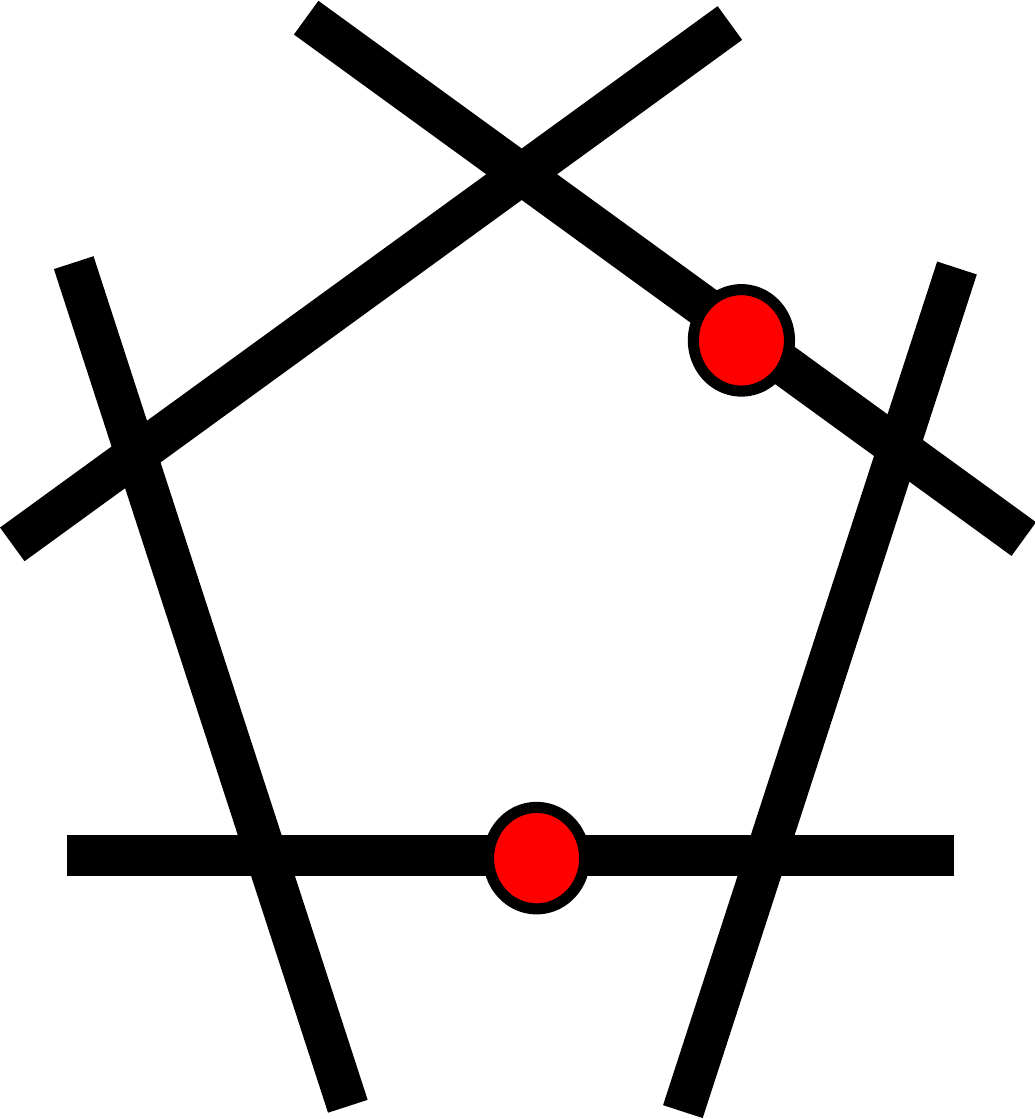}
  \caption{From left to right, showing the $I_5^{(01)}$, $I_5^{(0|1)}$ and $I_5^{(0||1)}$ fibers, respectively, with sections marked in red.\label{fig:I5Fibs}}
\end{figure}

\begin{figure}
  \centering
  \includegraphics[width=8cm]{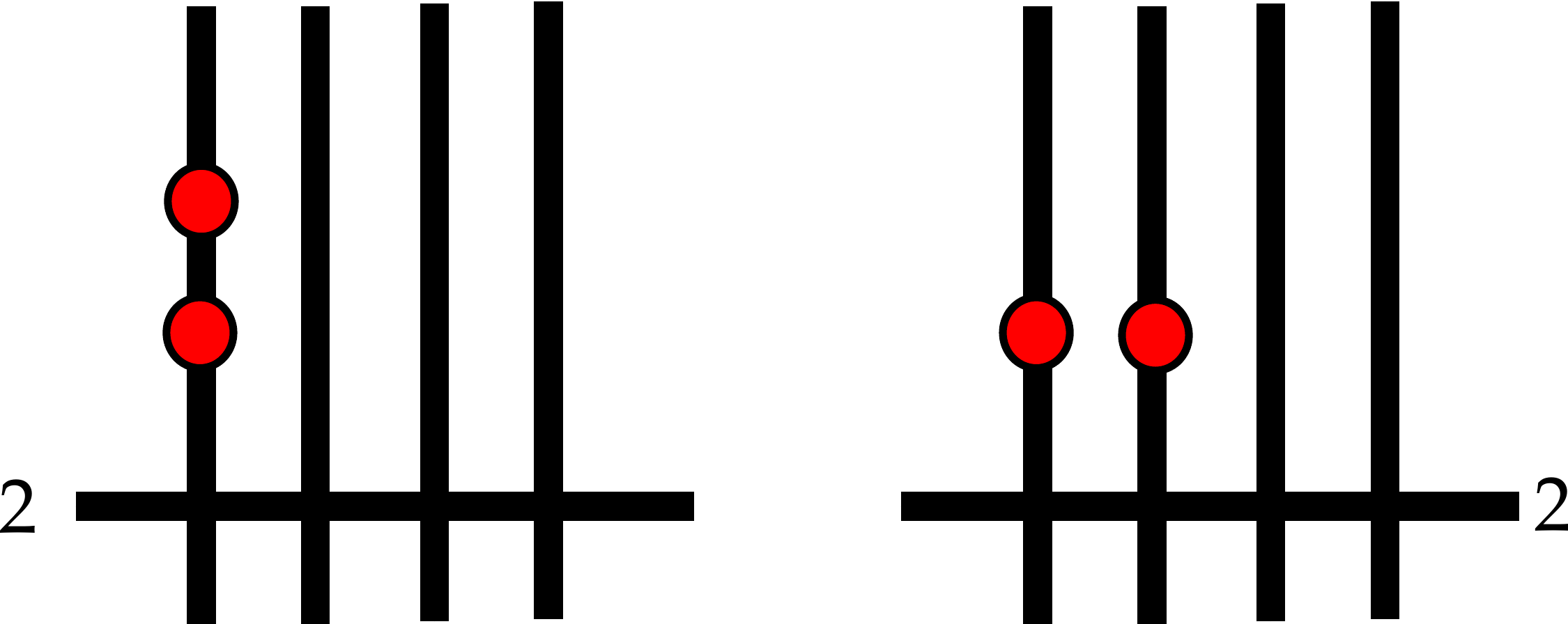}
  \caption{$I_0^{*(01)}$ and $I_0^{*(0|1)}$ fibers, where the 2 next to a black lines indicates multiplicity two of the fiber component, all other components are multiplicity one. The extra sections can only be on the multiplicity one fiber components. \label{fig:I0sFibs}}
\end{figure}

\begin{itemize}
  \item $P_0=0$: $I_5^{(01)}$ \\
  This polynomial term can be solved as in  (\ref{P0Sol}), and in fact allows for a shift to a canonical model, corresponding to   $c_{0,4}=c_{1,2}=0$. The fiber type is shown on the left of figure  \ref{fig:I5Fibs}, and after the shift this is realized as a canonical model
    \begin{equation} \label{Q5310002}  
      I_5^{(01)}:\qquad    \CQ (5, 3, 1, 0, 0, 0, 2) \,.
    \end{equation}
Note that this model also appears from the other branch, starting with $I_4^{(0|1)}$, i.e.  $\CQ(3,2,1,1,0,0,1)$, where we set $b_{0,0}=0$, and by the lopping we identify these models automatically. 
  \item $c_{3,0}=0$: $I_5^{(0|1)}$ \\
    This enhancement yields a fiber with canonical form, shown in the middle of figure  \ref{fig:I5Fibs}, which is realized by
    \begin{equation}\label{Q4211002}
   I_5^{(0|1)}:\qquad   \CQ(4,2,1,1,0,0,2) \,.
    \end{equation}

  \item $b_{1,0}=0$: $I_0^{*(01)}$ \\
    Again $b_{1,0}$ moves out of the $I_n$ branch, and at this order starts entering the $I_n^*$ branch, which realizes the $SO(2n)$ gauge groups, shown in figure \ref{fig:I0sFibs},
    \begin{equation}
      I_0^{*(01)}:\qquad    \CQ (4,2,1,0,0,1,2) \,.
    \end{equation} 
    with the sequence $(z,x,y,\zeta_1)$, $(\zeta_1,y,\zeta_2)$, $(\zeta_1,\zeta_2,\zeta_3)$, $(\zeta_2,x,\zeta_4)$. After these blow-ups, the divisor $z\zeta_1=0$ does not intersect the elliptic fibration anymore, and the fiber components are $z$, $\zeta_2$, $\zeta_3$ and $\zeta_4$, with all curves only intersecting $\zeta_4$. However, $\zeta_2=0$ here is a doubled curve, in the sense that it has self-intersection $-4$ and intersects $\zeta_4$ twice. Furthermore, the fibration over $\zeta_2=0$ is given by
    \begin{equation}
      c_{1,1} z^2 + c_{2,1} z \zeta_4 + c_{3,1} \zeta_4^2 = 0 \,.
    \end{equation}
    This equation will factor into two parts if its discriminant is a perfect square, that is, if
    \begin{equation}
      c_{2,1}^2 - 4 c_{1,1}c_{3,1} = p^2
      \label{eq:I0sMonodromy}
    \end{equation}
    for some section $p$. If this is the case, $\zeta_2=0$  splits into two fiber components, and we obtain an intersection structure like the one in figure \ref{fig:I0sFibs} on the left. If the discriminant is not a perfect square, the fiber  will still locally look like the one in the figure, but there will be a monodromy relating two of the multiplicity 1 fiber curves on which there is no section. The former case is denoted $I_0^{*s}$ in the literature (with associated gauge group $SO(8)$), and the latter one $I_0^{*ss}$ (with associated gauge group $SO(7)$).
    
Although we specialized to split-type models at the beginning of this section, let us also note that Tate's algorithm yields a realization of the $I_0^{*ns(01)}$ fiber type with associated gauge group $G_2$. Its equation is given by
    \begin{equation}
   I_0^{*ns(01)}:\qquad    \CQ(4,2,0,0,0,0,2) \,,
    \end{equation}
    with the additional monodromy condition that 
    \begin{equation}
      c_{2,0}=-b_{1,0}^2/4 \,.
      \label{eq:I0nsMonodromy}
    \end{equation}      
    Note that the fiber indeed becomes  $I_0^{*ss(01)}$ for $b_{1,0}=0$.
\end{itemize}


\subsubsection{$I_4^{(0|1)}$ Branch}
\label{sec:I40s1}

The second branch at order $z^4$ emanates from $I_4^{(0|1)}$, realized in terms of $\CQ(3,2,1,1,0,0,1)$ in (\ref{eq:3211001}), which has leading order discriminant
\begin{equation}
  \Delta_{I_4^{(0|1)}} = b_{1,0}^4 b_{0,0} P_0 P_1 z^4 + O(z^5)\,,
\end{equation}
with polynomial terms
\begin{align}
  P_0 &= b_{0,0} c_{2,1} - b_{1,0} c_{3,1} \,, \label{eq:D3211001P0}\\
  P_1 &= b_{1,0}^2 c_{0,3} - b_{1,0} b_{2,1} c_{1,2} + b_{2,1}^2 c_{2,1} \label{eq:D3211001P1}\,.
\end{align}
The case $b_{0,0}$ already was reached in the other branch by (\ref{Q4210002}), and is lopped out. 
\begin{itemize}
   \item $P_0=0$: $I_{5, nc}^{(0||1)}$ \\
    Using appendix \ref{sec:TwoTermPoly}, one can solve for $P_0=0$, which results in a non-canonical form with the sections located on twice removed fiber components, see the right most fiber in figure \ref{fig:I5Fibs}, 
    \begin{equation}
    I_{5, nc}^{(0||1)}:\qquad   \CQ(3,2,1,1,0,0,1)|_{(\ref{eq:D3211001P0})} \,.
      \label{eq:Q3211001P0}
    \end{equation}
  \item $P_1=0$: $I_{5}^{(0|1)}$, $I_{5, nc}^{(0|1)}$ \\
    This fibration, too, has a non-canonical form, and solving $P_1=0$ using appendix \ref{sec:ThreeTermPoly} yields the canonical model, which is exactly already reached by alternative route in (\ref{Q4211002}), as well as the non-canonical
    \begin{equation}
  I_{5, nc}^{(0|1)}:\qquad     \CQ(3,2,1,1,0,0,1)|_{(\ref{eq:D3211001P1})} \,.
      \label{eq:Q3211001|P1}
    \end{equation}
  \item $b_{1,0}=0$: $I_0^{*(0|1)}$ \\
    This enhancement yields a fiber with canonical form
    \begin{equation}
     I_0^{*(0|1)}: \qquad \CQ(3,2,1,1,0,1,1) \,.
    \end{equation}
   The $I_0^{*(0|1)}$-fiber also has a semi-split version, given by the form
    \begin{equation}
      I_0^{*ss(0|1)}: \qquad \CQ(2,2,1,1,0,1,1) \,.
    \end{equation}
    This form can be obtained e.g., as an enhancement of the $IV^{(0|1)}$ fiber (\ref{eq:IV0s1}).
\end{itemize}


\subsection{Codimension two fibers for canonical $I_5$}
\label{sec:I5s}

After working through Tate's algorithm starting from the $I_1$ fibration of $\mathrm{Bl}_{[0,1,0]}\bbP^{(1,1,2)}$, one finds two canonical $I_5$ models: (\ref{Q5310002}) and (\ref{Q4211002}). 
These two models will be of particular interest for  applications in F-theory model building. Therefore we will provide a few more details for these fiber types. First of all, we can determine the next order discriminant, and thereby the codimension 2 fiber types of these models. This allows computation also of the matter and corresponding $U(1)$ charges induced by the additional section, using the methods outlined in section \ref{sec:Fibration}. The results are given in table \ref{tab:I5Canw=1}, and correspond to top 1 and 2 of \cite{Borchmann:2013jwa}, respectively. A detailed discussion of the map to tops is given in appendix \ref{sec:tops}.

We argued in appendix \ref{app:Deligne}, that $b_0\not=0$ in codimension one. However, it can vanish in codimension two or higher. This leads to sections ``wrapping" entire fiber components. Nevertheless there is a
characterization of codimension two fibers with a specific $U(1)$ charge, similar to the notation used for codimension one fibers, in terms of section separation \cite{LSW}.

The non-canonical models will be discussed later, and go beyond the top models.  
We find that the codimension two fiber structure does not depend on a specific realization of a codimension one fiber type. Furthermore, the fiber structure in codimension one and two determines uniquely the matter and $U(1)$ charges.

\begin{table}
\centering
  \begin{tabular}{c|c|c|c|c}
Fiber &  Model & Codim 2 locus & Representation & Codim 2 fiber \\ \hline \hline                    
   $ I_5^{(01)}$& $\CQ(5, 3, 1, 0, 0, 0, 2)$ & 
   			      $b_{1,0}$ & $\mathbf{10}_0 + \overline{\mathbf{10}}_{0}$ & ${I_1^*}^{(01)}$ \\
               &          & $c_{3,0}$ & $\mathbf{5}_{-1} + \overline{\mathbf{5}}_{1}$ & $I_6^{(0|1)}$\\
                &         & $c_{3,0}+b_{0,0}b_{1,0}$ & $\mathbf{5}_{1} + \overline{\mathbf{5}}_{-1}$ & $I_6^{(1|0)}$ \\
                 &        & $b_{1,0}^2c_{0,5} - b_{1,0}b_{2,2}c_{1,3} + b_{2,2}^2c_{2,1}$ & $\mathbf{5}_{0} + \overline{\mathbf{5}}_{0}$ & $I_6^{(01)}$ \\ \hline
                    
 $I_5^{(0|1)}$    &   $\CQ(4,2,1,1,0,0,2)$ & 
 			$b_{1,0}$ & $\mathbf{10}_2 + \overline{\mathbf{10}}_{-2}$ &  ${I_1^*}^{(0|1)}$\\
     &                      & $b_{0,0}$ & $\mathbf{5}_6 + \overline{\mathbf{5}}_{-6}$ &$I_6^{(01)}$ \\
     &                     & $b_{0,0} c_{2,1} - b_{1,0} c_{3,1}$ & $\mathbf{5}_{-4} + \overline{\mathbf{5}}_{4}$ &$I_6^{(0||1)}$ \\
     &                    & $ b_{1,0}^2 c_{0,4} - b_{1,0} b_{2,2} c_{1,2} - c_{1,2}^2$ & $\mathbf{5}_{1} + \overline{\mathbf{5}}_{-1}$ &$I_6^{(0|1)}$
  \end{tabular}
  \caption{There are two canonical $I_5$ models, for which we tabulate the vanishing order, codimension 2 enhancement loci, and the corresponding matter with $U(1)$ charges and codimension two fiber types. The models $I_5^{(01)}$ and $I_5^{(0|1)}$   agrees with the top 1 and 2, respectively, in the toric nomenclature of \cite{Borchmann:2013jwa}. 
}
  \label{tab:I5Canw=1}
\end{table}


Note that the fiber type in codimension two uniquely corresponds to a given matter locus, with the exception of the two loci $c_{3,0}$ and $c_{3,0}+b_{0,0}b_{1,0}$. This is not very surprising however, as the set of $U(1)$ charges in this model has a charge symmetry $q\rightarrow -q$, interchanging the charges of the two matter loci in question and leaving the other charges unchanged.


\begin{figure}
  \centering
  \includegraphics[width=12cm]{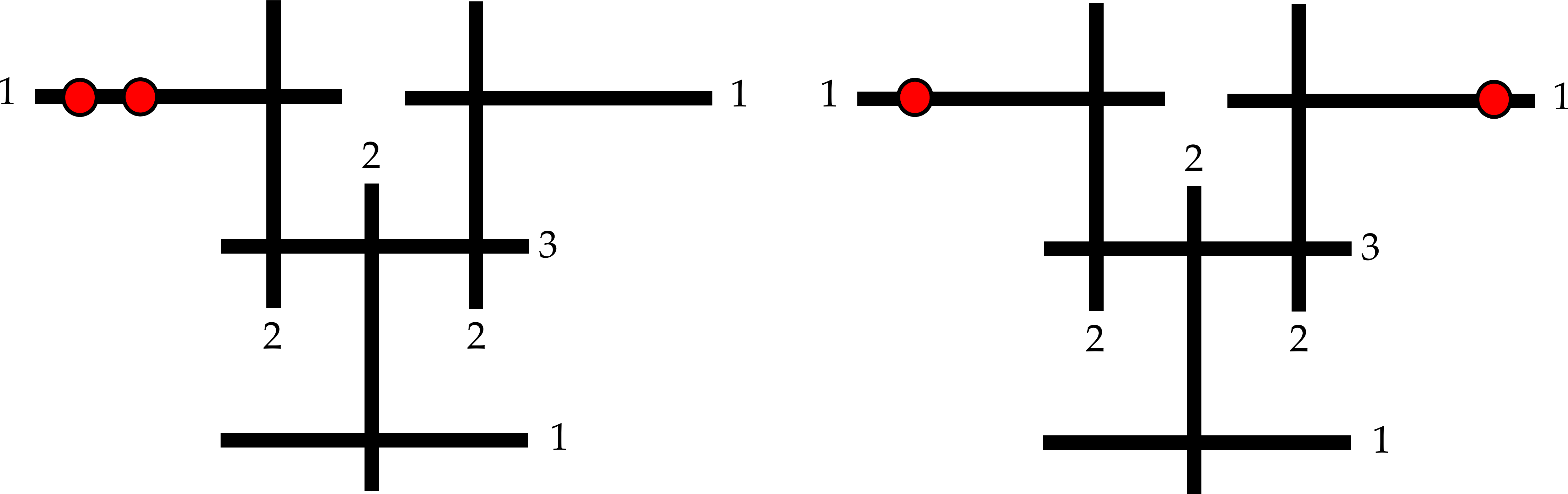} \qquad
   \caption{ Modulo the $\mathbb{Z}_3$ symmetry, there are two type $IV^{*}$ fibers with two sections: $IV^{*(01)}$ and $IV^{*(0|1)}$.  Numerical labels indicate the multiplicity of the fiber components. The sections can again only meet the multiplicity one fiber components. }
  \label{fig:IV*}
\bigskip
\bigskip
  \centering
  \includegraphics[width=12cm]{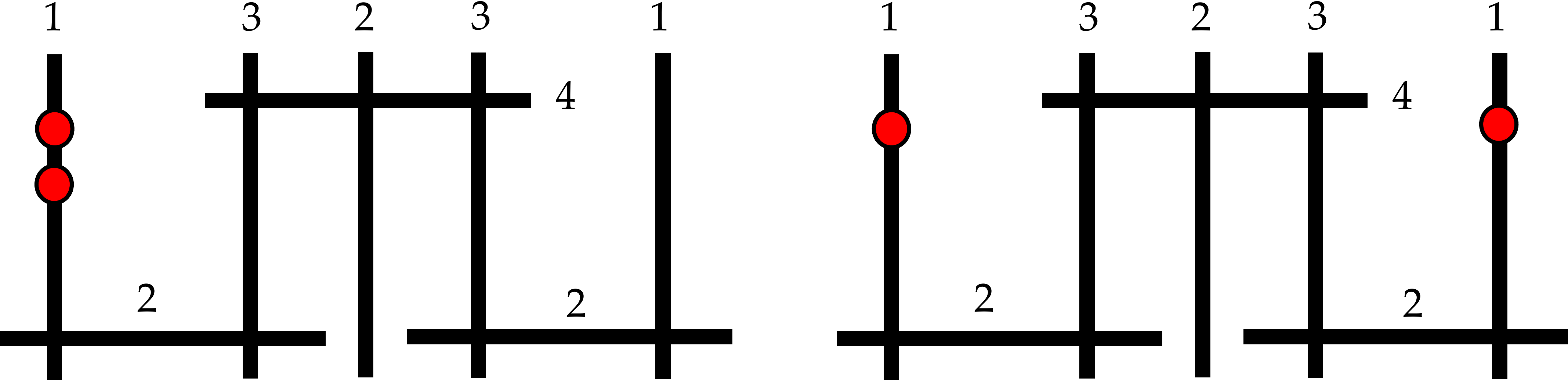} 
  \caption{$ III^{*(01)}$ and $ III^{*(0|1)}$ fibers with the sections passing through on the two multiplicity one fiber components only. Numerical labels specify the multiplicity of the fiber components. }
  \label{fig:III*}

\bigskip
\bigskip
  \centering
  \includegraphics[width=9cm]{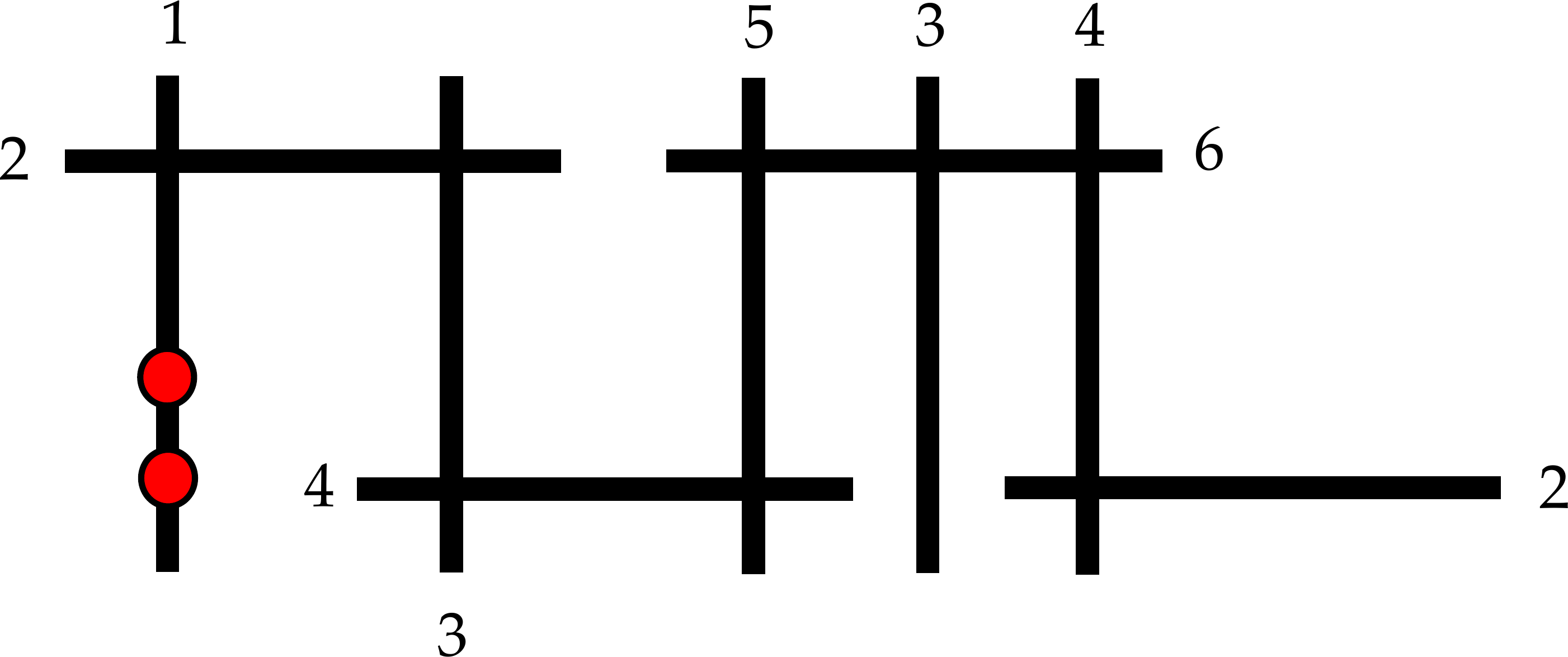} \qquad
  \caption{There is exactly one $II^{*(01)}$ fiber type, with both sections on the single multiplicity one fiber component. }
  \label{fig:II*}
\end{figure}


\section{Tate tree tops and infinite branches}
\label{sec:TreeTopsInfty}

Despite the Tate algorithm being somewhat more involved in the present case, one can determine the remaining branches. 
Even with the  lopping transformation taken into account to reduce the number of presentations of a given fiber type, it is still a tour de force to prove the algorithm by induction for all $I_n, I_n^*$. The existence of (multiply) non-canonical forms further complicate this matter. Here we determine the ``tree tops", i.e. fibers realizing exceptional gauge groups models, which enhance to non-minimal models (and are thus endpoints of the algorithm). We also present canonical forms for $I_n$ and $I_n^*$ with any section distribution.

\subsection{Tate tree tops}
\label{sec:TreeTops}

Some branches of the Tate tree stop, as one reaches fibers with exceptional type gauge groups, such as $II^*$, $III^*$, $IV^*$. 
Further enhancement of the discriminant beyond these ``tree tops" yields non-minimal models. 

\begin{itemize}
  \item The $IV^*$ fiber types are depicted in figure \ref{fig:IV*}. From Tate's algorithm, there are two forms, which are canonical, given by\footnote{The nomenclature for these fibers is as explained in section \ref{sec:TateSetup}. I.e. the superscript on the standard Kodaira-Neron fiber label refers only to the multiplicity one components in the fiber, where the sections can meet.}
  \be
    \ba
      IV^{*(01)}: &\qquad \CQ(5,3,2,0,0,1,2) \cr
      IV^{*(0|1)}: &\qquad \CQ(3,2,2,1,0,1,2)\,.
    \ea
  \ee
 \item There are two $III^*$ fiber types, shown in figure \ref{fig:III*}. From the algorithm, these are realized by the canonical models
  \be
    \ba
      III^{*(01)}: &\qquad  \CQ(5,3,2,0,0,1,3) \cr
      III^{*(0|1)}: &\qquad \CQ(3,3,2,1,0,1,2) \,.
    \ea
  \ee
  \item There is only a single $II^{*(01)}$ fiber is given in figure \ref{fig:II*}, realized for instance in terms of 
  \be 
    II^{*(01)}: \qquad  \CQ(5,4,2,0,0,1,3) \,.
  \ee
\end{itemize}
Note that all these models are such that the sections intersect the multiplicity one fiber components only, confirming our earlier general argument. There are no non-canonical forms present in the algorithm for these fiber types.\footnote{At first this is an empirical observation, i.e., the discriminant in those cases does not have non-trivial polynomial terms. 
This is largely due to the fact that these branches have $b_{1,0}=0$, which turns all the non-trivial polynomials encountered in section \ref{sec:Tate} into simple one-term discriminant factors.  This seems to be closely related to the issue that arises in codimension 2 enhancements, where  a non-abelian commutant  in the codimension 2 enhanced symmetry group results in monodromy, rather than multiple enhancement loci, like in \cite{Hayashi:2014kca}.}
The codimension two enhancements and spectra of these models are summarized in table \ref{tab:E67charges}.

\begin{table}
\centering
  \begin{tabular}{c|c|c|c|c}
  Fiber &  Model & Codim 2 locus & Representation & Codim 2 fiber \\ \hline \hline                    
   $IV^{*(01)}$ & $\CQ(5,3,2,0,0,1,2)$ 
          & $b_{2,2}$ & $\mathbf{27}_0 + \overline{\mathbf{27}}_{0}$ & ${III^*}^{(01)}$ \\
       &  & $c_{3,0}$ & --- & non-minimal \\ \hline
   $IV^{*(0|1)}$ & $\CQ(3,2,2,1,0,1,2)$ 
          & $b_{0,0}$ & $\mathbf{27}_2 + \overline{\mathbf{27}}_{-2}$ & ${III^*}^{(01)}$ \\
       &  & $c_{1,2}$ & $\mathbf{27}_{-1} + \overline{\mathbf{27}}_{1}$ & ${III^*}^{(0|1)}$ \\ \hline
   $III^{*(01)}$ & $\CQ(5,3,2,0,0,1,3)$ 
          & $c_{1,3}$ & $\mathbf{56}_{0} + \overline{\mathbf{56}}_{0}$ & ${II^*}^{(01)}$ \\
       &  & $c_{3,0}$ & --- & non-minimal \\ \hline
   $III^{*(0|1)}$ & $\CQ(3,3,2,1,0,1,2)$ 
          & $b_{0,0}$ & $\mathbf{56}_1 + \overline{\mathbf{56}}_{-1}$ & ${II^*}^{(01)}$ \\
       &  & $c_{0,3}$ & --- & non-minimal
  \end{tabular}
  \caption{Codimension two fiber types and $U(1)$ charges for the $IV^*$ and $III^*$ models.  }
  \label{tab:E67charges}
\end{table}

Finally, the $IV^{*ns}$ fiber, which realizes the group $F_4$, is given by
\begin{equation}
  IV^{*ns}: \qquad \CQ(4, 3, 2, 0, 0, 1, 2) \,.
\end{equation}
This model degenerates to the split case $IV^*$ with $c_{0,4}=0$. 
The vanishing of the leading order discriminant for the non-split case 
\begin{equation}
  \Delta_{IV^{*ns}} = c_{3,0}^4 \left( b_{2,2}^2 + 4 c_{0,4} \right) z^8 + O(z^9) \,,
\end{equation}
 enhance either to a non-minimal model ($c_{3,0}=0$) or back to $III^{*(01)}$.


\begin{figure}
  \centering
  \includegraphics[width=9cm]{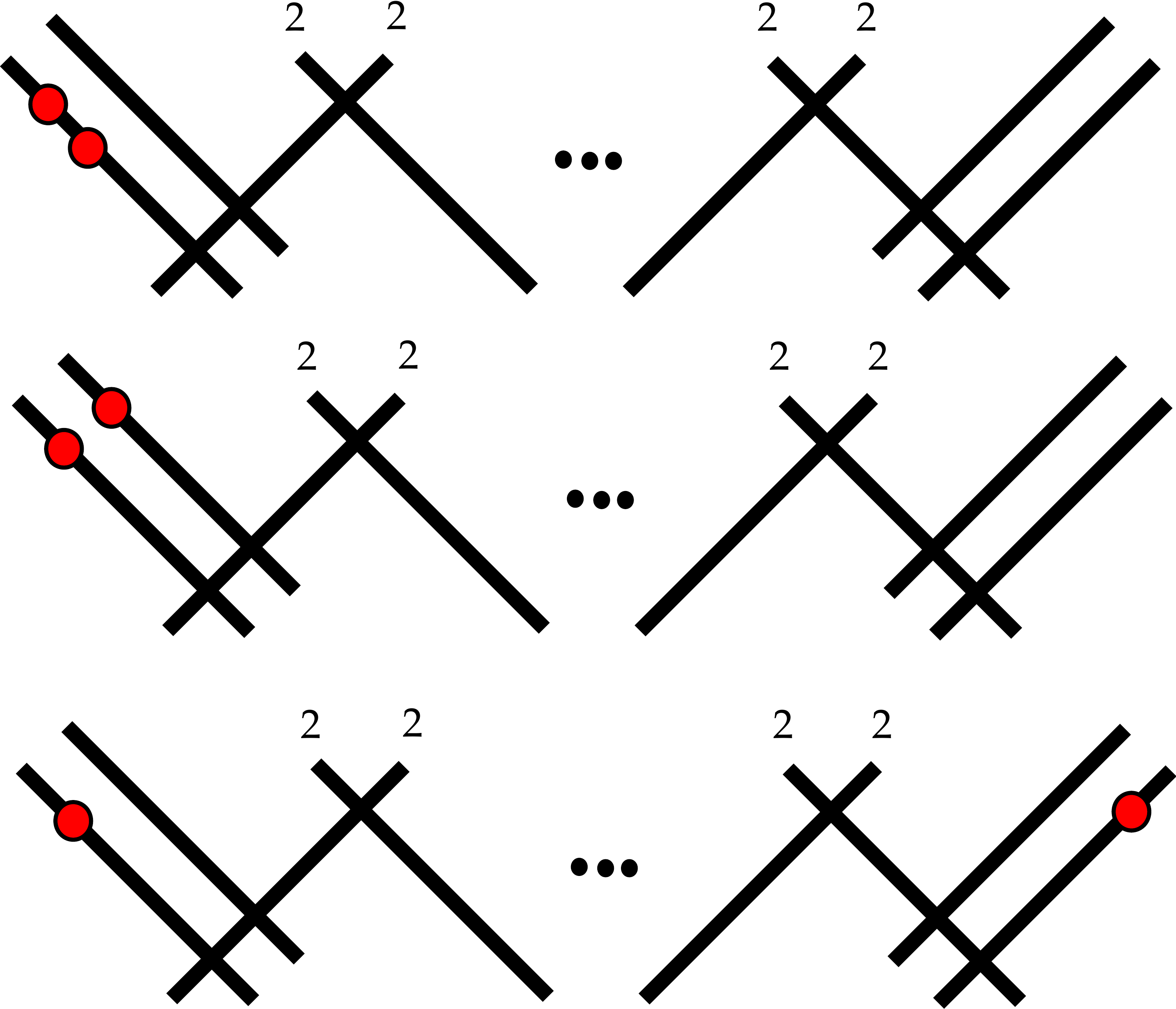} 
  \caption{From top to bottom: $I_n^{*(01)}$, $I_n^{*(0|1)}$ and $I_n^{*(0||1)}$ fiber types with sections, which can intersect only the  multiplicity one fiber components. Modulo the symmtries of the affine $D_n$ Dynkin diagram, there are three distinct such distributions, all of which occur in the Tate tree. }
  \label{fig:In*}
\end{figure}



\subsection{$I_n^*$ Branch}

The fibers of Kodaira type $I_n^*$ only have four multiplicity one fibers components, which can intersect the sections by the argument presented in section \ref{sec:Constr}.  Therefore, modulo the symmetries of the affine $D_n$ Dynkin diagram, there are three distinct fiber types, denoted $I_n^{*(01)}$, $I_n^{*(0|1)}$ and $I_n^{*(0||1)}$, respectively, which are shown in  figure \ref{fig:In*}.
We list  three infinite series of realizations, which together realize all these fiber types for $n\geq 1$ in table \ref{tab:FiberTypeTable}.

The two $I_0^*$ fiber types have been described in section \ref{sec:Tate}.
In agreement with the general argument from section \ref{sec:Constr}, there is no $I_n^{*ns(0||1)}$ fiber.
The resolved geometries and Cartan divisors of all split-type fibrations are given in appendix \ref{sec:ResolvedGeometries}.
For purposes of model building in F-theory, the codimension two structure and $U(1)$ charges of the $I_1^*$ fibers are tabulated in table \ref{tab:I1starcharges}.


\begin{table}
\centering
  \begin{tabular}{c|c|c|c|c}
  Fiber &  Model & Codim 2 locus & Representation & Codim 2 fiber \\ \hline \hline                    
   $I_1^{*(01)}$ & $\CQ(5,3,1,0,0,1,2)$ 
          & $c_{2,1}$ & $\mathbf{16}_{0} + \overline{\mathbf{16}}_{0}$ & $IV^{*(01)}$ \\
       &  & $c_{3,0}$ & $\mathbf{10}_{1} + \mathbf{10}_{-1}$ & $I_2^{*(0|1)}$ \\
       &  & $b_{2,2}$ & --- & $I_2^{*ns(01)}$ \\ \hline

   $I_1^{*(0|1)}$ & $\CQ(4,2,1,1,0,1,2)$ 
          & $c_{2,1}$ & $\mathbf{16}_{1} + \overline{\mathbf{16}}_{-1}$ & $IV^{*(0|1)}$ \\
       &  & $b_{0,0}$ & $\mathbf{10}_{-2} + \mathbf{10}_{2}$ & $I_2^{*(01)}$ \\
       &  & $c_{1,2}$ & --- & $I_2^{*ns(0|1)}$ \\ \hline
       
   $I_1^{*(0||1)}$ & $\CQ(3,2,2,1,0,1,1)$ 
          & $b_{2,1}$ & $\mathbf{16}_{-1} + \overline{\mathbf{16}}_{1}$ & $IV^{*(0|1)}$ \\
       &  & $b_{0,0}$ & $\mathbf{16}_{3} + \overline{\mathbf{16}}_{-3}$ & $IV^{*(01)}$ \\
       &  & $b_{0,0}c_{1,2}-b_{2,1}c_{3,1}$ & $\mathbf{10}_{2} + \mathbf{10}_{-2}$ & $I_2^{*(0||1)}$
  \end{tabular}
  \caption{Codimension two fiber types and $U(1)$ charges for the $I_1^*$ models. Note that the $\mathbf{10}$ and $\overline{\mathbf{10}}$ representations in $SO(10)$ are identical. Enhancements from split-type fibers to non-split-type fibers do not yield additional localized matter.}
  \label{tab:I1starcharges}
\end{table}


\subsection{$I_n$ Branch}
\label{sec:In}
The Kodaira $I_n$ fibers have multiplicity one for each fiber component, and so the fibers that arise when taking into account the structure of extra sections can be characterized by the number of fiber components $k$, that separate the two components  which intersect the sections $\sigma_0$ and $\sigma_1$. Such an  $I_n^{(0|^k1)}$ fiber  is realized by 
  \begin{equation}\label{Ininfty}
  \ba
I_{2m}^{(0|^k1)}:\qquad & 
   \CQ\bigg( 2m-k, \max \left\{ \left\lceil m - \frac{k}{2} \right\rceil, k \right\}, \max \left\{ 1, k \right\}, k, \\ 
   & \qquad \qquad \qquad 0, 0, \min \left\{ \left\lfloor m - \frac{k}{2} \right\rfloor, \max \left\{ 1, 2(m-k) \right\} \right\} \bigg) \cr
    I_{2m+1}^{(0|^k1)}:\qquad   & \CQ\bigg( 2m+1-k, \max \left\{ \left\lceil m+1 - \frac{k}{2} \right\rceil, k \right\}, \max \left\{ 1, k \right\}, k, \\
   & \qquad \qquad 0, 0, \max \left\{ \left\lfloor m+1 - \frac{k}{2} \right\rfloor, \max \left\{ 1, 2(m-k) + 1 \right\} \right\} \bigg) \,.
    \ea
\end{equation}
 In both cases, we assume $m\geq 3$, with the lower cases having been given in section \ref{sec:Tate}. The full ordered sequence of resolutions, the resolved geometries, and the Cartan divisors can again be found in appendix \ref{sec:ResolvedGeometries}.  An alternative representation of these is given in the summary table \ref{tab:FiberTypeTable}.

Note that these forms are only canonical. However, for $I_n$ with $n>9$, the general solutions to the polynomial discriminant factors of these $I_n$ models introduce codimension two loci which the fibration enhances to a non-minimal form. Requiring the absence of such loci allows us to always find coordinate shifts that put these models into canonical form. This is similar to the situation in $\mathbb{P}^{(1,2,3)}$. 

\begin{figure}
\centering
  \includegraphics[width=9cm]{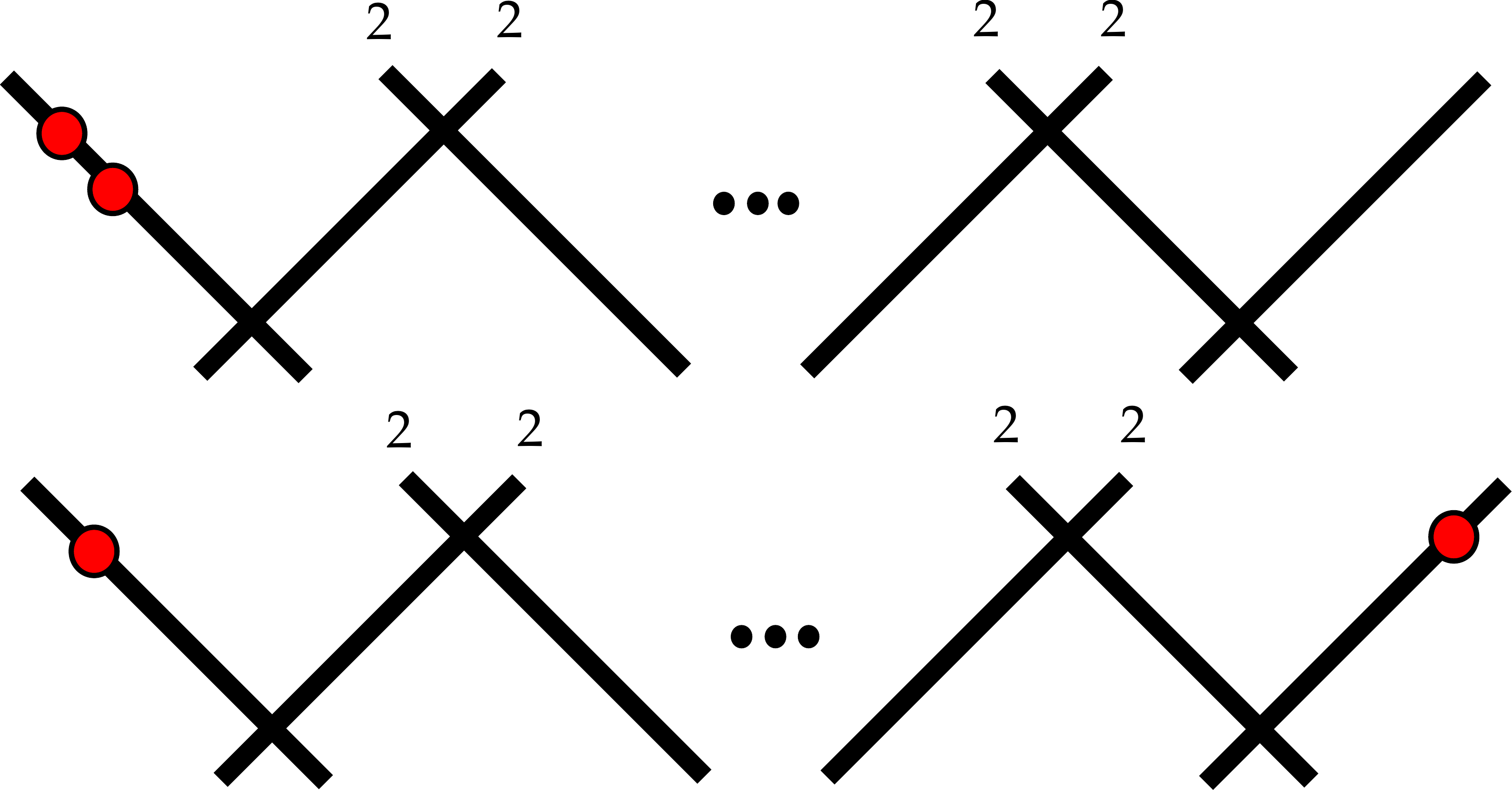}
  \caption{There are two non-split $I_{n}^{ns(01)}$ and $I_{n}^{ns(0|1)}$ fibers with two sections, which intersect the multiplicity one fiber components. }
  \label{fig:Innsfibers}
\end{figure}

Furthermore, there are fibrations realizing the non-split fiber types $I_{2m}^{ns(01)}$, $I_{2m+1}^{ns(01)}$, and $I_{2m}^{ns(0|1)}$. They are given by
\begin{equation}
  \begin{aligned}
    I_{2m}^{ns(01)}: \qquad & \CQ(2m, m, 0, 0, 0, 0, m) \\
    I_{2m+1}^{ns(01)}: \qquad & \CQ(2m+1, m+1, 0, 0, 0, 0, m+1) \\
    I_{2m}^{ns(0|1)}: \qquad & \CQ(m, m, m, m, 0, 0, 0) \,.
  \end{aligned}
\end{equation}
While the first two of these forms arise as enhancements of the $I_1$ starting point fiber, the last series is obtained by enhancing the third starting point fibration from (\ref{ExtraStart}), which was not contained within a single patch of the ambient space $X$ over the entire locus $z=0$ in the base manifold $B$. The fiber types are depicted in figure \ref{fig:Innsfibers}.

\section{Tate Trees: Non-Canonical Forms}
\label{sec:P112NonCanonical}

In the last two section we considered only the canonical enhancement patterns in the Tate tree, i.e. those that are characterized by vanishing orders of the coefficient sections alone. There are several branches which are however non-canonical: there is no shift or simple coordinate change that is locally well-defined and will put the forms into a canonical form. These arise whenever the discriminant has a factor that is a quadratic or higher degree polynomial in the sections $c_i$ and $b_j$. 
Studying these branches amounts to finding solutions to polynomial equations in UFD, which can be done explicitly in simple instances and is summarized in appendix \ref{sec:PolySols}. 

For $\mathbb{P}^{(1,2,3)}$ a similar analysis was performed for the standard Tate's algorithm in \cite{Katz:2011qp}, 
where one cannot achieve the standard (i.e. canonical) Tate form in only a few outlier cases.
In $\mathbb{P}^{(1,1,2)}$, the situation is quite different: 
non-canonical forms are very common. In fact, 
each non-canonical form gives rise to a new branch of the algorithm, with multiply-non-canonical forms, 
e.g. a canonical $I_n$ model can enhance to a non-canonical $I_{n+1}$ model, which in turn can have a non-trivial polynomial term in the discriminant, which yields a doubly non-canonical $I_{n+2}$ model etc.  
In section \ref{sec:Tate} we only summarized the fiber types of these non-canonical models and will now provide details for these, as well as some studies of doubly non-canonical models. 
Multiply non-canonical models can be quite involved, we leave this for future work. 

From the point of view of model building in F-theory, these non-canonical forms open up some exciting model building prospects. The types of codimension one fibers that can be realized in terms of non-canonical models are the same as in the canonical branch. 
However, the codimension two structure is very different, and allows for instance to have multiple enhancement loci from $I_n$ to $I_{n-4}^*$. Concretely, for $I_5$ models realizing $SU(5)$ gauge theories, this means there are multiple, distinct loci with ${\bf 10}$ matter, charged differently under the $U(1)$ that arises from the extra section. 
In the following we will concentrate on the  non-canonical $I_5$ fibers, either arising from canonical $I_4$ or non-canonical $I_4$. 


\subsection{Non-canonical $I_5$ from canonical $I_4$}
\label{sec:NonCanI5}

Starting with the canonical $I_4$ models, there are two non-canonical enhancements to $I_5$, both of which emanate from $I_4^{(0|1)}$ realized in terms of $\CQ(3,2,1,1,0,0,1)$, and are part of the branch discussed in section \ref{sec:I40s1}:
The codimension 2 fibers, and matter with $U(1)$ charge spectrum of these non-canonical $I_5$ models are summarized in table \ref{tab:I5NonCanCharges}. Note that the fiber structure in codimension 2 follows the pattern discussed in section \ref{sec:Constr}.

\begin{table}
\centering
  \begin{tabular}{c|c|c|c|c}
Fiber &  Model & Codim 2 locus & Representation & Codim 2 fiber \\ \hline \hline  
    $I_{5, nc}^{(0||1)}$ & $Q(3,2,1,1,0,0,1)|_{(\ref{eq:D3211001P0})}$ 
     & $\sigma_3$ & $\mathbf{10}_1 + \overline{\mathbf{10}}_{-1}$ & $I_1^{*(0||1)}$ \\
    && $\sigma_1$ & $\mathbf{10}_{-4} + \overline{\mathbf{10}}_{4}$ & $I_1^{*(01)}$ \\
    && $\sigma_2$ & $\mathbf{5}_{-7} + \overline{\mathbf{5}}_{7}$ & $I_6^{(0|1)}$ \\
    && (\ref{eq:NC1P2}) & $\mathbf{5}_{-2} + \overline{\mathbf{5}}_{2}$ & $I_6^{(0||1)}$ \\
    && (\ref{eq:NC1P3}) & $\mathbf{5}_{3} + \overline{\mathbf{5}}_{-3}$ & $I_6^{(0|||1)}$ \\\hline
    $I_{5, nc}^{(0|1)}$ & $Q(3,2,1,1,0,0,1)|_{(\ref{eq:D3211001P1})}$
     & $\sigma_1$ & $\mathbf{10}_{2} + \overline{\mathbf{10}}_{-2}$ & $I_1^{*(0|1)}$ \\
    && $\sigma_2$ & $\mathbf{10}_{-3} + \overline{\mathbf{10}}_{3}$ & $I_1^{*(0||1)}$ \\
    && $b_{0,0}$ & $\mathbf{5}_{6} + \overline{\mathbf{5}}_{-6}$ & $I_6^{(01)}$ \\
    && (\ref{eq:NC2P2}) & $\mathbf{5}_{-4} + \overline{\mathbf{5}}_{4}$ & $I_6^{(0||1)}$ \\
    && (\ref{eq:NC2P3}) & $\mathbf{5}_{1} + \overline{\mathbf{5}}_{-1}$ & $I_6^{(0|1)}$
  \end{tabular}
   \caption{Codimension two loci, fiber types, and matter and  $U(1)$ charges for non-canonical $I_5$ models arising from canonical $I_4$ models. These models generalize top 4, and tops 2 and 3 respectively.}
  \label{tab:I5NonCanCharges}
\end{table}

\subsubsection{$I_{5,nc}^{(0||1)}$}

This model arises in the algorithm in (\ref{eq:Q3211001P0}), as a specialization of the canonical $I_4$ model $\CQ(3,2,1,1,0,0,1)$, which has a component in the discriminant given by   
\begin{equation}
   P = b_{0,0} c_{2,1} - b_{1,0} c_{3,1} =0 \,.
\end{equation}
As explained in appendix \ref{sec:TwoTermPoly}, over a UFD, this requires the existence of new sections $\sigma_i$
satisfying
   \be\label{P0example}
   b_{0,0} = \sigma_1 \sigma_2 \,,\quad 
   c_{2,1} = \sigma_3 \sigma_4  \,, \quad 
   b_{1,0} = \sigma_1 \sigma_3 \,,\quad 
   c_{3,1} = \sigma_2 \sigma_4 \,,
   \ee
with $\sigma_2$ and $\sigma_3$ coprime, which automatically solves $P=0$.  The resulting model has fiber type
         \be
  I_{5,nc}^{(0||1)}:\qquad  \CQ(3,2,1,1,0,0,1)|_{(\ref{eq:D3211001P0})} \,,
   \ee    
and  leading order discriminant 
    \begin{equation}
      \Delta_{I_5^{(0||1)}} = \sigma_1^4 \sigma_3^4 \sigma_2 P_2 P_3 z^5 + O(z^6) \,,
    \end{equation}
    where
    \begin{align}
        \label{eq:NC1P2}
        P_2 =&\, \sigma_4 b_{2,1}^2 + \sigma_1^2 \sigma_3 c_{0,3} - \sigma_1 b_{2,1} c_{1,2} \\
        \label{eq:NC1P3}
        P_3 =&\, \sigma_1 \sigma_2^2 \left(\sigma_1 c_{1,2}-\sigma_4 b_{2,1}\right)+\sigma_3 \sigma_2 \left(\sigma_4
        \sigma_1 b_{1,1}-\sigma_1^2 c_{2,2} +\sigma_4^2\right) \notag\\
        &+ \sigma_1 \sigma_3^2 \left(\sigma_1 c_{3,2}-\sigma_4 b_{0,1}\right) \,.
    \end{align}
Note that  the standard $b_{1,0}=0$ locus is now reducible due to (\ref{P0example}), which gives
rise to two codimension two enhancements to $I_1^*$ (or ${\bf 10}$ matter loci), shown in figure \ref{fig:I1*0|1}.     
The spectrum is summarized in table \ref{tab:I5NonCanCharges}. {In appendix \ref{sec:tops}, it is shown that if the section $\sigma_1$ never vanishes on $B$, one can perform a coordinate shift to obtain the canonical model $\CQ(3,2,2,2,0,0,1)$, which is also known as top 4 in the literature.} This non-canonical model is therefore a generalization of the top 4 of  \cite{Borchmann:2013jwa}. 

 \begin{figure}
  \centering
  \includegraphics[width=5.5cm]{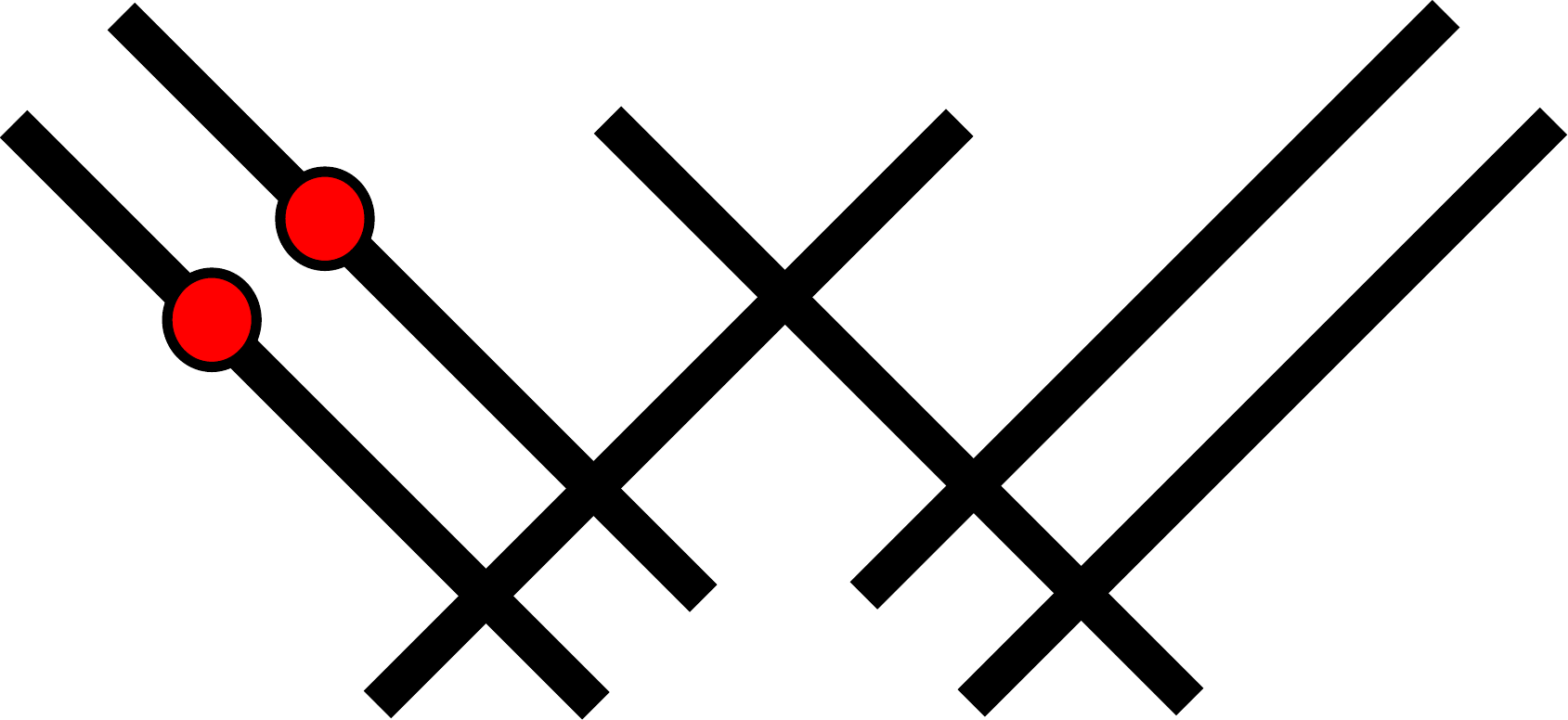} \qquad
  \includegraphics[width=5.5cm]{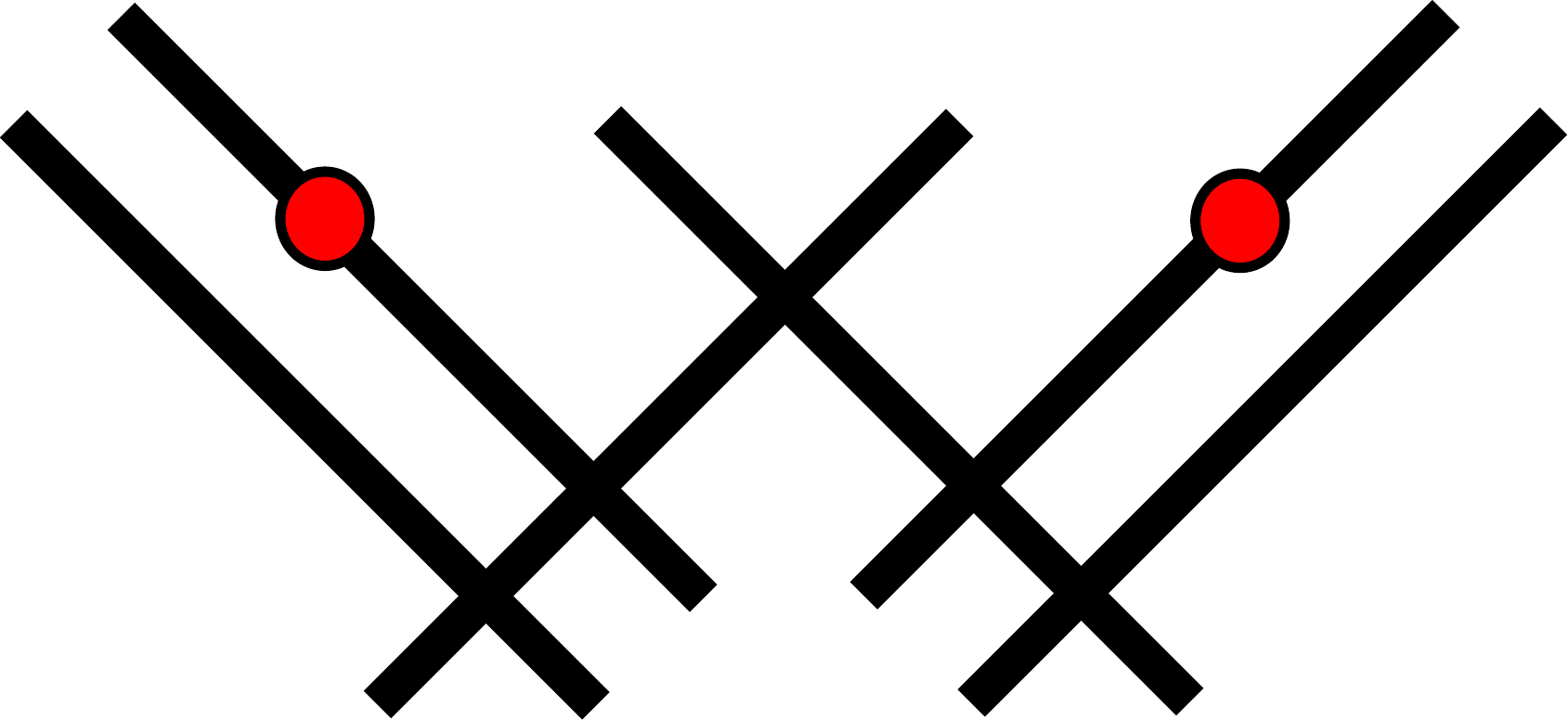}
  \caption{$I_1^{*(0|1)}$ and $I_1^{*(0||1)}$ fibers obtained in codimension two  of the $I_{5, nc}^{(0|1)}$ fiber over the curves $\sigma_1=0$ and $\sigma_2=0$, respectively.}
  \label{fig:I1*0|1}
\end{figure}

\begin{figure}
  \centering
  \includegraphics[width=5.5cm]{I1star0ssss1.pdf} \qquad
  \includegraphics[width=5.5cm]{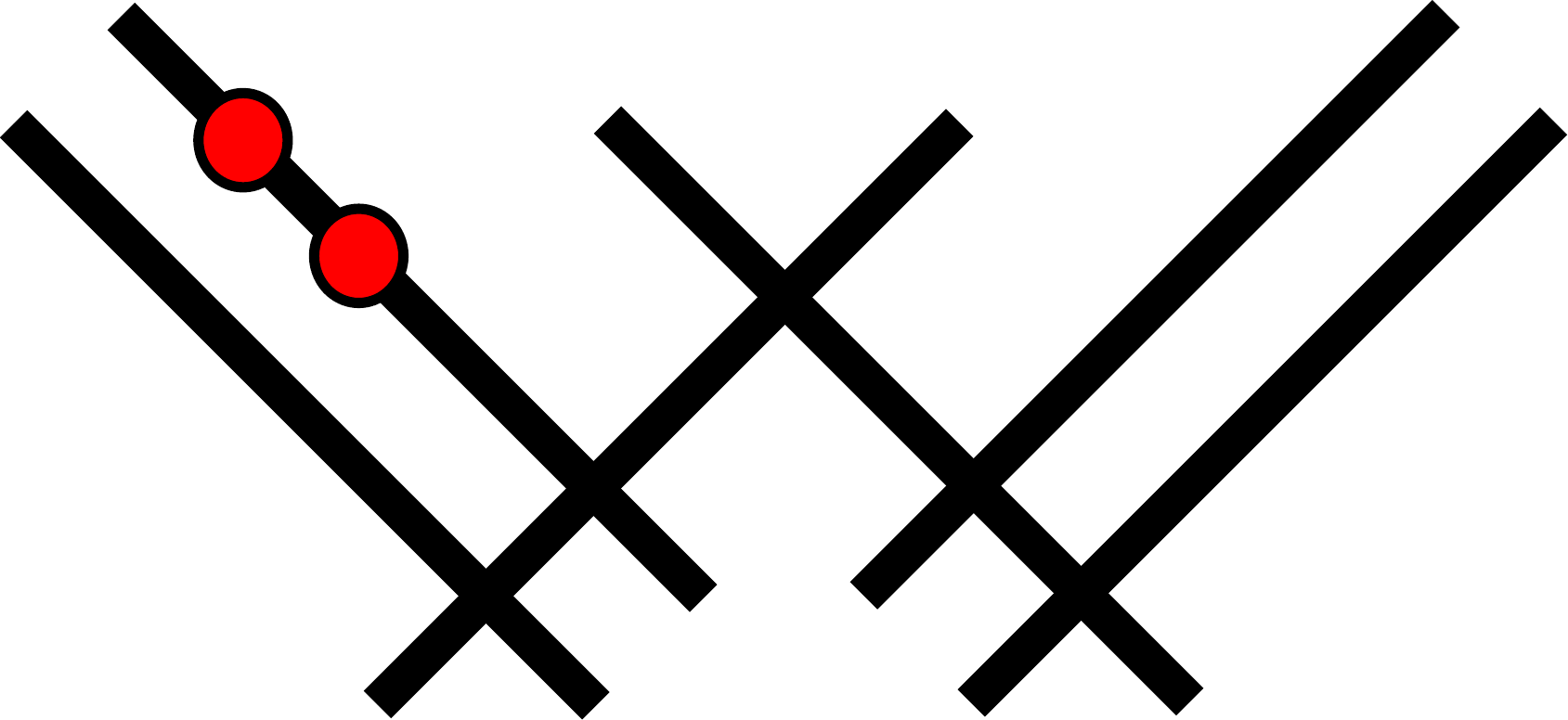}
  \caption{$I_1^{*(0||1)}$ and $I_1^{*(01)}$ fibers obtained in codimension two of the $I_{5. nc}^{(0||1)}$ fiber over the curves $\sigma_3=0$ and $\sigma_1=0$, respectively.}
  \label{fig:I1*0||1}
\end{figure}

\subsubsection{${I_{5, nc}^{(0|1)}}$}

Starting with the same canonical $I_4$ there is a non-canonical $I_5$ obtained by setting 
\be
P=  b_{1,0}^2 c_{0,3} - b_{1,0} b_{2,1} c_{1,2} + b_{2,1}^2 c_{2,1}  =0\,,
\ee
which has a general solution obtained in appendix \ref{sec:ThreeTermPoly}, if there exist sections $\sigma_i$ such that
\be
b_{1,0}=  \sigma_1 \sigma_2 \,,\quad 
 b_{2,1} = \sigma_1 \sigma_3 \,,\quad 
c_{0,3} = \sigma_3 \sigma_4 \,,\quad 
c_{2,1} = \sigma_2 \sigma_5 \,,\quad 
c_{1,2}= \sigma_2 \sigma_4 + \sigma_3 \sigma_5 \,,
\ee
again with $\sigma_2$ and $\sigma_3$ coprime. This model has fiber type 
\be 
I_{5, nc}^{(0|1)}:\qquad \CQ(3,2,1,1,0,0,1)|_{(\ref{eq:D3211001P1})}
\ee
and discriminant     
\begin{equation}
         \Delta_{I_5^{(0|1)}} = \sigma_1^4 \sigma_2^4 b_{0,0} P_2 P_3 z^5 + O(z^6) \,,
\end{equation}
where
 \begin{align}
        \label{eq:NC2P2}
        P_2 =&\, \sigma_1 c_{3,1} - \sigma_5 b_{0,0} \\
        \label{eq:NC2P3}
        P_3 =&\, - \sigma_2^3 \left(\sigma_4 \sigma_1 b_{2,2}-\sigma_1^2 c_{0,4} + \sigma_4^2 \right) + \sigma_3 \sigma_2^2 \left(\sigma_1 \left(\sigma_5 b_{2,2}-\sigma_1
        c_{1,3}\right)+\sigma_4 \left(\sigma_1 b_{1,1}+2 \sigma_5\right)\right) \notag \\
        & -\sigma_3^2 \sigma_2 \left(\sigma_1 \left(\sigma_4 b_{0,0}-\sigma_1 c_{2,2}\right)+\sigma_1 \sigma_5 b_{1,1}+\sigma_5^2\right)+\sigma_1 \sigma_3^3 \left(\sigma_5 b_{0,0}-\sigma_1 c_{3,1}\right) \,.
    \end{align}
Again $b_{1,0}= \sigma_1 \sigma_2$, factors and yields two codimension two loci of $I_1^*$ type, shown in figure \ref{fig:I1*0||1}, and the spectrum is listed in  table \ref{tab:I5NonCanCharges}. 

{If either of the two sections $\sigma_1$, $\sigma_2$ do not vanish on the base manifold, there is a coordinate shift, explicitly given in appendix \ref{sec:tops}, into the canonical models $\CQ(4,3,2,1,0,0,1)$ (if $\sigma_1$ does not vanish), or $\CQ(4,2,1,1,0,0,2)$ (if $\sigma_2$ is always nonzero). These two models are sometimes referred to in the literature as tops 3 and 2, respectively, and the non-canonical model is a generalization of these two toric models.}


\subsection{Non-canonical $I_5$ from non-canonical $I_4$}

Non-canonical $I_5$ models can arise also from non-canonical $I_4$ models, which in turn are enhancements of (canonical\footnote{We consider the branch starting from a  non-canonical $I_3$ case in the next subsection.}) $I_3$ models.  Solving in full generality for these discriminant loci is quite complicated, and we will present here only example solutions for these  doubly non-canonical forms. The key feature is that these models potentially allow for three enhancement loci to ${I_1^*}$.  
The models we present here will be special solutions to the discriminant equation for doubly non-canonical $ncnc$ models, i.e. enhancements along non-trivial polynomial factors in the discriminant, which arise in non-canonical models. We will address $ncnc$ type models also in section \ref{sec:P123Reloaded} in $\mathbb{P}^{(1,2,3)}$. 

\subsubsection{$I_{4, nc}^{(01)}$ Branch}
\label{sec:I5ncnc}
The form $\CQ(3,2,1,0,0,0,1)|_{(\ref{eq:D3210001P0})}$ from (\ref{eq:Q3210001P0}) has leading-order discriminant
\begin{equation}
  \Delta_{I_4^{(01)}} = \sigma_1^4 \sigma_2^4 \sigma_5 \left( \sigma_5 + \sigma_1 b_{0,0} \right) P_2 z^4 + O(z^5) \,,
\end{equation}
with
\begin{equation}
  \begin{aligned}
    P_2 &=\sigma _1  \left(\sigma _4 \sigma _2^2-\sigma _3 \sigma _6 \sigma _2+\sigma _3^2 \sigma _5\right) \left(\sigma _2^2 b_{2,2}+\sigma _3 \left(\sigma
   _3 b_{0,0}-\sigma _2 b_{1,1}\right)\right)\cr
&   +\sigma _2 \sigma _1^2 \left(\sigma _3 \left(\sigma _3 \left(\text{c31} \sigma _3-\sigma _2
   c_{2,2}\right)+\sigma _2^2 c_{1,3}\right)-\sigma _2^3 c_{0,4}\right)+\left(\sigma _4 \sigma _2^2-\sigma _3 \sigma _6 \sigma _2+\sigma _3^2
   \sigma _5\right){}^2 \,,
  \end{aligned}
\end{equation}
and $\sigma_2$, $\sigma_3$ coprime. $\sigma_1$ and $\sigma_2$ divide $b_{1,0}$, hence the corresponding enhancements are not considered here.
\begin{itemize}
  \item $\sigma_5=0$: $I_5^{(0|1)}$ \\
    $\sigma_5$ divides $c_{3,0}$. Thus, by first enhancing $c_{3,0}$ and then reconsidering the non-canonical enhancement that led to $\CQ(3,2,1,0,0,0,1)|_{(\ref{eq:D3210001P0})}$, one arrives at the non-canonical fibration $\CQ(3,2,1,1,0,0,1)|_{(\ref{eq:D3211001P1})}$.

  \item $P_2=0$: ${I_{5,ncnc}^{(01)}}$ \\
  Solving this in full generality is rather difficult. However, the main goal here is to obtain a class of solutions, that result in three charged ${\bf 10}$
matter loci, i.e. three loci where the fiber enhances to ${I_1^*}$.
  
    This enhancement yields a doubly non-canonical fibration. Note that $P_2$ is of schematic form
    \begin{equation}
      P_2 = \sigma_2^4 A - \sigma_2^3 \sigma_3 B + \sigma_2^2 \sigma_3^2 C - \sigma_2 \sigma_3^3 D + \sigma_3^4 (\sigma_5 (\sigma_5 + \sigma_1 b_{0,0})) \,.
    \end{equation}
    As in the solutions in appendix \ref{sec:PolySols}, one notes that necessary conditions for $P_2=0$ are $\sigma_3|\sigma_2^4 A$ and $\sigma_2|\sigma_3^4 ((\sigma_5 (\sigma_5 + \sigma_1 b_{0,0})))$. However, $\sigma_2=\sigma_3=0$ results directly in a non-minimal enhancement, and so we can discard this. Therefore we have $(\sigma_2, \sigma_3)=1$. 
Then one other alternative is that $\sigma_2 | \sigma_5$ (or $\sigma_2|(\sigma_5 + \sigma_1 b_{0,0})$ --we will not consider this and thereby the solution is only an example solution for this doubly non-canonical case), so that 
     \begin{equation}
      \chi =(\sigma_2,\sigma_5)\,, \qquad \sigma_2 = \chi  \chi_2 \,, \qquad \sigma_5 = \chi \chi_5 \,.
      \label{eq:DoublyNonCanDivisionConditions}
    \end{equation}
From this $b_{1,0} =\sigma_1 \sigma_2= \sigma_1 \chi \chi_2$, one expects that  this twice non-canonical model therefore should have three $\mathbf{10}$ curves.
Inserting this results in the polynomial, we obtain
\be
P_2= \chi( \chi \tilde{A}^2 + \tilde{B}  \tilde{A} \sigma_1 + \sigma_1^2  \chi_2 \tilde{C})\,,
\ee
where $\tilde{A} = \sigma_3^2 \chi_5 - \sigma_3 \sigma_6 \chi_2 + \sigma_4\chi \chi_2^2 $. At this point we specialize to the solution where we solve $\tilde{A}=0$ and then the remaining terms to vanish. This is not the general solution, however it will exemplify the feature that this model has three {\bf 10} matter loci. 
Solving $\tilde{A} =0$ by the three-term polyonimal solution  in appendix \ref{sec:ThreeTermPoly} results in 
\be
\sigma_2 = s_1 s_2 \,,\quad \sigma_3 = s_1 s_3 \,,\quad
\chi \sigma_4 = s_3 s_4 \,,\quad \chi_5 = s_2 s_5 \,,\quad 
\sigma_6 = s_2 s_4 + s_3 s_5 \,.
\ee
By the co-primeness of $\sigma_1$ and $\sigma_2$ we have that $s_1=1$.  Furthermore, the middle equation implies 
\be
\chi = \lambda_1\lambda_2 \,,\quad 
\sigma_4= \sigma_3 \lambda_4 \,,\quad 
s_3 = \lambda_1 \lambda_3 \,,\quad 
s_4 = \lambda_2 \lambda_4 \,.
\ee
Furthermore solving $\tilde{C}=0$ for $c_{3,1}$ results in a complete solution of $P_2=0$ where
\be
b_{1,0} = \lambda_1 \lambda_2 \sigma_1 s_2 \,.
\ee
The codimension two locus $\lambda_1=0$ is non-minimal, whereas all remaining ones give rise to $I_1^*$ fibers. 
Thus, one would naively think that this is a model with three ${\bf 10}$ matter curves. However, $\lambda_2$ and $s_2$ appear in the exact same pattern in the hypersurface equation of this form, and therefore behave similarly. One is thus left with only two ${\bf 10}$ curves, namely $s_2=0$ and $\sigma_1=0$. The leading coefficients in summary are (setting $\lambda_1=1$ to avoid the non-minimal locus)
\be
\ba
c_{0,3}  &= \lambda_3^2 \lambda_4 \cr
c_{1,2} &= \lambda_3 (s_5 \lambda_3 + 2 s_2 \lambda_2 \lambda_4) \cr
c_{2,1} &=  s_2 \lambda_2 (2 s_5 \lambda_3 + s_2 \lambda_2 \lambda_4)\cr
c_{3,0} &= s_2^2 s_5 \lambda_2^2 \cr
c_{3,1} &= \frac{1}{\lambda _3^3} \lambda _2 s_2  \left(\lambda _2^2 s_2^2 c_{0,4}+\lambda _3 \left(\lambda _3 c_{2,2}-\lambda _2 s_2 c_{1,3}\right)\right) \cr
b_{0,0} &= b_{0,0} \cr
b_{1,0} &= s_2 \lambda_2 \sigma_1 \cr
b_{2,1} &= \lambda_3 \sigma_1 \,.
\label{eq:I5ncncconditions}
\ea
\ee
Note that we can set $\lambda_3=1$ without any loss of matter loci. 
The discriminant of this equation is given by
\begin{equation}
  \Delta_{I_5^{(01)}} = s_2^4 s_5 \lambda_2^4 \sigma_1^4 \left( s_2 s_5 \lambda_2 + b_{0,0} \sigma_1 \right) P_3
\end{equation}
with
\begin{equation}
  \begin{aligned}
    P_3 =& b_{0,0}^2 \left( - s_5 + s_2 \lambda_2 \lambda_4 \right) \\
    &+ b_{0,0} s_2 \lambda_2 \bigg[ 2 s_5 \left( b_{1,1} - s_2 \lambda_2 b_{2,2} \right) \\
    &+ s_2 \lambda_2 \left( - 2 \lambda_4 b_{1,1} + 2 s_2 \lambda_2 \lambda_4 b_{2,2} - 3 s_2 \lambda_2 \sigma_1 c_{0,4} + 2 \sigma_1 c_{1,3} \right) - \sigma_1 c_{2,2} \bigg] \\
    &+ s_2^2 \lambda_2^2 \bigg[ - s_5 \left( b_{1,1} - s_2 \lambda_2 b_{2,2} \right)^2 + s_2^3 \lambda_2^3 \left( \lambda_4 b_{2,2}^2 + \sigma_1 \left( - 3 b_{2,2} c_{0,4} + \sigma_1 c_{0,5} \right) \right) \\
    &+ s_2^2 \lambda_2^2 \left( -2 \lambda_4 b_{1,1} b_{2,2} + \sigma_1 \left( 3 b_{1,1} c_{0,4} + 2 b_{2,2} c_{1,3} - \sigma_1 c_{1,3} \right) \right) \\
    &+ s_2 \lambda_2 \left( \lambda_4 b_{1,1}^2 + \sigma_1 \left( -2 b_{1,1} c_{1,3} - b_{2,2} c_{2,2} + \sigma_1 c_{2,3} \right) + \sigma_1 \left( b_{1,1} c_{2,2} - \sigma_1 c_{3,2} \right) \right) \,.
    \bigg]
  \end{aligned}
  \label{eq:ncncdiscrpoly}
\end{equation}
The matter curves and $U(1)$ charges of this model are shown in table \ref{tab:I5ncncCharges}.
Note that this is in fact a new fiber type, as there is no $I_{5,nc}^{(01)}$. 

\begin{table}
\centering
  \begin{tabular}{c|c|c|c|c}
Fiber &  Model & Codim 2 locus & Representation & Codim 2 fiber \\ \hline \hline  
    $I_{5, ncnc}^{(01)}$ & $Q(3,2,1,0,0,0,1)|_{(\ref{eq:I5ncncconditions})}$ 
    & $\sigma_1$ & $\mathbf{10}_{0} + \overline{\mathbf{10}}_{0}$ & $I_1^{*(01)}$ \\
    && $s_2$ & $\mathbf{10}_{0} + \overline{\mathbf{10}}_{0}$ & $I_1^{*(01)}$ \\
    && $s_5$ & $\mathbf{5}_{1} + \overline{\mathbf{5}}_{-1}$ & $I_6^{(1|0)}$ \\
    && $s_2 s_5 \lambda_2 + b_{0,1} \sigma_1$ & $\mathbf{5}_{-1} + \overline{\mathbf{5}}_{1}$ & $I_6^{(0|1)}$ \\
    && (\ref{eq:ncncdiscrpoly}) & $\mathbf{5}_{0} + \overline{\mathbf{5}}_{0}$ & $I_6^{(01)}$ \\
  \end{tabular}
   \caption{Codimension two loci, fiber types, and matter and  $U(1)$ charges of the twice non-canonical $I_5$ model. }
  \label{tab:I5ncncCharges}
\end{table}


  \item $P_2=0$: (alternative solution) {$I_5^{(01)}$}\\
  An alternative way to solve for this doubly non-canonical model is to consider  $P_2$ with the subleading $c_{i,j}$ terms set to zero
\be
P_2|_{c_{i,j}=0} =\left(\sigma_2^2 \sigma_4-\sigma_2 \sigma_3 \sigma_6+\sigma_3^2 \sigma_5\right) 
\left( \sigma_3^2 (b_{0,0} \sigma_1+\sigma_5)-\sigma_2 \sigma_3 (b_{1,1} \sigma_1+\sigma_6)+\sigma_2^2 (b_{2,2} \sigma_1+\sigma_4) \right)  \,.
\ee
The first factor will not give an $SU(5)$ model, as the discriminant goes up to $O(z^7)$. However the second factor 
\be
\tilde{P}_2 = \sigma_3^2 (b_{0,0} \sigma_1+\sigma_5)-\sigma_2 \sigma_3 (b_{1,1} \sigma_1+\sigma_6)+\sigma_2^2 (b_{2,2} \sigma_1+\sigma_4) 
\ee
can be solved by
\be
\ba
\sigma_4 &= \sigma_3 \rho _4-\sigma _1 b_{2,2} \cr
\sigma_5&= \sigma_2 \rho _5-\sigma _1 b_{0,0}\cr
\sigma_6&= -\sigma _1 b_{1,1}+\sigma_2 \rho _4+\sigma_3 \rho_5 \,.
\ea
\ee
The resulting model has discriminant
\be\ba
&\Delta_{\tilde{P}_2} = \rho _5 \sigma _1^4 \sigma _2^4 
\left(\sigma _2^2 b_{2,2}+\sigma _3  \left(\sigma _3 b_{0,0}-\sigma _2 b_{1,1}\right)\right) 
   \left(\sigma _1 b_{0,0}-\rho _5 \sigma _2\right) \times\cr 
& \times  (\rho _4 \sigma _2^3 b_{2,2}-\sigma
   _3 \sigma _2^2 (\rho _4 b_{1,1}+\rho _5 b_{2,2})+\sigma _3^2
   \sigma _2 (\rho _4 b_{0,0}+\rho _5 b_{1,1}+\sigma _1^2
   b_{0,1})-\rho _5 \sigma _3^3 b_{0,0}) z^5  + O(z^6) \,.
\ea\ee
There are three loci that result in codimension two loci $\sigma_1=0$ and $\sigma_3=0$ with $I_1^*$ fibers, however  
the third factor $\sigma _2^2 b_{2,2}+\sigma _3  \left(\sigma _3 b_{0,0}-\sigma _2 b_{1,1}\right) =0$ is in fact again  $I_6$. 
It would be quite exciting to solve these ncnc enhacements in generality and determine models with three distinctly charged ${\bf 10}$ matter loci. 

\end{itemize}


\subsubsection{$I_{4, nc}^{(0||1)}$ Branch}

The non-canonical fibration $\CQ(2,1,1,1,0,0,1)|_{(\ref{eq:D2111001P1})}$ obtained from (\ref{eq:Q2111001P1}) has discriminant
\begin{equation}
  \Delta_{I_4^{(0||1)}} = \sigma_1^4 \sigma_2 \sigma_3^4 P_2 P_3 z^4 + O(z^5)
\end{equation}
with
\begin{equation}
  \begin{aligned}
    P_2 =&\, \sigma_1^2 c_{0,2} - \sigma_1 \sigma_4 b_{2,1} - \sigma_4^2 \,, \\
    P_3 =&\, \sigma_2^3 \left(-\left(\sigma_4 \sigma_1 b_{2,1} - \sigma_1^2 c_{0,2} + \sigma_4^2\right)\right)+\sigma_3 \sigma_2^2 \left(\sigma_1 \left(\sigma_5 b_{2,1}-\sigma_1
   c_{1,2}\right)+\sigma_4 \left(\sigma_1 b_{1,1}+2 \sigma_5\right)\right) \\
   &-\sigma_3^2 \sigma_2 \left(\sigma_1 \left(\sigma_4 b_{0,1}-\sigma_1 c_{2,2}\right)+\sigma_1 \sigma_5 b_{1,1}+\sigma_5^2\right)+\sigma_1 \sigma_3^3 \left(\sigma_5 b_{0,1}-\sigma_1 c_{3,2}\right) \,,
  \end{aligned}
\end{equation}
and $\sigma_2$, $\sigma_3$ coprime. $\sigma_1$ and $\sigma_3$ divide $b_{1,0}$, hence their enhancements do not yield $I_n$ singularities and are irrelevant to us here.
\begin{itemize}
  \item $\sigma_2=0$: $I_5^{(0|1)}$ \\
 Recall that $\sigma_2$ divides $b_{0,0}$. Thus, by first enhancing along this locus and then reconsidering the polynomial that led to $\CQ(2,1,1,1,0,0,1)|_{(\ref{eq:D2111001P1})}$, one obtains the canonical fibration 
    $\CQ(2,1,1,2,1,0,1)$.
  \item $P_2=0$: $I_5^{(0||1)}$ \\
    Similarly (and by a suitable coordinate shift), one finds that $P_2=0$ yields the non-canonical fiber $\CQ(3,2,1,1,0,0,1)|_{(\ref{eq:D3211001P0})}$.
  \item $P_3=0$: \\
   This results in a doubly non-canonical fiber, which can be studied along the lines of the example in section \ref{sec:I5ncnc}. 
\end{itemize}


\subsection{Non-canonical $I_4$ from non-canonical $I_3$}

\subsubsection{$I_3^{(0|1)}$ Branch}

The discriminant of $\CQ(1,1,1,1,0,0,1)|_{(\ref{eq:D1111001P0})}$ obtained from (\ref{eq:Q1111001P0}) at leading order reads
\begin{equation}
  \Delta_{I_3^{(0|1)}} = \sigma_1^6 \sigma_2 \sigma_3^2 \sigma_4 P_2 P_2 z^3 + O(z^4) \,,
\end{equation}
with
\begin{equation}
  \begin{aligned}
    P_2 =& -\sigma_3 \sigma_4 \sigma_2^4 \left(\sigma_1 b_{2,1}+2 \alpha \right)+\sigma_3^2 \sigma_2^3 \left(\alpha
    \sigma_1 b_{2,1}+\sigma_4 \left(\sigma_1 b_{1,1}+2 \sigma_5\right) - \sigma_1^2 c_{0,2}+\alpha^2\right) \\
    &-\sigma_3^3 \sigma_2^2 \left(\sigma_5 \left(\sigma_1 b_{2,1}+2 \alpha \right)+\sigma_1 \left(\alpha  b_{1,1}+\sigma_4 b_{0,1}-\sigma_1 c_{1,2}\right)\right) \\
    &+ \sigma_3^4 \sigma_2 \left(\sigma_1 \left(\alpha  b_{0,1}-\sigma_1 c_{2,2}\right)+\sigma_1 \sigma_5 b_{1,1}+\sigma_5^2\right)+\sigma_1 \sigma_3^5 \left(\sigma_1 c_{3,2}-\sigma_5 b_{0,1}\right)+\sigma_4^2 \sigma_2^5 \,.
  \end{aligned}
\end{equation}
Again, $\sigma_2$ and $\sigma_3$ are coprime. Since $\sigma_1$ and $\sigma_3$ are divisors of $b_{1,0}$, $\sigma_1=0$ and $\sigma_3=0$ are enhancements leaving the $I_n$ branch and are thus not considered here.
\begin{itemize}
  \item $\sigma_2=0$: $I_4^{(0|1)}$ \\
    Since $\sigma_2$ divides $b_{0,0}$, one can, instead of first considering the enhancement $P_0$ of the $I_2^{(0|1)}$ form leading to $\CQ(1,1,1,1,0,0,1)|_{(\ref{eq:D1111001P0})}$, alternatively first enhance $b_{0,0}=0$ to $\CQ(1,1,1,1,1,0,1)$, and consider the $P_0=0$ enhancement with $\sigma_2=0$ imposed there. This imposition yields $P_0=c_{3,1}$, hence the resulting fibration is canonical and given by
    \begin{equation}
      \CQ(1,1,1,2,1,0,1) \,,
    \end{equation}
    which is {lop-equivalent to $\CQ(3,2,1,1,0,0,2)$, a special case of the canonical $I_4^{(0|1)}$ fiber type discussed in section \ref{sec:Tate}.}
  \item $\sigma_4=0$: $I_4^{(0||1)}$ \\
    Similarly, this enhancement yields the non-canonical fibration $\CQ(2,1,1,1,0,0,1)|_{(\ref{eq:D2111001P1})}$.
  \item $P_2=0$: \\
    This enhancement yields a doubly non-canonical fibration.
\end{itemize}


\subsection{Non-canonical forms for $I_n$}
{
It is natural to ask whether the non-canonical forms exist for higher values of $n$ for $I_n$ fibers, or whether, for high enough $n$, as in the case of standard Weierstrass models realize in $\mathbb{P}^{(1,2,3)}$, the models are all canoncial, once one forbids all non-minimal loci in codimension two  as shown in \cite{Katz:2011qp}. The requirement for  the absence of non-minimal loci in codimension two implies in terms of the low energy physics, that all codimension two loci have an interpretation in terms of matter fields in the gauge theory. In practice this means that coordinate changes, that require divisions by sections, whose vanishing correspond to non-minimal codimension two loci, are allowed. For $\mathbb{P}^{(1,2,3)}$ it was shown inductively in \cite{Katz:2011qp}, that for large enough $n$ a local coordinate change always exists that brings any $I_n$ model back to a standard Tate form. 

The situation in $\mathbb{P}^{(1,1,2)}$ is in this sense more complicated, as not all models have such coordinate change, that is locally well-defined that would bring the $I_n$ back to canonical Tate-like forms. For instance consider 
\be
I_{2m+k}^{(0|^k1)}: \qquad \mathcal{Q}(2 m, m, k, k, 0, 0, m) \,,
\ee
which has leading order discriminant
\be
\Delta =  b_0 b_1^4 (b_1^2 c_0 - b_1 b_2 c_1 - c_1^2) (b_0 c_2 - b_1 c_3) z^{2 m + k} + O(z^{2m+k+1})\,.
\ee
Starting with this canonical $I_n$ model, we can ask whether the enhancements for suitably large $n$ can always be brought back to canonical form, allowing for shifts that can have divisions by sections that would yield non-minimal codimension two loci. 

The first term $b_0=0$ enhances further (to a lop equivalent model) to  
$I_{2(m+1)+k-1}^{(0|^{k-1}1)}$.
The second factor, $b_1=0$ yields a non-minimal enhancement if the sections are separated by $k\geq 2$. For $k=1$ it enters the $I^*$ branch and yields the canonical fiber type $I_{2(m-2)}^{*(0|1)}$.
The third factor is shiftable to $c_0=c_1=0$ and yields canonical $I_{2m+k+1}^{(0|^k1)}$. 
Finally, the last term in the discriminant, for $k\geq2$ can be shifted into a canonical model as $b_1=0$ is non-minimal, $\mathcal{Q}(2m,m,k+1,k+1,0,0,m)$.
However for $k=1$ this shift is not allowed, as $b_1=0$ is not a non-minimal locus. In fact, solving the two-term polynomial $b_0 c_2 - b_1 c_3=0$ yields a non-canonical
\be
k=1:\qquad I_{2m+1}^{(0|1)} \quad \rightarrow \quad {I_{2m+2, nc}^{(0||1)}} \,.
\ee
This cannot be shifted back to a canonical model as all codimension two loci are perfectly fine minimal enhancements. 
This demonstrates that non-canonical forms exist for any $I_n$ fiber with extra section. If the separation between the section and the zero section is large, the $b_1=0$ locus is generically non-minimal (as there is no suitable enhancement to the $I_m^*$ branch). However, for small separation, these non-canonical fibers exist for all $n$, and are distinct from the canonical forms.}


\section{Non-canonical forms in $\bbP^{(1,2,3)}$}
\label{sec:P123Reloaded}

\begin{figure}
  \centerline{
	\xymatrix{
	\dots \ar@{->}[r] & I_4 \ar@{->}[r] & I_5\ \ar@{->}[r] \ar@{->}[dr] & I_6\ \ar@{->}[r] & I_7 \ar@{->}[r] \ar@{->}[dr] & I_8\ \ar@{->}[r] & I_9 \ar@{->}[r] \ar@/_2pc/[r] & I_{10} \ar@{->}[r] & I_{11} \ar@{->}[r] \ar@/_2pc/[r] & \dots \\
     & & & I_{6,nc}\ \ar@{->}[ur] \ar@{->}[dr] & & I_{8,nc}\ \ar@{->}[ur] \ar@{->}[dr] & & &\\
     & & & & I_{7,nc\,nc}\ \ar@{->}[ur] \ar@{->}[dr] & & I_{9,nc\,nc}\ \ar@{->}[uur] & & &\\
     & & & & & I_{8, nc\,nc\,nc} \ar@{->}[ur] \ar@{->}[dr]&&  \\
     & & & & &  & I_{9, nc\,nc\,nc\,nc} \ar@{->}[uuuur]& 
	}}
  \caption{The schematic Tate tree for the $I_n$-type enhancements of the $I_5$ fiber in $\bbP^{(1,2,3)}$, including the non-canonical forms for $I_6$ to $I_9$. The $I_9$ enhancement which would normally lead to $I_{10,nc}$ can be shifted to the canonical $I_{10}$, and the same applies to all $I_{2m+1}\rightarrow I_{2m+2}$ enhancements that follow  \cite{Katz:2011qp}. Below we will show explicitly that there is a multiply non-canonical enhancement, starting from $I_7$, which yields an $I_{11}$ model, that can then be brought back into canonical Tate form.  }
  \label{fig:p123TateTree}
\end{figure}
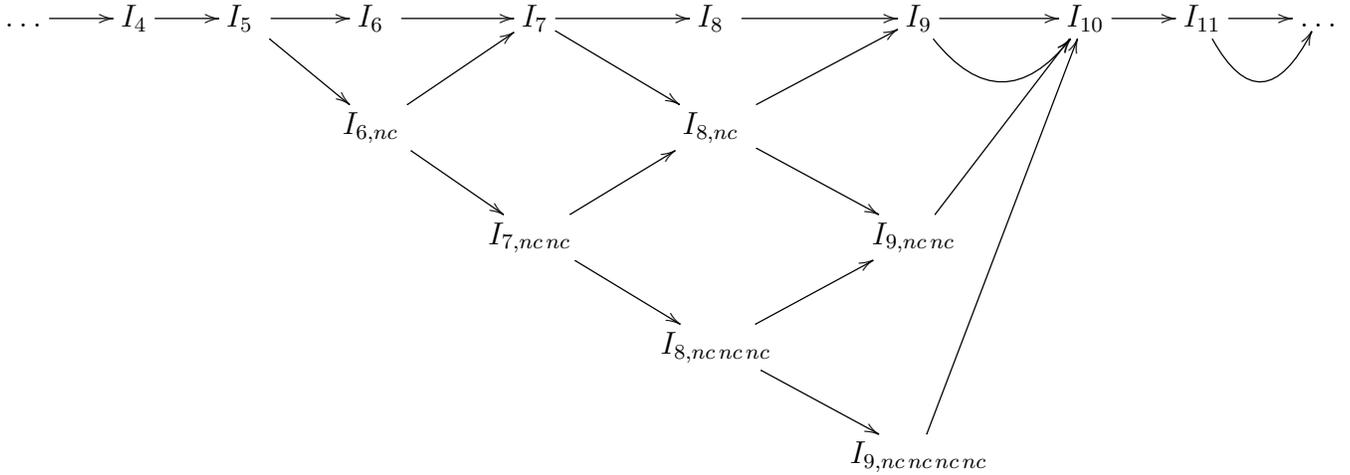

While non-canonical forms appear very commonly in the Tate tree of $\bbP^{(1,1,2)}$, they also occur in Tate's algorithm for Weierstrass forms embedded in $\bbP^{(1,2,3)}$. For most parts in $\mathbb{P}^{(1,2,3)}$ the canonical Tate forms can be reached through locally well-defined coordinate changes, 
i.e. those that do not require any divisions. 
In \cite{Katz:2011qp}, it was already observed that  a generalized ansatz is required  for the fiber types $I_n$ for $n= 2m+1$ except $n=7,9$. 
There are furthermore outlier cases, where neither the Tate form nor the new ansatz of \cite{Katz:2011qp} can be achieved, $I_n$ for $n= 6,7,8,9$. These models can be discussed with the same type of methods that we have used for the non-canonical models in  $\mathbb{P}^{(1,1,2)}$, and we will now derive explicit forms for the non-canonical models and examples for the multiply non-canonical cases. Note that one of the key assumptions in \cite{Katz:2011qp} is that there are no codimension two non-minimal loci, i.e. once put into Weierstrass form, the vanishing orders of $(f, g, \Delta)$ in codimension one and two stay below $(4,6,12)$. This allowed shifts back to canonical forms for the infinite series $I_n$ without monodromy. 
The structure of the non-canonical forms in the Tate tree of $\bbP^{(1,2,3)}$ is depicted in figure \ref{fig:p123TateTree}.  

Recall that a generic elliptic fibration with section can be embedded into the projective space $\bbP^{(1,2,3)}$ by the hypersurface equation
\begin{equation}
  \CP: \qquad y^2  +\fkb_1 y x  + \fkb_3 y = x^3 + \fkb_2 x^2  + \fkb_4 x  + \fkb_6  \,,
\end{equation}
where $w$, $x$ and $y$ are the coordinates of $\bbP^{(1,2,3)}$ with weights $1$, $2$, and $3$, respectively, and we have set $w=1$. Tate's algorithm in $\bbP^{(1,2,3)}$ then provides the vanishing orders $(i_1,i_2,i_3,i_4,i_6)$ of the sections $\fkb_i$ of canonical Tate forms, which we will denote by 
\begin{equation}
  \begin{aligned}
    \CP(i_1,i_2,i_3,i_4,i_6): \qquad & y^2  +\fkb_{1,i_1} z^{i_1} y x  + \fkb_{3,i_3} z^{i_3} y  \\
    &\qquad = x^3 + \fkb_{2,i_2} z^{i_2} x^2  + \fkb_{4,i_4} z^{i_4} x  + \fkb_{6,i_6} z^{i_6}  \,.
  \end{aligned}
\end{equation}

\subsection{Non-canonical $I_{6}$ from canonical $I_5$}

The first non-canonical Tate forms in  the $\bbP^{(1,2,3)}$ Tate tree arise from the enhancement of the $I_5$ canonical Tate form
\begin{equation}
\CP(0,1,2,3,5):  \qquad y^2  + \fkb_1 y x  + \fkb_{3,2} z^2 y  = x^3 + \fkb_{2,1} z x^2  + \fkb_{4,3} z^3 x  + \fkb_{6,5} z^5  \,,
\end{equation}
which has discriminant 
\begin{equation}
  \Delta_{I_5}: \qquad b_{1,0}^4 \left( b_{3,2}^2 b_{2,1} - b_{3,2} b_{1,0} b_{4,3} + b_{1,0}^2 b_{6,5} \right) z^5 + O(z^6) \,.
\end{equation}
The locus $b_{1,0}=0$ yields an  $I_1^*$ model $\CP(1,1,2,3,5)$.

The interesting part of the discriminant enhancements arise in the $I_n$ branch, where 
$P=0$ gives rise to a canonical $I_6$ and a non-canonical $I_{6,nc}$ model.
The polynomial term in this discriminant is exactly of the form discussed in appendix \ref{sec:ThreeTermPoly}. 
The only solutions that do not also set $b_{1,0}=0$, which is already a factor in the discriminant, are: $b_{3,2}=b_{6,5}=0$, and the non-trivial solution from (\ref{ThreeTermSol1}). The former solution indeed yields the canonical  form $\CP(0,1,3,3,6)$ of the $I_6$ fiber in $\bbP^{(1,2,3)}$. The latter solution, however, has not been discussed in the literature so far, and it leads to a non-canonical $I_6$ Tate form, namely  
\begin{equation}  \label{GenSolI5}
  \begin{aligned}
    I_{6,nc}: &\qquad \CP(0,1,2,3,5)  \hbox{ with }  \cr
    & b_{1,0} = \sigma_1 \sigma_3, \
    b_{3,2} = \sigma_1 \sigma_2, \
    b_{6,5} = \sigma_2 \sigma_5, \
    b_{2,1} = \sigma_3 \sigma_4, \ 
    b_{4,3} = \sigma_3 \sigma_5 + \sigma_2 \sigma_4
  \end{aligned}
\end{equation}
or explicitly 
\be\label{eq:p123i6ncmodel}
I_{6,nc}:\qquad   y^2  +\sigma _2^2 x y+\zeta _0^2 \sigma _2 \sigma _3 y=  
x^3+ \zeta_0 \sigma _2 \sigma _5 x^2+ \zeta _0^3 \left(\sigma _2 \sigma _4+\sigma _3 \sigma _5\right) x +\zeta _0^5 \sigma _3 \sigma _4 \,.
\ee
Note that in general there are no shifts that bring this into canonical form. 
The discriminant of this non-canonical $I_6$ model is
\begin{equation}
  \Delta_{I_{6, nc}} = \sigma_1^4 \sigma_3^3 P_6 z^6 + O(z^7)
  \label{eq:p123i6ncdiscr}
\end{equation}
with
\begin{equation}
  \begin{aligned}\label{P6P123}
    P_6 =& \, \sigma_3 \left( \sigma_2 \sigma_4 - \sigma_3 \sigma_5 \right)^2 + \sigma_1 \sigma_3 \left( \sigma_3 \sigma_5 - \sigma_2 \sigma_4 \right) \left( \sigma_3 b_{3,3} - \sigma_2 b_{1,1} \right) \\
    & + \sigma_1^2 \left( \sigma_2^3 - \sigma_2^2 \sigma_3 b_{2,2} + \sigma_2 \sigma_2^2 b_{4,4} - \sigma_3^3 b_{6,6} \right) \,.
  \end{aligned}
\end{equation}
For this form the ${\bf 15}$ matter locus $b_{1,0}=0$ splits into two components: 
\begin{equation}
  b_{1,0} = \sigma_1 \sigma_3 \,.
\end{equation}
Setting either of $\sigma_1$ or $\sigma_3$ to zero yields an enhancement to type $I_2^*$. Therefore, this model has two ${\bf 15}$ curves, plus the single ${\bf 6}$ curve already present in the canonical model and here obtained from setting the long polynomial to zero.

In \cite{Katz:2011qp}, there was a non-canonical $I_6$ model originating from \cite{Katz:1996xe} presented as evidence that not all $I_6$ singularities can be brought into canonical form. The model reads
\begin{equation}
  y^2 - \frac 9 4 t^2 x y + z^2 y = x^3 \,.
\end{equation}
It is a special case of the general non-canonical $I_6$ model (\ref{eq:p123i6ncmodel}), which can be obtained by $\sigma_1=\sigma_2=1$, $\sigma_4=\sigma_5=0$, $\sigma_3 = -9/4 t^2$, and also setting all subleading terms to zero.

\subsection{Non-canonical  $I_{7}$ from non-canonical $I_{6}$}

Continuing on from this $I_{6, nc}$ non-canonical model, the polynomial term $P_6$ in the discriminant allows for a doubly non-canonical enhancement to an $I_{7, ncnc}$ model, $I_{6, nc}|_{P_6=0}$, i.e.
\be\label{I7ncncP123}
I_{7, ncnc}:\qquad
\left( y^2  +\sigma _2^2 x y+\zeta _0^2 \sigma _2 \sigma _3 y=  
x^3+ \zeta _0 \sigma _2 \sigma _5 x^2+ \zeta _0^3 \left(\sigma _2 \sigma _4+\sigma _3 \sigma _5\right) x +\zeta _0^5 \sigma _3 \sigma _4 \right)|_{P_6=0}
\ee
with $P_6$ in  (\ref{P6P123}). 
Solving the condition $P_6=0$ in general is rather tricky, however several subcases of solutions can be obtained. For instance (setting the subleading $b_{i, j}=0$ in $P_6$, as well as $\sigma_5=0$) for 
\be
b_{1,0}=  u^5\,,\quad 
b_{2,0} = 0 \,,\quad 
b_{3,0}= - u^3 v^2\,,\quad 
b_{4,0} =   u^2 v^3 \,,\quad 
b_{6,0}= -  v^5 \,,\qquad (u, v)=1\,.
\ee
An alternative solution, again setting the higher $b_{i, j}=0$ in $P_6$, as well as $\sigma_4=0$ yields
\be\label{I7ncncSpec}
b_{1,0}=  \sigma_1^3 \,,\quad 
b_{2,0} = \sigma_1^2 \sigma_5 \,,\quad 
b_{3,0}= -\sigma_1 \sigma_5^2\,,\quad 
b_{4,0} = -\sigma_5^3   \,,\quad 
b_{6,0}= 0 \,,\qquad (\sigma_1, \sigma_5)=1\,.
\ee
Another model from \cite{Katz:1996xe}, presented in \cite{Katz:2011qp} as an outlier case, is $I_7$ fibration which could not be brought into canonical form:
\begin{equation}
  y^2 - 54 t^3 x y + 24 t z^2 y = x^3 + 36 t^2 z x^2 - 16 z^3 x \,.
\end{equation}
This model is a special case of the doubly non-canonical $I_7$ model (\ref{I7ncncP123}), and the special solution (\ref{I7ncncSpec}), which enhances the non-canonical $I_6$ model by solving the polynomial term in (\ref{eq:p123i6ncdiscr}). One can find it by starting with the non-canonical $I_6$ model from (\ref{eq:p123i6ncmodel}), and setting
\begin{equation}
  \sigma_4 =0 \,, \quad
  \sigma_5 = 1 \,, \quad
  \sigma_1 = - \frac 3 4 t \,, \quad
  \sigma_2 = 36 t^2 \,, \quad
  \sigma_3 = -16 \,,
\end{equation}
as well as having all subleading terms equal to zero.

\subsection{Non-canonical $I_{8}$ from canonical $I_7$}

Another non-canonical form, which is similar to the one above, can be obtained by starting from the canonical $I_7$ form $\CP(0,1,3,4,7)$, with discriminant
\begin{equation}
  \Delta_{I_7}: \qquad b_{1,0}^4 \left( b_{3,3}^2 b_{2,1} - b_{3,3} b_{1,0} b_{4,4} + b_{1,0}^2 b_{6,7} \right) z^7 + O(z^8) \,,
\end{equation}
and solving the three-term polynomial. It reads
\begin{equation}
  \begin{aligned}
  I_{8,nc}: &\qquad \mathcal{P}(0,1,3,4,7)  \hbox{ with }  \cr
  & b_{1,0}= \sigma_1 \sigma_2 , \
    b_{3,3} = \sigma_1 \sigma_3, \
    b_{6,7}= \sigma_3 \sigma_4, \
    b_{2,1} = \sigma_2 \sigma_5 ,\ 
     b_{4,4}= \sigma_2 \sigma_4 + \sigma_3 \sigma_5 \,.
  \end{aligned}
  \label{eq:p123i8ncmodel}
\end{equation}
By the same argument as above, this model has two ${\bf 28}$ curves (as opposed to the canonical $I_8$ model with a single ${\bf 28}$ curve), plus an ${\bf 8}$ curve.
Again there is a non-canonical enhancement starting from this 
\begin{equation}
P_8= \sigma _2^2 \left(\sigma _3 \left(\sigma _3 b_{2,2}-b_{4,5} \sigma _2\right)+\sigma _2^2 b_{6,8}\right)-\left(\sigma _2 \sigma _4-\sigma _3 \sigma _5\right) \sigma _2 \left(\sigma _2 b_{3,4}-\sigma _3 b_{1,1}\right)-\left(\sigma _2 \sigma _4-\sigma _3 \sigma _5\right)^2 \,.
   \label{eq:p123I8ncdiscr}
\end{equation}

\subsection{Canonical $I_{11}$ model via non-canonical enhancements}

Finally, one can enhance beyond the outlier cases, and reach vanishing orders of the discriminant that are larger than $10$. In those cases, it was shown in \cite{Katz:1996xe} that these can be shifted back to canonical mocels, under the assumption of absence of non-minimal loci in codimension two. 

In practice, we can see this in the following enhancement of the non-canonical $I_8$ model from the previous subsection by finding solutions to (\ref{eq:p123I8ncdiscr}). While not completely general,
\begin{equation}
  b_{1,1}=b_{2,2}=b_{3,4}=b_{4,5}=b_{6,8}=0\,, \quad \sigma_5=\sigma_2 \alpha\,, \quad \sigma_4=\sigma_3 \alpha
\end{equation}
solves $P_8=0$. Note that here we only set the subleading $b_{i,j}=0$, not the full series. 
 The resulting enhancement is of Kodaira type $I_9$ and has leading order discriminant
\begin{equation}
  \Delta_{I_9} = \sigma_1^6 \sigma_2^3 \left( \sigma_2^3 b_{6,9} - \sigma_2^2 \sigma_3 b_{4,6} + \sigma_2 \sigma_3^2 b_{2,3} - \sigma_3^3 \right) z^9 + O(z^{10}) \,.
\end{equation}
The polynomial term in the discriminant of this specialized $I_9$ can in turn be solved by
\begin{equation}
  \sigma_2=1 \,, \quad b_{6,9} = \rho_4 \sigma_3 \,, \quad b_{4,6} = \rho_4 + \alpha \sigma_3 \,, \quad b_{2,3} = \alpha \sigma_3 \,.
\end{equation}
Also setting $\fkb_{1,3}=\fkb_{2,4}=\fkb_{3,6}=\fkb_{4,7}=\fkb_{6,10}=0$ yields a model of Kodaira type $I_{11}$, with equation
\begin{equation}
  \begin{aligned}
    y^2 &+ \left( \sigma_1 + z^2 b_{1,2} \right)  x y + \left( \sigma_3 \sigma_1 + z^2 b_{3,5} \right) z^3  y \\
    =&\, x^3 + \left( \alpha + \left( \alpha + \sigma_3 \right) z^2 \right) z x^2  + \left( 2 \alpha \sigma_3 + \left( \alpha \sigma_3 + \rho_5 \right) z^2 \right) z^4 x  + \left( \alpha \sigma_3^2 + z^2 \sigma_3\rho_5 \right) z^7 
  \end{aligned}
\end{equation}
and discriminant
\begin{equation}
  \Delta_{I_{11}} = \sigma_1^4 \left( \sigma_3 b_{1,2} - b_{3,5} \right) \left( \alpha \sigma_1 \sigma_3 - \sigma_3^2 \sigma_1 - \rho_5 \sigma_1 - \alpha \sigma_3 b_{1,2} + \alpha b_{3,5} \right) z^{11} + O(z^{12}) \,.
\end{equation}
The locus $\sigma_1=0$ enhances to $I_7^*$, while the two polynomials give enhancements to $I_{12}$, but are both minimal, i.e. the corresponding Weierstrass model does not have vanishing orders $\hbox{mult}(f,g,\Delta)\geq (4, 6, 12)$. 
Consistently with the result in \cite{Katz:1996xe}, there is a coordinate change, that brings this to canonical form\footnote{We thank Dave Morrison and Sheldon Katz for discussions on this point.}
\be
{ (x , y )\quad  \rightarrow \quad \left( x - z^3 \rho_3, \  y - {1\over 2}  x z^2 b_{1,2}\right) \,.}
\ee

\subsection{Comment on matter loci for Tate forms}

From our considerations of Tate's algorithm and non-canonical models in $\mathbb{P}^{(1,2,3)}$, 
we can infer various interesting points about the matter loci, i.e. codimension two loci, 
where the fiber enhances. Consider for instances $I_5$, which in applications in F-theory realizes $SU(5)$ gauge groups. 
The discriminant factor (or ${\bf 5}$ matter locus)
\be
P_{5}= b_{3,2}^2 b_{2,1} - b_{3,2} b_{1,0} b_{4,3} + b_{1,0}^2 b_{6,5} 
\ee
that enhances $I_5$ to $I_6$ was shown to have a general solution (\ref{GenSolI5}) derived in appendix \ref{sec:ThreeTermPoly}
\be
 b_{1,0} = \sigma_1 \sigma_3, \
    b_{3,2} = \sigma_1 \sigma_2, \
    b_{6,5} = \sigma_2 \sigma_5, \
    b_{2,1} = \sigma_3 \sigma_4, \ 
\ee
which implies
\be
P_5 = \sigma_1^2 \sigma_2 \sigma_3 (-b_4+\sigma_2 \sigma_4+\sigma_3 \sigma_5) \,.
\ee
There are four branches yielding $P_5=0$, which correspond to the enhacements
\be
\ba
\sigma_1=0: & \qquad  IV^* (E_6) \cr
\sigma_2=0: & \qquad  I_2^* (SO(12))\cr
\sigma_3=0: & \qquad  I_6 (SU(6)) \cr
b_4=\sigma_2 \sigma_4+\sigma_3 \sigma_5  : & \qquad  I_6 (SU(6))\,.
\ea
\ee
There are two loci which correspond to $I_6$ enhancements,  and yield two different ways of realizing  {\bf 5} matter loci.  


\section*{Acknowledgements}

We thank Sheldon Katz, Craig Lawrie, and Dave Morrison for important discussions on various aspects of Tate's algorithm. We also thank Andreas Braun, Denis Klevers, Damiano Sacco, Martin Weidner and Jenny Wong for discussions at various stages of this work.  STFC supported this work in part by the rolling grant ST/J002798/1.

\newpage
\appendix

\section{Normal forms for elliptic curves}
\label{app:Deligne}

In this appendix, we rederive the Weierstrass normal form for elliptic curves and the quartic normal form for elliptic curves with Mordell-Weil rank one using Riemann Roch as in \cite{Silverman}. Related arguments can be found in \cite{Park:2011ji}, although we provide here a detailed argument why the coefficient of $y^2$ is one, and the additional constraint that $\mathfrak{b}_0$ cannot vanish. 

Let us first recall a few definitions. For an algebraic curve $\CC$ over some algebraically closed field $K$, one defines a divisor $D$ of $\CC$ to be a formal sum of the points in $\CC$,
\begin{equation}
  D = \sum_{P \in \CC} a_p (P) \,,
\end{equation}
where $a_p \in \mathbb{Z}$ and $a_p=0$ for almost all $P$. The divisors, together with addition, form an abelian group, which is called $\mathrm{Div}(\CC)$. Furthermore, the degree of a divisor is given by
\begin{equation}
  \mathrm{deg} D = \sum_{P \in \CC} a_p \,.
\end{equation}
For any function $f \in K(\CC)$, one can associate a divisor to $f$ given by
\begin{equation}
  \mathrm{div}(f) = \sum_{P \in \CC} \mathrm{ord}_P(f) (P) \,,
\end{equation}
where $\mathrm{ord}_P(f)$ is zero if $f(P)$ takes on a non-zero value at $P$. If $f(P)=0$, then $\mathrm{ord}_P(f)$ gives the order of the zero of $f$ at $P$, and if $f$ has a pole at $P$, then $-\mathrm{ord}_P(f)$ is the order of the pole. Furthermore, a divisor $D =  \sum a_p (P)$ is positive (or effective), written as $D \geq 0$, if $a_P \geq 0$ for all points $P$. Similarly, $D_1 \geq D_2$ if $D_1 - D_2 \geq 0$.

One can associate to any divisor $D \in \mathrm{Div}(\CC)$ the set of functions
\begin{equation}
  \CL(D) = \left\{ f \in K(\CC): \mathrm{div}(f) \geq - D \right\} \cup \left\{ 0 \right\} \,.
\end{equation}
$\CL(D)$ is a finite-dimensional vector space, with its dimension being called $l(D)$. Also, if $\mathrm{deg}(D) < 0$, then $l(D) = 0$. Roughly speaking, $\CL(D)$ is the set of functions whose poles are no worse than $D$.

We can now state the Riemann-Roch theorem: Let $\CC$ be a smooth curve with canonical divisor $K_\CC$. Then there is an integer $g \geq 0$, the genus of $\CC$, such that for every divisor $D$:
\begin{equation}
  l(D) - l(K_C - D) = \mathrm{deg} (D) - g + 1 \,.
\end{equation}
For $D=0$, one immediately finds $l(K_C) = g$, and for $D=K_C$, one finds $\mathrm{deg} K_C = 2g - 2$. Also, if $\mathrm{deg} (D) > 2g - 2$, then the Riemann-Roch theorem reduces to
\begin{equation}
  l(D) = \mathrm{deg} D - g + 1 \,.
  \label{eq:RRsimplified}
\end{equation}

Let us now specialize to elliptic curves, i.e.\ algebraic curves of genus $1$ with a marked point $P$. In this case and for $\mathrm{deg}(D) \geq 1$, (\ref{eq:RRsimplified}) simply states that $l(D) = \mathrm{deg} (D)$.

For $D = (P)$, one thus finds $l(P) = 1$. Since the identity function $1$ has no poles anywhere on $\CC$, it is certainly in $\CL(P)$. Since $l(P)=1$, the identity function in fact spans all of $\CL(P)$. For $D = 2(P)$, one has $l(P)=2$, and there must be a second generator of $\CL(P)$ besides the identity, which we will call $x$. One can now ask about the order of the pole of $x$ at $P$. Since $x \in \CL(2P)$, it cannot be worse than $2$. But since $x \notin \CL(P)$ ($x$ and $1$ are linearly independent, and $1$ generates $\CL(P)$), it must be more than $1$. Therefore, $x$ must have a pole of order exactly $2$ at $P$. Next,  $\CL(3P)$ is three-dimensional and contains $1$, $x$, and a new generator $y$ with $\mathrm{ord}_P(y) = -3$, by similar reasoning.

$\CL(4P)$ is four-dimensional with $1$, $x$, $y$ and $x^2$, and $\CL(5P)$ is five-dimensional with $1$, $x$, $x^2$, $y$ and $xy$. $\CL(6P)$ should be six-dimensional, but we know of seven generators: $1$, $x$, $x^2$, $x^3$, $y$, $y^2$, $xy$. Therefore, there mus be a relation of the form
\begin{equation}
  A_1 y^2 + b_1 x y + b_3 y = A_2 x^3 + b_2 x^2 + b_4 x + b_6 \,.
  \label{eq:WSRelation}
\end{equation}
This relation must hold on all of $\CC$, so it also must hold on $P$. We want to prove by contradiction that $A_1 A_2 \neq 0$. So let us first assume that $A_1 = 0$. Then the only term that has a pole of order six at $P$ is $A_2 x^3$. But the relation cannot have a pole at $P$ at all, since it must be zero at $P$. Therefore, one also needs $A_2=0$. But then, the only term with a pole of order five is $b_1 x y$, and thus $b_1 = 0$. Similarly, one finds $b_2=0$, $b_3=0$, $b_4=0$, and $b_6=0$. The same happens if one starts with $A_1=0$. Thus, the only way for the relation (\ref{eq:WSRelation}) to exist non-trivially, is if $A_1$ and $A_2$ are non-zero, so that their poles of order six cancel each other, leaving a residual term with a pole of order five, which then cancels the pole of $b_1 x y$, leaving a residual of order four, and so on.

One then rescales $y \rightarrow A_1 A_2^2 y$, $x \rightarrow A_1 A_2 x$ and divides the whole equation by $A_1^3 A_2^4$, and obtains
\begin{equation}
  y^2 + b_1 x y + b_3 y = x^3 + b_2 x^2 + b_4 x + b_6 \,,
\end{equation}
from which the standard Weierstrass form can be found by completing the square in $y$ and the cube in $x$.

To find the normal form for an elliptic curve with Mordell-Weil rank one, one starts with a genus $1$ curve $\CC$ and two marked points $P$ and $Q$. Riemann-Roch tells us that $\CL(P+Q)$ is two-dimensional, that is, it is generated by $1$ and a non-constant function $x$. Note that $x$ must have a pole of order $1$ both at $P$ and at $Q$, since if it would have a pole only at, say, $P$, it would be in $\CL(P)$. But $\CL(P)$ is $1$-dimensional, and thus consists of only the constant functions.

Next, consider $\CL(2(P+Q))$. This four-dimensional vector space contains $1$, $x$, $x^2$ and a new generator $y$. The pole structure of $y$ is slightly intricate: It needs to have a pole of order $2$ either at $P$ or at $Q$, otherwise it would be in $\CL(P+Q)$. Without loss of generality, this pole can be at $P$. Its pole order at $Q$ is not uniquely defined, but there is a basis where $y$ has a pole of order $1$ at $Q$. To see this, note that there is a function $x_P$ that has a second order pole exclusively at $P$ (namely, this is the second generator of $\CL(2P)$), and that $y=x+x_P$ is in $\CL(2(P+Q))$ with the desired pole structure, and linearly independent of $1$, $x$, and $x^2$.

Next, $\CL(3(P+Q))$ is six-dimensional, with generators $1$, $x$, $x^2$, $x^3$, $y$, and $yx$, and $\CL(4(P+Q))$ is eight-dimensional, with a relation among the nine naively independent functions:
\begin{equation}
  a y^2 + b_0 x^2 y + b_1 x y + b_2 y = c_0 + c_1 x + c_2 x^2 + c_3 x^3 + c_4 x^4 \,.
\end{equation}
The only term with a pole of order $4$ at $P$ is $c_4 x^4$, hence $c_4=0$. Furthermore, the only two terms with a third-order pole at $P$ now are $a y^2$ and $b_0 x^2 y$, and by an argument similar to the one for Weierstrass forms, one has $a \neq 0$, $b_0 \neq 0$. Dividing the entire equation by $a$ and relabeling the $b_i$ and $c_i$ results in the quartic form
\begin{equation}
  y^2 + b_0 x^2 y + b_1 x y + b_2 y = c_0 + c_1 x + c_2 x^2 + c_3 x^3 + c_4 x^4
\end{equation}
that is the starting point for the analysis in the main text.


\section{Polynomial equations in UFDs}
\label{sec:PolySols}

In this appendix, we summarize how  to solve polynomial equations over unique factorization domain (UFD),  which appear recurrently  in the discriminant. The sections $b_i$ and $c_j$ of the quartic equation realizing the elliptic curve with two sections, take values in a UFD, given by the ring of local functions on the base of the fibration \cite{MR0103906}. Similar methods were used for $\mathbb{P}^{(1,2,3)}$ in \cite{Katz:2011qp} for the Tate's algorithm. 

\subsection{Two-term Polynomial}
\label{sec:TwoTermPoly}

The first such recurring polynomial is given by
\begin{equation}
  P = s_\alpha s_\beta - s_\gamma s_\delta \,.
\end{equation}
The condition $P=0$ then amounts to the fact that $s_\alpha s_\beta$ and $s_\gamma s_\delta$ have identical factorizations into irreducibles. Therefore, the most general Ansatz compatible with $P=0$ is
\begin{equation}
  \begin{aligned}
    s_\alpha = \sigma_1 \sigma_2\,, \qquad s_\beta = \sigma_3 \sigma_4 \,, \\
    s_\gamma = \sigma_1 \sigma_3\,, \qquad s_\delta = \sigma_2 \sigma_4 \,.
  \end{aligned}
\end{equation}
Furthermore, with $(a,b)$ denoting the greatest common divisor of $a$ and $b$, it is possible to choose these sections such that $\sigma_1=(s_\alpha,s_\gamma)$ and $\sigma_4=(s_\beta,s_\delta)$, thereby making the pairs $\{\sigma_2, \sigma_3\}$ and $\{\sigma_1, \sigma_4\}$ coprime.

\subsection{Four-term Polynomial}
\label{sec:FourTermPoly}

A second recurring four-term polynomial in the discriminants of the fiber is of the form
\begin{equation}
  P = b_{i}^3 c_0 - b_{i}^2 b_{j} c_1 + b_{i} b_{j}^2 c_2 - b_{j}^3 c_3 \,.
\end{equation}
To solve $P=0$, first note, that $b_j$ divides the last three terms, it also has to divide the first, hence $b_j|b_i^3 c_0$. Analogously, $b_i|b_j^3 c_3$.

Next, decompose $b_i=\sigma_1 \sigma_2$ and $b_j=\sigma_1 \sigma_3$, where $\sigma_1=(b_i,b_j)$. $\sigma_1$ being the greatest common divisor implies that all irreducibles in $\sigma_2$ do not divide $\sigma_3$ and vice versa -- any such irreducibles would be subsumed within $\sigma_1$. Then, one can rewrite the polynomial
\begin{equation}
  P = \sigma_1^3 \left( \sigma_2^3 c_0 - \sigma_2^2 \sigma_3 c_1 + \sigma_2 \sigma_3^2 c_2 - \sigma_3^3 c_3 \right) \,.
\end{equation}
Now one has $\sigma_2 | \sigma_3^3 c_3$. But since $\sigma_2$ and $\sigma_3$ do not share any irreducibles, this condition amounts to $\sigma_2 | c_3$. Thus $c_3=\sigma_2 \sigma_5$. By a similar argument $c_0=\sigma_3 \sigma_4$. Applying these decompositions, the polynomial reads
\begin{equation}
  P = \sigma_1^3 \sigma_2 \sigma_3 \left( \sigma_2 \left( \sigma_2 \sigma_4 - c_1 \right) + \sigma_3 \left( c_2 - \sigma_3 \sigma_5 \right) \right) \,.
\end{equation}
The factors $\sigma_1^3 \sigma_2 \sigma_3$ can at times give rise to new canonical enhancements, which have not appeared elsewhere in the algorithm and therefore have to be always checked as well. So these solutions are
\be\label{FourTermSolCan}
  \ba
    \sigma_1=0 &:\qquad  b_i =b_j=0 \cr
    \sigma_2 =0 &: \qquad b_i = c_3=0\cr
    \sigma_3=0 &:\qquad b_j = c_0 =0
  \ea
\ee
and correspond to canonical enhancements. 

The other solutions are characterized in the tems of the vanishing of the remaining factor in $P$. 
Define $\tilde\alpha=  \sigma_2 \sigma_4 - c_1$ and  $\tilde\beta= c_2 - \sigma_3 \sigma_5 $. As again $\sigma_2$ and $\sigma_3$ do not share any irreducibles, one has $\sigma_3|\tilde\alpha$ and $\sigma_2|\tilde\beta$, hence there are decompositions $\tilde\alpha = \sigma_3 \alpha$ and $\tilde\beta = \sigma_2 \beta$. With the non-vanishing of $b_i$ and $b_j$, $P=0$ is now reduced to $\alpha = - \beta$. Solving for the coefficients, one obtains
\begin{equation}\label{FourTermSol}
  \begin{aligned}
    c_0 &= \sigma_3 \sigma_4 \\
    c_1 &= \sigma_2 \sigma_4 + \sigma_3 \alpha \\
    c_2 &= \sigma_3 \sigma_5 + \sigma_2 \alpha \\
    c_3 &= \sigma_2 \sigma_5 \\
    b_i &= \sigma_1 \sigma_2 \\
    b_j &= \sigma_1 \sigma_3
  \end{aligned}
\end{equation}
as the final solution set to $P=0$ with six free functions $\sigma_1$, $\sigma_2$, $\sigma_3$, $\sigma_4$, $\sigma_5$ and $\alpha${, with $\sigma_2$ and $\sigma_3$ coprime}.

There are in summary four branches of the solution set to $P=0$: (\ref{FourTermSolCan}) and (\ref{FourTermSol}).

\subsection{Three-term Polynomials}
\label{sec:ThreeTermPoly}

Another recurring polynomial with three terms is given by
\begin{equation}
  P = b_i^2 c_\alpha - b_i b_j c_\beta + b_j^2 c_\gamma \,.
\end{equation}
In the same vein as above, one can use the divisibility conditions to find the general Ansatz 
\begin{equation}
  \begin{aligned}
    b_i &= \sigma_1 \sigma_2 \\
    b_j &= \sigma_1 \sigma_3 \\
    c_\alpha &= \sigma_3 \sigma_4 \\
    c_\gamma &= \sigma_2 \sigma_5 \,.
  \end{aligned}
\end{equation}
Then, the polynomial equation reduces to
\be\label{ThreeTerm1}
P = \sigma_1^2 \sigma_2 \sigma_3 (  \sigma_2 \sigma_4 + \sigma_3 \sigma_5  - c_\beta )
\ee
This has solutions
\begin{equation}\label{ThreeTermSolCan}
\ba
\sigma_1=0: &\qquad b_i =b_j=0 \cr
\sigma_2=0: & \qquad b_i =c_\gamma =0 \cr
\sigma_3=0: & \qquad b_j = c_\alpha =0 \,.
\ea
\end{equation}
as well as the solution where $b_i, b_j \not=0$ 
\be\label{ThreeTermSol1}
\ba
  b_i &= \sigma_1 \sigma_2 \cr
    b_j &= \sigma_1 \sigma_3 \cr
    c_\alpha &= \sigma_3 \sigma_4 \cr
    c_\gamma &= \sigma_2 \sigma_5 \cr
c_\beta &= \sigma_2 \sigma_4 + \sigma_3 \sigma_5 \,.
\ea
\ee
The most general solution to the three-term polynomial thus has five free functions $\sigma_1$, $\sigma_2$, $\sigma_3$, $\sigma_4$ and $\sigma_5${, and the two functions $\sigma_2$, $\sigma_3$ are coprime}. In summary, there are four solution sets to (\ref{ThreeTerm1}): (\ref{ThreeTermSolCan}) and (\ref{ThreeTermSol1}).

Another three-term polynomial that one encounters while working through the algorithm is
\begin{equation}
  P = b_1^2 c_0 - b_1 b_2 c_1 - c_1^2 \,.
  \label{eq:AnotherThreeTerm}
\end{equation}
Since $b_1$ divides the first two terms, one has $b_1|c_1^2$, the most general ansatz compatible with which is $b_1 = \alpha^2 \beta$, $c_1= \alpha \beta \gamma$. Let furthermore $\delta=(\alpha,\gamma)$, and $\alpha = \delta \tilde \alpha$, $\gamma = \delta \tilde c_1$. Then, $(\tilde \alpha, \tilde c_1)=1$, and one has
\begin{equation}
  \begin{aligned}
    b_1 &= \tilde \alpha^2 \delta^2 \beta \,, \\
    c_1 &= \delta^2 \tilde \alpha \beta \tilde c_1 \,.
  \end{aligned}
\end{equation}
The polynomial is now given by
\begin{equation}
  P = \tilde \alpha^2 \delta^4 \beta^2 \left( \tilde \alpha^2 c_0 - \tilde \alpha b_2 \tilde c_1 - \tilde c_1^2 \right) \,.
\end{equation}
Immediately there is the solution, which gives rise to a canonical model 
\be
 b_1 = c_1 = 0  \,.
\ee
The remaining polynomial term gives another solution:
as $\tilde \alpha$ divides the first two terms in the bracket, $\tilde \alpha | \tilde c_1^2$ holds. However, since also  $(\tilde \alpha, \tilde c_1)=1$, it follows that $\tilde \alpha=1$. Therefore, $b_1 = \delta^2 \beta$ and the second solution to $P=0$ is given by
\begin{equation}
\ba
  b_1 &= \delta^2 \beta \cr
    c_1 &= \delta^2 \beta \tilde c_1 = b_1 \tilde c_1 \cr
    c_0 &= b_2 \tilde c_1 + \tilde c_1^2 \,.
 \ea
 \end{equation}


\section{Alternative forms for $I_5$}
\label{sec:AppP112}

In this section, the matter content and $U(1)$ charges of all canonical and singly non-canonical $I_5$ models is summarized. We have already seen that there are multiple, equivalent ways of realizing $I_5$ fibers with two sections, and at times it might be useful to have all the realizations. 
Our focus on $I_5$ is purely motivated from its application in F-theory model building, but similar forms can be obtained in the algorithm for any $I_n$ following our results in  section \ref{sec:In} for the $I_n$ and the symmetries in section \ref{sec:Sym} and the lop transformations in \ref{sec:Lop}. 
The forms presented in the main part of the paper, are a minimal set, realizing each fiber type, as well as being equivalent to the ones in this appendix by the arguments in sections \ref{sec:Sym} and \ref{sec:Lop}.  

In table \ref{tab:I5NonCanChargesAppendix} the following definitions were used:
\begin{align}
    P_0 =&\, b_{0,1} c_{2,1} - b_{1,0} c_{3,2} \label{eq:D1112100P0}\\
    P_1 =&\, b_{1,0}^2 c_{0,1} - b_{1,0} b_{2,0} c_{1,1} + b_{2,0}^2 c_{2,1} \label{eq:D1112100P1}\\
        \label{eq:NC3P2}
    P_2 =&\, \sigma_4 b_{2,0}^2 + \sigma_1^2 \sigma_3 c_{0,1} - \sigma_1 b_{2,0} c_{1,1} \\
        \label{eq:NC3P3}
    P_3 =&\, \sigma_1 \sigma_2^2 \left(\sigma_1 c_{1,1}-\sigma_4 b_{2,0}\right)+\sigma_3 \sigma_2 \left(\sigma_4
        \sigma_1 b_{1,1} - \sigma_1^2 c_{2,2} + \sigma_4^2\right) \notag\\
        &+ \sigma_1 \sigma_3^2 \left(\sigma_1 c_{3,3}-\sigma_4 b_{0,2}\right) \\
        \label{eq:NC4P2}
    P_4 =&\, \sigma_1 c_{3,2} - \sigma_5 b_{0,1} \\
        \label{eq:NC4P3}
    P_5 =&\, - \sigma_2^3 \left(\sigma_4 \sigma_1 b_{2,1} - \sigma_1^2 c_{0,2}+\sigma_4^2 \right)+\sigma_3 \sigma_2^2
        \left(\sigma_1 \left(\sigma_5 b_{2,1}-\sigma_1 c_{1,2}\right)+\sigma_4 \left(\sigma_1 b_{1,1}+2 \sigma_5\right)\right) \notag\\
        &-\sigma_3^2 \sigma_2 \left(\sigma_1 \left(\sigma_4 b_{0,1}-\sigma_1 c_{2,2}\right)+\sigma_1 \sigma_5 b_{1,1}+\sigma_5^2\right) + \sigma_1 \sigma_3^3 \left(\sigma_5 b_{0,1}-\sigma_1 c_{3,2}\right) \,.
\end{align}


\begin{table}\centering
  \begin{tabular}{c|c|c|c}
    Model & Matter locus & Representation & Fiber type \\ \hline \hline
    

    $Q(1,1,1,2,1,0,0)|_{(\ref{eq:D1112100P0})}$  & $\sigma_3$ & $\mathbf{10}_{1} + \overline{\mathbf{10}}_{-1}$ & $I_5^{(0||1)}$ \\
                               & $\sigma_1$ & $\mathbf{10}_{-4} + \overline{\mathbf{10}}_{4}$ & \\
                               & $\sigma_2$ & $\mathbf{5}_{-7} + \overline{\mathbf{5}}_{7}$ & \\
                               & (\ref{eq:NC3P2}) & $\mathbf{5}_{-2} + \overline{\mathbf{5}}_{2}$ & \\
                               & (\ref{eq:NC3P3}) & $\mathbf{5}_{3} + \overline{\mathbf{5}}_{-3}$ & \\ \hline
                              
                              
    $Q(1,1,1,2,1,0,0)|_{(\ref{eq:D1112100P1})}$  & $\sigma_1$ & $\mathbf{10}_{2} + \overline{\mathbf{10}}_{-2}$ & $I_5^{(0|1)}$ \\
                               & $\sigma_2$ & $\mathbf{10}_{-3} + \overline{\mathbf{10}}_{3}$ & \\
                               & $b_{0,1}$ & $\mathbf{5}_{6} + \overline{\mathbf{5}}_{-6}$ & \\
                               & (\ref{eq:NC4P2}) & $\mathbf{5}_{-4} + \overline{\mathbf{5}}_{4}$ & \\
                               & (\ref{eq:NC4P3}) & $\mathbf{5}_{1} + \overline{\mathbf{5}}_{-1}$ & 

  \end{tabular}
  \caption{Matter curves and $U(1)$ charges for non-canonical $I_5$ models arising from canonical $I_4$ models through Tate's algorithm,  which are related to those in section \ref{sec:P112NonCanonical} under lopping transformation.}
  \label{tab:I5NonCanChargesAppendix}
\end{table}

\section{Relation to Top Models and Spectral Covers}
\label{sec:tops}

Previously models with extra sections were constructed based on toric tops \cite{Candelas:2012uu, Borchmann:2013jwa, Borchmann:2013hta, Braun:2013nqa}. Furthermore, there are models in the standard $\mathbb{P}^{(1,2,3)}$ that realize extra sections  \cite{Grimm:2010ez, Mayrhofer:2012zy}, 
 and were constructed inspired by and are related to factored spectral cover models  \cite{Marsano:2009gv, Marsano:2009wr, Dolan:2011iu}. We should finally comment on the relation of the Tate models found here to these top models. 

The short summary is: all top models feature in the tree, in terms of canonical models. However the tree gives rise to more models, namely, the non-canonical models. The non-canonical models have the same fiber types as canonical ones, however their codimension 2 structure is different: in particular the canonical models, and thus the tops, have only one type of codimension 2 locus that is of type $I_1^*$, i.e. gives rise to ${\bf 10}$ matter. 

The top models 1 and 2   (in the nomenclature  of \cite{Borchmann:2013jwa}) were already mentioned and are  exactly the two canonical $I_5$ models, obtained in section \ref{sec:I5s}. 
The other tops are obtained from non-canonical forms as specializations.
Consider the non-canonical $I_5$ model, $\CQ(3,2,1,1,0,0,1)|_{(\ref{eq:D3211001P1})}$, and specialize by assuming that $\sigma_2$ never vanishes. This has two effects: The matter curve above $\sigma_2=0$ will not be present in the spectrum anymore, and $b_{1,0}|b_{2,1}$. Therefore, an expression of the form $\frac{b_{2,1}}{b_{1,0}}=\frac{\sigma_3}{\sigma_2}$ is now well-defined over the whole base manifold. Then apply the coordinate shift
\begin{equation}
  \left( \begin{array}{c} x \\ y \end{array} \right) \rightarrow \left( \begin{array}{c} x - \frac{\sigma_3}{\sigma_2}z s w \\ y \end{array} \right) \,,
\end{equation}
which gives a new fibration that has canonical form
\begin{equation}
  \CQ(4,2,1,0,0,0,2)
\end{equation}
and is known in the literature as Top 2. Similarly, if $\sigma_1=1$, then $b_{1,0}|c_{2,1}$ and $b_{2,1}|c_{0,3}$, and the now well-defined shift
\begin{equation}
y \quad  \rightarrow \quad y + \frac{\sigma_4}{\sigma_1} z^2 s w^2 + \frac{\sigma_5}{\sigma_1} z w x 
\end{equation}
produces the canonical form
\begin{equation}
  \CQ(4,3,2,1,0,0,1) \,,
\end{equation}
also known as Top 3.

Next, the noncanonical form $\CQ(3,2,1,1,0,0,1)|_{(\ref{eq:D3211001P0})}$ has a canonical subform for $\sigma_1=1$, which is reachable by shifting
\begin{equation}
y \quad  \rightarrow \quad  y +  \frac{\sigma_4}{\sigma_1} z w x 
\end{equation}
and given by
\begin{equation}
  \CQ(3,2,2,2,0,0,1)
\end{equation}
or Top 4.

Furthermore, there is an identification between the non-canonical $I_5^{(0|1)}$ and $I_5^{(0||1)}$ models and the ones arising from mapping a factorised Tate form in $\bbP^{(1,2,3)}$ to $\mathrm{Bl}_{[0,1,0]}\bbP^{(1,1,2)}$ that have been discussed in \cite{Mayrhofer:2012zy}. There, the model corresponding to a $4+1$-factorized Tate model was found to be given by
\begin{equation}
  s y^2 + \fkb_0 x^2 y = \fkc_{0,2} s^3 w^4 + \fkc_{1,1} s^2 w^3 x + \fkc_{2,0} s w^2 x^2 + \fkc_{3,0} w x^3 \,,
\end{equation}
with coefficient specializations
\begin{equation}
  \begin{aligned}
    c_{0,2} &= \frac{1}{4} b_{2,1}^2 \\
    c_{1,1} &= - \frac{1}{2} \sigma_1 b_{2,1} \sigma_3 \\
    c_{2,0} &= \frac{1}{4} \sigma_1^2 \sigma_3^2 \\
    c_{2,1} &= \sigma_4 \sigma_3 - \frac{1}{2} \sigma_1 b_{2,1} \\
    c_{3,0} &= \frac{1}{2} \sigma_1^2 \sigma_3 \\
    c_{3,1} &= \sigma_4 \,.
  \end{aligned}
\end{equation}
After the coordinate shift
\begin{equation}
y  \quad \rightarrow\quad   y - \frac{1}{2} b_{2,1} z s w^2 + \frac{1}{2} \sigma_1 \sigma_3 w x \,,
\end{equation}
this model turns out to be identical to the non-canonical fibration $Q(3,2,1,1,0,0,1)|_{(\ref{eq:D3211001P0})}$, specialized with $\sigma_2=1$.

The $3+2$-factorized Tate model reads
\begin{equation}
  s y^2 + \fkb_0 x^2 y = \fkc_{0,2} s^3 w^4 + \fkc_{1,1} s^2 w^3 x + \fkc_{2,0} s w^2 x^2 + \fkc_{3,0} w x^3 \,,
\end{equation}
with coefficient specializations
\begin{equation}
  \begin{aligned}
    c_{0,2} &= \frac{1}{4} \sigma_1^2 \sigma_3^2 \\
    c_{0,3} &= \sigma_3 \sigma_4 \\
    c_{1,1} &= \frac{1}{2} \sigma_2 \sigma_1^2 \sigma_3 \\
    c_{1,2} &= \sigma_5 \sigma_3 + \sigma_2 \sigma_4 \\
    c_{2,0} &= \frac{1}{4} \sigma_2^2 \sigma_1^2 \\
    c_{2,1} &= \sigma_2 \sigma_5 - \frac{1}{2} \sigma_1 \sigma_3 b_{0,0} \\
    c_{3,0} &= - \frac{1}{2} \sigma_2 \sigma_1 b_{0,0} \,.
  \end{aligned}
\end{equation}
Here, the coordinate shift
\begin{equation}
y  \quad \rightarrow\quad   y - \frac{1}{2} \sigma_1 \sigma_3 z s w^2 - \frac{1}{2} \sigma_1 \sigma_2 w x 
\end{equation}
provides an identification of this fibration with the non-canonical model $Q(3,2,1,1,0,0,1)|_{(\ref{eq:D3211001P1})}$.


\section{Resolutions for $I_n^*$ and $I_n$ fibers}
\label{sec:ResolvedGeometries}

In this appendix, we present the resolved geometries and Cartan divisors for the fibrations with fiber types $I_n^{*(01)}$, $I_n^{*(0|1)}$, $I_n^{*(0||1)}$ and $I_n^{(0|^k1)}$ given in section \ref{sec:TreeTopsInfty}. All resolutions were implemented in mathematica using \cite{Smooth}. 


\tocless\subsection{$I_n^{*(01)}$}

The ordered set of resolutions that resolves the $I_n^{*(01)}$ fibration in all codimensions is given by
\begin{equation}
  \begin{aligned}
    (z, x, y; \zeta_1), \qquad (\zeta_1, y; \epsilon_0), \qquad (\zeta_1, \epsilon_0; \delta_0), & \\
    (\epsilon_0, x; \epsilon_1), \qquad (\epsilon_0, \epsilon_1; \delta_1), \qquad (\epsilon_1, y; \epsilon_2), & \\
    (\epsilon_{k-1}, \epsilon_k; \delta_k), \qquad (\epsilon_k, x; \epsilon_{k+1}) & \qquad k \text{ even} \\
    (\epsilon_{k-1}, \epsilon_k; \delta_k), \qquad (\epsilon_k, y; \epsilon_{k+1}) & \qquad k \text{ odd} \,.
  \end{aligned}
\end{equation}
for $k=2,\dots,n$. For $n$ odd, the fully resolved geometry is given by
\begin{equation}
  \begin{aligned}
    & y^2 s \left( \epsilon_0 \epsilon_2 \epsilon_4 \cdots \epsilon_{n+1} \right) + \fkb_0 x^2 y \zeta_1 \left( \delta_0 \delta_1^2 \delta_2^3 \cdots \delta_n^{n+1} \right) \\
    & + \fkb_{1,1} s w x y z \zeta_1 \left(\delta_0 \cdots \delta_n\right) \left( \epsilon_0 \cdots \epsilon_{n+1} \right) + \fkb_{2,2+\frac{n-1}{2}} y s^2 w^2 z^{2+\frac{n-1}{2}} \zeta_1^{1+\frac{n-1}{2}} \left( \delta_0^n \delta_1^{n-1} \cdots \delta_{n-1} \right) \\
    = & \, \fkc_{0,n+4} w^4 s^3 z^{4+n} \zeta_1^{2+n} \left( \delta_0^{2n+2} \delta_1^{2n} \cdots \delta_n^{2} \right) \left( \epsilon_0^{n+1} \epsilon_1^{n} \cdots \epsilon_n \right) \\
    & + \fkc_{1,2+\frac{n+1}{2}} w^3 s^2 x z^{2+\frac{n+1}{2}} \zeta_1^{1+\frac{n+1}{2}} \left( \delta_0^{2n+2} \delta_1^{2n} \cdots \delta_n^2 \right) \left( \left( \epsilon_0 \epsilon_1\right)^{\frac{n+1}{2}} \left( \epsilon_2 \epsilon_3\right)^{\frac{n-1}{2}} \cdots \left( \epsilon_{n-1} \epsilon_{n} \right) \right) \\
    & + \fkc_{2,1} s w^2 x^2 z \zeta_1 \left( \epsilon_1 \epsilon_3 \epsilon_5 \cdots \epsilon_n \right) \\
    & + \fkc_3 w x^3 \zeta_1 \left( \delta_1 \delta_2^2 \cdots \delta_n^n \right) \left( \left( \epsilon_1 \epsilon_2\right) \left(\epsilon_3 \epsilon_4\right)^2 \cdots \left(\epsilon_{n} \epsilon_{n+1}\right)^{\frac{n+1}{2}} \right) \left( \epsilon_1 \epsilon_3 \cdots \epsilon_n \right) \,.
  \end{aligned}
\end{equation}
The irreducible Cartan divisors are
\begin{equation}
  \begin{tabular}{c|c}
    Section & Equation in $Y_4$ \\ \hline
    $z$ & $- c_{3,0}w x^3 \zeta_1 + y \epsilon_0 \left( s y+b_{0,0} x^2 \zeta_1 \delta_0 \right)$ \\
    $\epsilon_0$ & $c_{2,1} z + c_{3,0} \delta_1$ \\
    $\delta_0$ & $\epsilon_0 - x^2 \zeta_1 \epsilon_1 \left( c_{2,1} z + c_{3,0} x \delta_1 \epsilon_1 \right)$ \\
    $\delta_{0<i<n}$, $i$ odd & $c_{2,1} \epsilon_i - y^2 \epsilon_{i-1} \epsilon_{i+1}$ \\
    $\delta_{0<i<n}$, $i$ even & $\epsilon_i - c_{2,1} x^2 \epsilon_{i-1} \epsilon_{i+1}$ \\
    $\epsilon_n$ & $b_{2,2+\frac{n-1}{2}}\delta_{n-1} + \epsilon_{n+1}$ \\
    $\epsilon_{n+1}$ & $b_{2,2+\frac{n-1}{2}} y - \epsilon_n \left( c_{2,1} x^2 + \delta_n\left( c_{1,2+\frac{n+1}{2}} x + c_{0,n+4} \delta_n \right) \right)$ \\
    $\delta_n$ & $c_{2,1} \epsilon_n - y \epsilon_{n-1} \left( y \epsilon_{n+1} + b_{2,2+\frac{n-1}{2}} \delta_{n-1} \right) $
  \end{tabular}
\end{equation}
The intersections follow from the projective relations induced by the blow-ups, can be computed as outlined in section \ref{sec:Fibration}.  They reproduce the affine $D_{n+4}$ Dynkin diagram if the divisors are ordered as $(z, \epsilon_0, \delta_0, \delta_1, \dots, \delta_n, \epsilon_n, \epsilon_{n+1})$. Both $\sigma_0$ and $\sigma_1$ intersect $z=0$.

For even $n$, the fully resolved geometry reads
\begin{equation}
  \begin{aligned}
    & y^2 s \left( \epsilon_0 \epsilon_2 \epsilon_4 \cdots \epsilon_n \right) + \fkb_0 x^2 y \zeta_1 \left( \delta_0 \delta_1^2 \delta_2^3 \cdots \delta_n^{n+1} \right) \\
    & + \fkb_{1,1} s w x y z \zeta_1 \left(\delta_0 \cdots \delta_n \right) \left( \epsilon_0 \cdots \epsilon_{n+1} \right) + \fkb_{2,2+\frac{n}{2}} y s^2 w^2 z^{2+\frac{n}{2}} \zeta_1^{1+\frac{n}{2}} \left( \delta_0^{n+1} \delta_1^n \cdots \delta_n \right) \\
    = & \, \fkc_{0,n+4} w^4 s^3 z^{4+n} \zeta_1^{2+n} \left( \delta_0^{2n+2} \delta_1^{2n} \cdots \delta_n^{2} \right) \left( \epsilon_0^{n+1} \epsilon_1^{n} \cdots \epsilon_n \right) \\
    & + \fkc_{1,2+\frac{n}{2}} w^3 s^2 x z^{2+\frac{n}{2}} \zeta_1^{1+\frac{n}{2}} \left( \delta_0^n \delta_1^{n-1} \cdots \delta_{n-1} \right) \left( \left( \epsilon_0 \epsilon_1\right)^{\frac{n}{2}} \left( \epsilon_2 \epsilon_3\right)^{\frac{n}{2}-1} \cdots \left( \epsilon_{n-2} \epsilon_{n-1} \right) \right) \\
    & + \fkc_{2,1} s w^2 x^2 z \zeta_1 \left( \epsilon_1 \epsilon_3 \epsilon_5 \cdots \epsilon_{n+1} \right) \\
    & + \fkc_3 w x^3 \zeta_1 \left( \delta_1 \delta_2^2 \cdots \delta_n^n \right) \left( \left( \epsilon_1 \epsilon_2\right) \left(\epsilon_3 \epsilon_4\right)^2 \cdots \left(\epsilon_{n-1} \epsilon_{n}\right)^{\frac{n}{2}} \right) \left( \epsilon_1 \epsilon_3 \cdots \epsilon_{n-1} \right) \epsilon_{n+1}^{2+\frac{n}{2}} \,,
  \end{aligned}
\end{equation}
and the irreducible Cartan divisors are
\begin{equation}
  \begin{tabular}{c|c}
    Section & Equation in $Y_4$ \\ \hline
    $z$ & $- c_{3,0} w x^3 \zeta_1 + y \epsilon_0 \left( s y + b_{0,0} x^2 \zeta_1 \delta_0 \right)$ \\
    $\epsilon_0$ & $c_{2,1} z + c_{3,0} \delta_1 $ \\
    $\delta_0$ & $\epsilon_0 - x^2 \zeta_1 \epsilon_1 \left( c_{2,1} z + c_{3,0} x \delta_1 \epsilon_1 \right) $ \\
    $\delta_{0<i<n}$, $i$ odd & $c_{2,1} \epsilon_i - y^2 \epsilon_{i-1} \epsilon_{i+1}$ \\
    $\delta_{0<i<n}$, $i$ even & $\epsilon_i - c_{2,1} x^2 \epsilon_{i-1} \epsilon_{i+1}$ \\
    $\epsilon_n$ & $c_{1,2+\frac{n}{2}}\delta_{n-1} + c_{2,1} \epsilon_{n+1}$ \\
    $\epsilon_{n+1}$ & $c_{1,2+\frac{n}{2}} x - \epsilon_n \left( y^2 + b_{2,2+\frac{n}{2}} y \delta_n - c_{0,4+n} \delta_n^2 \right)$ \\
    $\delta_n$ & $\epsilon_n - x \epsilon_{n-1} \left( c_{2,1} x \epsilon_{n+1} + c_{1,2+\frac{n}{2}} \delta_{n-1} \right) $
  \end{tabular}
\end{equation}
Again and with the same ordering as above, one finds that the intersections yield the affine $D_{n+4}$ Dynkin diagram, with $\sigma_0$ and $\sigma_1$ being located on $z=0$.

\tocless\subsection{$I_n^{*(0|1)}$}

The ordered set of resolutions that resolves the $I_n^{*(0|1)}$ fibration in all codimensions is given by
\begin{equation}
  \begin{aligned}
    (z, x, y; \zeta_1), \qquad (z, y; \epsilon_0), \qquad (\zeta_1, y; \epsilon_1), & \\
    (\epsilon_0, \zeta_1, \delta_0), \qquad (\zeta_1, \epsilon_1; \delta_1), \qquad (\epsilon_1, x; \epsilon_2), & \\
    (\epsilon_{k-1}, \epsilon_k; \delta_k), \qquad (\epsilon_k, y; \epsilon_{k+1}) & \qquad k \text{ even} \\
    (\epsilon_{k-1}, \epsilon_k; \delta_k), \qquad (\epsilon_k, x; \epsilon_{k+1}) & \qquad k \text{ odd} \,.
  \end{aligned}
\end{equation}
for $k=2,\dots,n$. The resolved geometry for odd $n$ reads
\begin{equation}
  \begin{aligned}
    & y^2 s \epsilon_0 \left( \epsilon_1 \epsilon_3 \cdots \epsilon_n \right) + \fkb_0 x^2 y \zeta_1 \left( \delta_1 \delta_2^2 \cdots \delta_n^n \right) \epsilon_1 \left( ( \epsilon_2 \epsilon_3 )^2 ( \epsilon_4 \epsilon_5 )^3 \cdots ( \epsilon_{n-1} \epsilon_n )^{\frac{n+1}{2}} \right) \epsilon_{n+1}^{\frac{n+3}{2}}  \\
    & + \fkb_{1,1} s w x y z \zeta_1 \left( \delta_0 \delta_1 \cdots \delta_n \right) \left( \epsilon_0 \epsilon_1 \cdots \epsilon_{n+1} \right) \\
    &+ \fkb_{2,1+\frac{n+1}{2}} y s^2 w^2 z^{1+\frac{n+1}{2}} \zeta_1^{\frac{n+1}{2}} \left( \delta_0^{n+1} \delta_1^n \cdots \delta_n \right) \epsilon_0^{1+\frac{n+1}{2}} \epsilon_1^{\frac{n+1}{2}} \left( (\epsilon_2 \epsilon_3)^{\frac{n-1}{2}} (\epsilon_4 \epsilon_5)^{\frac{n-3}{2}} \cdots ( \epsilon_{n-1} \epsilon_n ) \right) \\
    = & \, \fkc_{0,3+n} w^4 s^3 z^{3+n} \zeta_1^{1+n} \left( \delta_0^{2+2n} \delta_1^{2n} \cdots \delta_n^2 \right) \epsilon_0^{n+2} \left( \epsilon_1^n \epsilon_2^{n-1} \cdots \epsilon_n \right)  \\
    & + \fkc_{1,2+\frac{n-1}{2}} w^3 s^2 x z^{2+\frac{n-1}{2}} \zeta_1^{1+\frac{n-1}{2}} \left( \delta_0^n \delta_1^{n-1} \cdots \delta_{n-1} \right) \epsilon_0^{\frac{n+1}{2}} \left( (\epsilon_1 \epsilon_2)^{\frac{n-1}{2}} (\epsilon_3 \epsilon_4)^{\frac{n-3}{2}} \cdots (\epsilon_{n-2} \epsilon_{n-1}) \right) \\
    & + \fkc_{2,1} s w^2 x^2 z \zeta_1 \left( \epsilon_2 \epsilon_4 \cdots \epsilon_{n+1} \right) \\
    & + \fkc_{3,1} w x^3 z \zeta_1^2 \left( \delta_0 \delta_1^2 \cdots \delta_n^{n+1} \right) \epsilon_1 \left( (\epsilon_2 \epsilon_3)^2 (\epsilon_4 \epsilon_5)^3 \cdots (\epsilon_{n-1} \epsilon_n)^{\frac{n+1}{2}} \right) \left( \epsilon_2 \epsilon_4 \cdots \epsilon_{n-1} \right) \epsilon_{n+1}^{\frac{n+3}{2}} \,.
  \end{aligned}
\end{equation}
The irreducible Cartan divisors are given by
\begin{equation}
  \begin{tabular}{c|c}
    Section & Equation in $Y_4$ \\ \hline
    $z$ & $b_{0,0} x^2 \zeta_1 + s \epsilon_0$ \\
    $\epsilon_0$ & $b_{0,0} y \epsilon_1 - z \left( c_{2,1} s + c_{3,1} \delta_0 \epsilon_1 \right)$ \\
    $\delta_0$ & $c_{2,1} z \zeta_1 - \epsilon_1 \left( \epsilon_0 + b_{0,0} \zeta_1 \delta_1 \right)$ \\
    $\delta_{0<i<n}$, $i$ odd & $\epsilon_i - c_{2,1} x^2 \epsilon_{i-1} \epsilon_{i+1}$ \\
    $\delta_{0<i<n}$, $i$ even & $c_{2,1} \epsilon_i - y^2 \epsilon_{i-1} \epsilon_{i+1}$ \\
    $\epsilon_n$ & $c_{1,2+\frac{n-1}{2}} \delta_{n-1} + c_{2,1} \epsilon_{n+1}$ \\
    $\epsilon_{n+1}$ & $c_{1,2+\frac{n-1}{2}} x + \epsilon_n \left( y^2 + b_{2,1+\frac{n+1}{2}} y \delta_n - c_{0,3+n} \delta_n^2 \right)$ \\
    $\delta_n$ & $\epsilon_n - x \epsilon_{n-1}\left(c_{1,2+\frac{n-1}{2}} \delta_{n-1} + c_{2,1} x \epsilon_{n+1}\right)$
  \end{tabular}
\end{equation}
Ordering the cartan divisors as $(z, \epsilon_0, \delta_0, \delta_1, \dots, \delta_n, \epsilon_n, \epsilon_{n+1})$, one reproduces the affine $D_{n+4}$ Dynkin diagram, with $w=0$ intersecting $z$, and $s=0$ intersecting $\epsilon_0$.

For even $n$, the geometry is
\begin{equation}
  \begin{aligned}
    & y^2 s \epsilon_0 \left( \epsilon_1 \epsilon_3 \cdots \epsilon_{n+1} \right) + \fkb_0 x^2 y \zeta_1 \left( \delta_1 \delta_2^2 \cdots \delta_n^n \right) \epsilon_1 \left( ( \epsilon_2 \epsilon_3 )^2 ( \epsilon_4 \epsilon_5 )^3 \cdots ( \epsilon_{n} \epsilon_{n+1} )^{\frac{n+2}{2}} \right) \\
    & + \fkb_{1,1} s w x y z \zeta_1 \left( \delta_0 \delta_1 \cdots \delta_n \right) \left( \epsilon_0 \epsilon_1 \cdots \epsilon_{n+1} \right) \\
    &+ \fkb_{2,1+\frac{n}{2}} y s^2 w^2 z^{1+\frac{n}{2}} \zeta_1^{\frac{n}{2}} \left( \delta_0^{n} \delta_1^{n-1} \cdots \delta_{n-1} \right) \epsilon_0^{1+\frac{n}{2}} \epsilon_1^{\frac{n}{2}} \left( (\epsilon_2 \epsilon_3)^{\frac{n-2}{2}} (\epsilon_4 \epsilon_5)^{\frac{n-4}{2}} \cdots ( \epsilon_{n-2} \epsilon_{n-1} ) \right) \\
    = & \, \fkc_{0,3+n} w^4 s^3 z^{3+n} \zeta_1^{1+n} \left( \delta_0^{2+2n} \delta_1^{2n} \cdots \delta_n^2 \right) \epsilon_0^{n+2} \left( \epsilon_1^n \epsilon_2^{n-1} \cdots \epsilon_n \right)  \\
    & + \fkc_{1,2+\frac{n}{2}} w^3 s^2 x z^{2+\frac{n}{2}} \zeta_1^{1+\frac{n}{2}} \left( \delta_0^{n+1} \delta_1^{n} \cdots \delta_n \right) \epsilon_0^{\frac{n+2}{2}} \left( (\epsilon_1 \epsilon_2)^{\frac{n}{2}} (\epsilon_3 \epsilon_4)^{\frac{n-2}{2}} \cdots (\epsilon_{n-1} \epsilon_{n}) \right) \\
    & + \fkc_{2,1} s w^2 x^2 z \zeta_1 \left( \epsilon_2 \epsilon_4 \cdots \epsilon_{n} \right) \\
    & + \fkc_{3,1} w x^3 z \zeta_1^2 \left( \delta_0 \delta_1^2 \cdots \delta_n^{n+1} \right) \epsilon_1 \left( (\epsilon_2 \epsilon_3)^2 (\epsilon_4 \epsilon_5)^3 \cdots (\epsilon_{n} \epsilon_{n+1})^{\frac{n+1}{2}} \right) \left( \epsilon_2 \epsilon_4 \cdots \epsilon_{n} \right) \,.
  \end{aligned}
\end{equation}
and the Cartan divisors are
\begin{equation}
  \begin{tabular}{c|c}
    Section & Equation in $Y_4$ \\ \hline
    $z$ & $b_{0,0} x^2 \zeta_1 + s \epsilon_0$ \\
    $\epsilon_0$ & $b_{0,0} y \epsilon_1 - z \left( c_{2,1} s + c_{3,1} \delta_0 \epsilon_1 \right)$ \\
    $\delta_0$ & $c_{2,1} z \zeta_1 - \epsilon_1 \left( \epsilon_0 + b_{0,0} \zeta_1 \delta_1 \right)$ \\
    $\delta_{0<i<n}$, $i$ odd & $\epsilon_i - c_{2,1} x^2 \epsilon_{i-1} \epsilon_{i+1}$ \\
    $\delta_{0<i<n}$, $i$ even & $c_{2,1} \epsilon_i - y^2 \epsilon_{i-1} \epsilon_{i+1}$ \\
    $\epsilon_n$ & $b_{2,1+\frac{n}{2}} \delta_{n-1} + \epsilon_{n+1}$ \\
    $\epsilon_{n+1}$ & $b_{2,1+\frac{n}{2}} y - \epsilon_n \left( c_{2,1} x^2 + c_{1,2+\frac{n}{2}} x \delta_n + c_{0,3+n} \delta_n^2 \right)$ \\
    $\delta_n$ & $c_{2,1} \epsilon_n - y \epsilon_{n-1} \left( b_{2,1+\frac{n}{2}} \delta_{n-1} + y \epsilon_{n+1}\right)$
  \end{tabular}
\end{equation}
The ordering of the Cartan divisors and intersection structure is equivalent to the odd case.

\tocless\subsection{$I_n^{*(0||1)}$}

The ordered set of resolutions to desingularize the $I_n^{*(0||1)}$ fibration is
\begin{equation}
  \begin{aligned}
    (z, x, y; \zeta_1), \,\, (z, y; \zeta_2), \,\, (\zeta_1, y; \zeta_3), & \\
    (\zeta_1, \zeta_2; \zeta_4), \,\, (\zeta_2, \zeta_3; \delta_1), & \\
    (\delta_{k-1}, y; \delta_{k}) & \qquad k=2, \dots, n\,.
  \end{aligned}
\end{equation}
The resolved geometry reads, for odd $n$,
\begin{equation}
  \begin{aligned}
    & y^2 s + \fkb_0 x^2 y \zeta_2 \zeta_3 \left( \delta_1 \delta_2^2 \cdots \delta_n^n \right) \\
    & + \fkb_{1,1} s w x y z \zeta_0 \zeta_1 \zeta_2 \zeta_3 \zeta_4 \left( \delta_1 \cdots \delta_n \right) + \fkb_{2,1} y s^2 w^2 z \zeta_2 \\
    = & \, \fkc_{0,2+\frac{n+1}{2}} w^4 s^3 z^{2+\frac{n+1}{2}} \zeta_1^{\frac{n+1}{2}} \zeta_2^{1+\frac{n+1}{2}} \zeta_3^{\frac{n-1}{2}} \zeta_4^{n+1} \left( \delta_1^n \delta_2^{n-1} \cdots \delta_n \right) \\
    & + \fkc_{1,2+\frac{n-1}{2}} w^3 s^2 x z^{2+\frac{n-1}{2}} \zeta_1^{1+\frac{n-1}{2}} \zeta_2^{1+\frac{n-1}{2}} \zeta_3^{\frac{n-1}{2}} \zeta_4^n \left( \delta_1^{n-1} \delta_2^{n-2} \cdots \delta_{n-1} \right) \\
    & + \fkc_{2,1+\frac{n+1}{2}} s w^2 x^2 z^{1+\frac{n+1}{2}} \zeta_1^{1+\frac{n+1}{2}} \zeta_2^{\frac{n+1}{2}} \zeta_3^{\frac{n+1}{2}} \zeta_4^{n+1} \left( \delta_1^n \delta_2^{n-1} \cdots \delta_n \right)  \\
    & + \fkc_{3,1+\frac{n-1}{2}} w x^3 z^{1+\frac{n-1}{2}} \zeta_1^{2+\frac{n-1}{2}} \zeta_2^{\frac{n-1}{2}} \zeta_3^{1+\frac{n-1}{2}} \zeta_4^n \left( \delta_1^{n-1} \delta_2^n \cdots \delta_{n-1} \right) \,.
  \end{aligned}
\end{equation}
and the corresponding irreducible Cartan divisors are
\begin{equation}
  \begin{tabular}{c|c}
    Section & Equation in $Y_4$ \\ \hline
    $z$ & $b_{0,0}x^2 \zeta_1 + s \zeta_2$ \\
    $\zeta_1$ & $b_{2,1} z + \zeta_3$ \\
    $\zeta_2$ & $b_{0,0} y$ \\
    $\zeta_3$ & $b_{2,1} y$ \\
    $\zeta_4$ & $b_{2,1} z \zeta_2 + b_{0,0} \zeta_1 \zeta_3 + \zeta_2 \zeta_3 \delta_1$ \\
    $\delta_i<n$ & $b_{2,1} \zeta_2 + b_{0,0} \zeta_3$ \\
    $\delta_n$ & $b_{2,1} y \zeta_2 + b_{0,0} y \zeta_3 - \zeta_2^{\frac{n-1}{2}} \zeta_3^{\frac{n-1}{2}} \delta_{n-1} \left( c_{2,1+\frac{n+1}{2}} \zeta_2 + c_{3,1+\frac{n-1}{2}} \zeta_3 \right)$
  \end{tabular}
\end{equation}
Again, the intersections reproduce the affine $D_{n+4}$ Dynkin diagram. The required ordering is $(z, \zeta_1, \zeta_4, \delta_1, \delta_2, \dots, \delta_n, \zeta_2, \zeta_3)$. Here, the section $w=0$ intersects $z$, and $s=0$ intersects $\zeta_2$.

For even $n$, the geometry is
\begin{equation}
  \begin{aligned}
    & y^2 s + \fkb_0 x^2 y \zeta_2 \zeta_3 \left( \delta_1 \delta_2^2 \cdots \delta_n^n \right) \\
    & + \fkb_{1,1} s w x y z \zeta_0 \zeta_1 \zeta_2 \zeta_3 \zeta_4 \left( \delta_1 \cdots \delta_n \right) + \fkb_{2,1} y s^2 w^2 z \zeta_2 \\
    = & \, \fkc_{0,2+\frac{n}{2}} w^4 s^3 z^{2+\frac{n}{2}} \zeta_1^{\frac{n}{2}} \zeta_2^{1+\frac{n}{2}} \zeta_3^{-1+\frac{n}{2}} \zeta_4^{n} \left( \delta_1^{n-1} \delta_2^{n-2} \cdots \delta_{n-1} \right) \\
    & + \fkc_{1,2+\frac{n}{2}} w^3 s^2 x z^{2+\frac{n}{2}} \zeta_1^{1+\frac{n}{2}} \zeta_2^{1+\frac{n}{2}} \zeta_3^{\frac{n}{2}} \zeta_4^{1+n} \left( \delta_1^n \delta_2^{n-1} \cdots \delta_{n} \right) \\
    & + \fkc_{2,1+\frac{n}{2}} s w^2 x^2 z^{1+\frac{n}{2}} \zeta_1^{1+\frac{n}{2}} \zeta_2^{\frac{n}{2}} \zeta_3^{\frac{n}{2}} \zeta_4^{n} \left( \delta_1^{n-1} \delta_2^{n-2} \cdots \delta_{n-1} \right)  \\
    & + \fkc_{3,1+\frac{n}{2}} w x^3 z^{1+\frac{n}{2}} \zeta_1^{2+\frac{n}{2}} \zeta_2^{\frac{n}{2}} \zeta_3^{1+\frac{n}{2}} \zeta_4^{1+n} \left( \delta_1^{n} \delta_2^{n-1} \cdots \delta_{n} \right) \,.
  \end{aligned}
\end{equation}
The irreducible Cartan divisors read
\begin{equation}
  \begin{tabular}{c|c}
    Section & Equation in $Y_4$ \\ \hline
    $z$ & $b_{0,0} x^2 \zeta_1 + s \zeta_2$ \\
    $\zeta_1$ & $b_{2,1} z + \zeta_3$ \\
    $\zeta_2$ & $b_{0,0} y$ \\
    $\zeta_3$ & $b_{2,1} y$ \\
    $\zeta_4$ & $b_{2,1} z \zeta_2 + b_{0,0} \zeta_1 \zeta_3 + \zeta_2 \zeta_3 \delta_1$ \\
    $\delta_{i<n}$ & $b_{2,1} \zeta_2 + b_{0,0} \zeta_3$ \\
    $\delta_n$ & $b_{2,1} y \zeta_2 + b_{0,0} y \zeta_3 - \zeta_2^{\frac{n}{2}} \zeta_3^{\frac{n-2}{2}} \delta_{n-1} \left( c_{0,2+\frac{n}{2}} \zeta_2 + c_{2,1+\frac{n}{2}} \zeta_3 \right)$
  \end{tabular}
\end{equation}
The intersection structure is equivalent to the odd case.

\tocless\subsection{$I_{2m+k}^{(0|^k 1)}$}

For this fiber type, the resolution sequence reads (with $z=\zeta_0$)
\begin{equation}
  \begin{aligned}
    (\zeta_i, x, y, \zeta_{i+1}) & \qquad i = 0, \dots, m-1, \\
    (\zeta_i, y, \delta_i) & \qquad i = 1, \dots, m-1, \\
    (z, y, \epsilon_1) & \qquad \text{if } k>0, \\
    (\epsilon_i, y, \epsilon_{i+1}) & \qquad i = 1, \dots, k-1 \,.
  \end{aligned}
\end{equation}
The resolved geometry is
\begin{equation}
  \begin{aligned}
    & y^2 s \left( \delta_1 \delta_2 \cdots \delta_{m-1} \right( \left( \epsilon_1 \epsilon_2^2 \cdots \epsilon_k^k \right) \\
    & + \fkb_0 x^2 y \left( \delta_1 \delta_2^2 \cdots \delta_{m-1}^{m-1} \right) \left( \zeta_1 \zeta_2^2 \cdots \zeta_m^m \right) \\
    & + \fkb_1 s w x y z  + \fkb_{2,m} y s^2 w^2 z^{m} \left( \delta_1^{m-1} \delta_2^{m-2} \cdots \delta_{m-1} \right) \left( \epsilon_1^m \epsilon_2^m \cdots \epsilon_k^m \right) \left( \zeta_1^{m-1} \zeta_2^{m-2} \cdots \zeta_{m-1} \right) \\
    = & \, \fkc_{0,2m} w^4 s^3 z^{2m} \left( \zeta_1^{2(m-1)} \zeta_2^{2(m-2)} \cdots \zeta_{m-1}^2 \right) \left( \delta_1^{2m-3} \delta_2^{2m-5} \cdots \delta_{m-1} \right) \left( \epsilon_1^{2m-1} \epsilon_2^{2m-3} \cdots \epsilon_k^{2(m-k)-1} \right) \\
    & + \fkc_{1,m} w^3 s^2 x z^{m} \left( \zeta_1^{m-1} \zeta_2^{m-2} \cdots \zeta_{m-1} \right) \left( \delta_1^{m-2} \delta_2^{m-3} \cdots \delta_{m-2} \right) \left( \epsilon_1^{m-1} \epsilon_2^{m-2} \cdots \epsilon_k^{m-k} \right) \\
    & + \fkc_{2,k} s w^2 x^2 z^{k} \left( \zeta_1^k \zeta_2^k \cdots \zeta_{m}^k \right) \left( \delta_1^{k-1} \delta_2^{k-1} \cdots \delta_{m-1}^{k-1} \right) \left( \epsilon_1^{k-1} \epsilon_2^{k-2} \cdots \epsilon_{k-1} \right) \\
    & + \fkc_{3,k} w x^3 z^{k} \left( \zeta_1^{k+1} \zeta_2^{k+2} \cdots \zeta_m^{k+m} \right) \left( \delta_1^k \delta_2^{k+1} \cdots \delta_{m-1}^{m+k-2} \right) \left( \epsilon_1^{k-1} \epsilon_2^{k-1} \cdots \epsilon_{k-1} \right) \,.
  \end{aligned}
\end{equation}
The Cartan divisors are
\begin{equation}
  \begin{tabular}{c|c}
    Section & Equation in $Y_4$ \\ \hline
    $z$ & $b_{1,0} s w x + \delta_1 \left( s \epsilon_1 + b_{0,0} x^2 \zeta_1 \right)$ \\
    $\zeta_1$ & $b_{1,0} x + \delta_1$ \\
    $\zeta_{2 \leq j \leq m-1}$ & $b_{1,0} x + \delta_{j-1} \delta_j$ \\
    $\zeta_m$ & $b_{1,0} x y - c_{1,m} x \zeta_{m-1} + \delta_{m-1} \left( y^2 + b_{2,m} y \zeta_{m-1} - c_{0,2m} \zeta_{m-1}^2 \right)$ \\
    $\delta_{j\leq m-2}$ & $b_{1,0} y$ \\
    $\delta_{m-1}$ & $b_{1,0} y - c_{1,m} \delta_{m-1} \zeta_m$ \\
    $\epsilon_{j \leq k-1}$ & $b_{1,0} s + b_{0,0} \delta_1$ \\
    $\epsilon_k$ & $b_{1,0} s y + b_{0,0} y \delta_1 - \delta_1^{k-1} \epsilon_{k-1} \left( c_{2,k} s + c_{3,k} \delta_1 \right)$
  \end{tabular}
\end{equation}
If one orders the Cartan divisors as $(z, \zeta_1, \dots, \zeta_m, \delta_{m-1}, \delta_{m-2}, \dots, \delta_1, \epsilon_k, \epsilon_{k-1}, \dots, \epsilon_1)$, then each Cartan divisor intersects exactly its neighbours, and $\zeta_0$ also intersects $\epsilon_1$. This replicates the affine $A_{2m+k-1}$ Dynkin diagram.

\tocless\subsection{$I_{2m+k+1}^{(0|^k 1)}$}

The resolution sequence here is (with $z=\zeta_0$)
\begin{equation}
  \begin{aligned}
    (\zeta_i, x, y, \zeta_{i+1}) & \qquad i = 0, \dots, m-1, \\
    (\zeta_i, y, \delta_i) & \qquad i = 1, \dots, m-1, \\
    (z, y, \epsilon_1) & \qquad \text{if } k>0, \\
    (\epsilon_i, y, \epsilon_{i+1}) & \qquad i = 1, \dots, k-1, \\
    (\zeta_m, y, \zeta_{m+1}) \,.
  \end{aligned}
\end{equation}
The resolved geometry is given by
\begin{equation}
  \begin{aligned}
    & y^2 s \left( \delta_1 \delta_2 \cdots \delta_{m-1} \right( \left( \epsilon_1 \epsilon_2^2 \cdots \epsilon_k^k \right) \zeta_{m+1} \\
    & + \fkb_0 x^2 y \left( \delta_1 \delta_2^2 \cdots \delta_{m-1}^{m-1} \right) \left( \zeta_1 \zeta_2^2 \cdots \zeta_m^m \right) \zeta_{m+1}^m \\
    & + \fkb_1 s w x y z  + \fkb_{2,m} y s^2 w^2 z^{m} \left( \delta_1^{m-1} \delta_2^{m-2} \cdots \delta_{m-1} \right) \left( \epsilon_1^m \epsilon_2^m \cdots \epsilon_k^m \right) \left( \zeta_1^{m-1} \zeta_2^{m-2} \cdots \zeta_{m-1} \right) \\
    = & \, \fkc_{0,2m} w^4 s^3 z^{2m} \left( \zeta_1^{2(m-1)} \zeta_2^{2(m-2)} \cdots \zeta_{m-1}^2 \right) \left( \delta_1^{2m-3} \delta_2^{2m-5} \cdots \delta_{m-1} \right) \left( \epsilon_1^{2m-1} \epsilon_2^{2m-3} \cdots \epsilon_k^{2(m-k)-1} \right) \\
    & + \fkc_{1,m} w^3 s^2 x z^{m} \left( \zeta_1^{m-1} \zeta_2^{m-2} \cdots \zeta_{m-1} \right) \left( \delta_1^{m-2} \delta_2^{m-3} \cdots \delta_{m-2} \right) \left( \epsilon_1^{m-1} \epsilon_2^{m-2} \cdots \epsilon_k^{m-k} \right) \\
    & + \fkc_{2,k} s w^2 x^2 z^{k} \left( \zeta_1^k \zeta_2^k \cdots \zeta_{m}^k \zeta_{m+1}^k \right) \left( \delta_1^{k-1} \delta_2^{k-1} \cdots \delta_{m-1}^{k-1} \right) \left( \epsilon_1^{k-1} \epsilon_2^{k-2} \cdots \epsilon_{k-1} \right) \\
    & + \fkc_{3,k} w x^3 z^{k} \left( \zeta_1^{k+1} \zeta_2^{k+2} \cdots \zeta_m^{k+m} \right) \left( \delta_1^k \delta_2^{k+1} \cdots \delta_{m-1}^{m+k-2} \right) \left( \epsilon_1^{k-1} \epsilon_2^{k-1} \cdots \epsilon_{k-1} \right) \zeta_{m+1}^{k+m-1} \,.
  \end{aligned}
\end{equation}
The Cartan divisors read
\begin{equation}
  \begin{tabular}{c|c}
    Section & Equation in $Y_4$ \\ \hline
    $z$ & $b_{1,0} s w x + \delta_1 \left( s \epsilon_1 + b_{0,0} x^2 \zeta_1 \right)$ \\
    $\zeta_1$ & $b_{1,0} x + \delta_1$ \\
    $\zeta_{2 \leq j \leq m-1}$ & $b_{1,0} x + \delta_{j-1} \delta_j$ \\
    $\zeta_m$ & $b_{1,0} x y + \delta_{m-1} \left( b_{2,m} \zeta_{m-1} + \zeta_{m+1} \right)$ \\
    $\zeta_{m+1}$ & $b_{1,0} x y + \delta_{m-1} \left( b_{2,m} y - \zeta_m \left( c_{1,m+1} x + c_{0,2m+1} \delta_{m-1} \right) \right)$ \\
    $\delta_{j\leq m-2}$ & $b_{1,0} y$ \\
    $\delta_{m-1}$ & $b_{1,0} y - c_{1,m} \delta_{m-1} \zeta_m$ \\
    $\epsilon_{j \leq k-1}$ & $b_{1,0} s + b_{0,0} \delta_1$ \\
    $\epsilon_k$ & $b_{1,0} s y + b_{0,0} y \delta_1 - \delta_1^{k-1} \epsilon_{k-1} \left( c_{2,k} s + c_{3,k} \delta_1 \right)$
  \end{tabular}
\end{equation}
Ordering the Cartan divisors as $(z, \zeta_1, \dots, \zeta_m, \zeta_{m+1}, \delta_{m-1}, \delta_{m-2}, \dots, \delta_1, \epsilon_k, \epsilon_{k-1}, \dots, \epsilon_1)$ yields the affine $A_{2m+k}$ Dynkin diagram.

\tocless\subsection{$I_{2(m+k)}^{(0|^m 1)}$}

In this case, the resolution sequence is, after identifying $z=\zeta_0$,
\begin{equation}
  \begin{aligned}
    (\zeta_i, y, \zeta_{i+1}) & \qquad i = 0, \dots, m-1, \\
    (\zeta_i, x, \delta_i) & \qquad i = 1, \dots, m-1, \\
    (\zeta_i, x, \zeta_{i+1}) & \qquad i = m, \dots, m+2k-1 \,.
  \end{aligned}
\end{equation}
The resolved geometry is
\begin{equation}
  \begin{aligned}
    & y^2 s \left(\delta_2 \delta_3^2 \cdots \delta_{m-1}^{m-2} \right) \left( \zeta_1 \zeta_2^2 \cdots \zeta_m^m \right) \left( \zeta_{m+1}^{m-1} \zeta_{m+2}^{m-2} \cdots \zeta_{m+2k}^{m-2k} \right) \\
    & + \fkb_0 x^2 y \left( \delta_1 \delta_2 \cdots \delta_{m-1} \right) \left( \zeta_{m+1} \zeta_{m+2}^2 \cdots \zeta_{m+2k}^{2k} \right) \\
    & + \fkb_1 s w x y z  + \fkb_{2,2k} y s^2 w^2 z^{2k} \left( \delta_1^{2k-1} \delta_2^{2k-1} \cdots \delta_{m-1}^{2k-1} \right) \left( \zeta_1^{2k} \zeta_2^{2k} \cdots \zeta_m^{2k} \right) \left( \zeta_{m+1}^{2k-1} \zeta_{m+2}^{2k-2} \cdots \zeta_{m+2k-1} \right) \\
    = & \, \fkc_{0,m+2k} w^4 s^3 z^{m+2k} \left( \zeta_1^{m+2k-1} \zeta_2^{m+2k-2} \cdots \zeta_{m+2k-1} \right) \left( \delta_1^{m+2k-2} \delta_2^{m+2k-3} \cdots \delta_{m-1}^{2k} \right) \\
    & + \fkc_{1,m} w^3 s^2 x z^{m} \left( \delta_1^{m-1} \delta_2^{m-2} \cdots \delta_{m-1} \right) \left( \zeta_1^{m-1} \zeta_2^{m-2} \cdots \zeta_{m-1} \right) \\
    & + \fkc_{2,m} s w^2 x^2 z^{m} \left( \delta_1^{m} \delta_2^{m-1} \cdots \delta_{m-1}^2 \right) \left( \zeta_1^{m-1} \zeta_2^{m-2} \cdots \zeta_{m-1} \right) \left( \zeta_{m+1} \zeta_{m+2}^2 \cdots \zeta_{m+2k}^{2k} \right) \\
    & + \fkc_{3,m} w x^3 z^{m} \left( \delta_1^{m+1} \delta_2^{m} \cdots \delta_{m-1}^3 \right) \left( \zeta_1^{m-1} \zeta_2^{m-2} \cdots \zeta_{m-1} \right) \left( \zeta_{m+1}^2 \zeta_{m+2}^4 \cdots \zeta_{m+2k}^{4k} \right) \,.
  \end{aligned}
\end{equation}
and the irreducible Cartan Divisors are
\begin{equation}
  \begin{tabular}{c|c}
    Section & Equation in $Y_4$ \\ \hline
    $z$ & $b_{1,0} s w x + b_{0,0} x^2 \delta_1 + s \zeta_1$ \\
    $\zeta_1$ & $b_{1,0} s + b_{0,0} \delta_1$ \\
    $\zeta_{2 \leq j \leq m-1}$ & $b_{1,0} s + b_{0,0} \delta_{j-1} \delta_k$ \\
    $\zeta_m$ & $b_{1,0} s y - \delta_{m-1} \left( c_{1,m} s^2 \zeta_{m-1} + \zeta_{m+1} \left( - b_{0,0} y + \delta_{m-1} \zeta_{m-1} \left( c_{2,m} s + c_{3,m} \delta_{m-1} \zeta_{m+1} \right) \right) \right)$ \\
    $\zeta_{m+1 \leq j \leq m+2k-1}$ & $b_{1,0} y - c_{1,m} \delta_{m-1}$ \\
    $\zeta_{m+2k}$ & $b_{1,0} x y - c_{1,m} x \delta_{m-1} + \delta_{m-1}^{2k-1} \zeta_{m+2k-1} \left( b_{2,2k} y - c_{0,m+2k} \delta_{m-1} \right)$ \\
    $\delta_1$ & $b_{1,0} x + \delta_2 \zeta_1 \zeta_2^2$ \\
    $\delta_{2 \leq j \leq m-1}$ & $ b_{1,0} x$ \\
  \end{tabular}
\end{equation}
With the ordering $(z, \zeta_1, \dots, \zeta_{m+2k}, \delta_{m-1}, \delta_{m-2}, \dots, \delta_1)$, the affine $A_{2(m+k)-1}$ Dynkin diagram is reproduced.

\tocless\subsection{$I_{2(m+k)+1}^{(0|^m 1)}$}

Here, the resolution sequence reads, after identifying $z=\zeta_0$,
\begin{equation}
  \begin{aligned}
    (\zeta_i, y, \zeta_{i+1}) & \qquad i = 0, \dots, m-1, \\
    (\zeta_i, x, \delta_i) & \qquad i = 1, \dots, m-1, \\
    (\zeta_i, x, \zeta_{i+1}) & \qquad i = m, \dots, m+2k \,.
  \end{aligned}
\end{equation}
The resolved geometry is given by
\begin{equation}
  \begin{aligned}
    & y^2 s \left(\delta_2 \delta_3^2 \cdots \delta_{m-1}^{m-2} \right) \left( \zeta_1 \zeta_2^2 \cdots \zeta_m^m \right) \left( \zeta_{m+1}^{m-1} \zeta_{m+2}^{m-2} \cdots \zeta_{m+2k+1}^{m-2k-1} \right) \\
    & + \fkb_0 x^2 y \left( \delta_1 \delta_2 \cdots \delta_{m-1} \right) \left( \zeta_{m+1} \zeta_{m+2}^2 \cdots \zeta_{m+2k+1}^{2k+1} \right) \\
    & + \fkb_1 s w x y z  + \fkb_{2,2k+1} y s^2 w^2 z^{2k+1} \left( \delta_1^{2k} \delta_2^{2k} \cdots \delta_{m-1}^{2k} \right) \left( \zeta_1^{2k+1} \zeta_2^{2k+1} \cdots \zeta_m^{2k+1} \right) \left( \zeta_{m+1}^{2k} \zeta_{m+2}^{2k-1} \cdots \zeta_{m+2k} \right) \\
    = & \, \fkc_{0,m+2k+1} w^4 s^3 z^{m+2k+1} \left( \zeta_1^{m+2k} \zeta_2^{m+2k-1} \cdots \zeta_{m+2k} \right) \left( \delta_1^{m+2k-1} \delta_2^{m+2k-2} \cdots \delta_{m-1}^{2k+1} \right) \\
    & + \fkc_{1,m} w^3 s^2 x z^{m} \left( \delta_1^{m-1} \delta_2^{m-2} \cdots \delta_{m-1} \right) \left( \zeta_1^{m-1} \zeta_2^{m-2} \cdots \zeta_{m-1} \right) \\
    & + \fkc_{2,m} s w^2 x^2 z^{m} \left( \delta_1^{m} \delta_2^{m-1} \cdots \delta_{m-1}^2 \right) \left( \zeta_1^{m-1} \zeta_2^{m-2} \cdots \zeta_{m-1} \right) \left( \zeta_{m+1} \zeta_{m+2}^2 \cdots \zeta_{m+2k+1}^{2k+1} \right) \\
    & + \fkc_{3,m} w x^3 z^{m} \left( \delta_1^{m+1} \delta_2^{m} \cdots \delta_{m-1}^3 \right) \left( \zeta_1^{m-1} \zeta_2^{m-2} \cdots \zeta_{m-1} \right) \left( \zeta_{m+1}^2 \zeta_{m+2}^4 \cdots \zeta_{m+2k+1}^{4k+2} \right) \,.
  \end{aligned}
\end{equation}
and the irreducible Cartan divisors by
\begin{equation}
  \begin{tabular}{c|c}
    Section & Equation in $Y_4$ \\ \hline
    $z$ & $b_{1,0} s w x + b_{0,0} x^2 \delta_1 + s \zeta_1$ \\
    $\zeta_1$ & $b_{1,0} s + b_{0,0} \delta_1$ \\
    $\zeta_{2 \leq j \leq m-1}$ & $b_{1,0} s + b_{0,0} \delta_{j-1} \delta_k$ \\
    $\zeta_m$ & $b_{1,0} s y - \delta_{m-1} \left( c_{1,m} s^2 \zeta_{m-1} + \zeta_{m+1} \left( - b_{0,0} y + \delta_{m-1} \zeta_{m-1} \left( c_{2,m} s + c_{3,m} \delta_{m-1} \zeta_{m+1} \right) \right) \right)$ \\
    $\zeta_{m+1 \leq j \leq m+2k}$ & $b_{1,0} y - c_{1,m} \delta_{m-1}$ \\
    $\zeta_{m+2k+1}$ & $b_{1,0} x y - c_{1,m} x \delta_{m-1} + \delta_{m-1}^{2k} \zeta_{m+2k} \left( b_{2,2k+1} y - c_{0,m+2k+1} \delta_{m-1} \right)$ \\
    $\delta_1$ & $b_{1,0} x + \delta_2 \zeta_1 \zeta_2^2$ \\
    $\delta_{2 \leq j \leq m-1}$ & $ b_{1,0} x$ \\
  \end{tabular}
\end{equation}
The divisor ordering $(z, \zeta_1, \dots, \zeta_{m+2k+1}, \delta_{m-1}, \delta_{m-2}, \dots, \delta_1)$ reproduces the affine $A_{2(m+k)}$ Dynkin diagram.


\tocless\subsection{$I_n^{ns(01)}$}

For the non-split-type fibers in the $I_n$ series, one again distinguishes between even $n=2m$ and odd $n=2m+1$. In both cases, the ordered resolution sequence is given by
\begin{equation}
  \begin{aligned}
    (z, x, y; \zeta_1), & \\
    (\zeta_i, x, y; \zeta_{i+1}), & \qquad i=1, \dots, m-1 & \,.
  \end{aligned}
\end{equation}
For $n$ even, the geometry is given by
\begin{equation}
  \begin{aligned}
    & \, y^2 s + \fkb_0 x^2 y \left( \zeta_1 \zeta_2^2 \cdots \zeta_m^m \right) + \fkb_1 s w x y + \fkb_{2,m} s^2 w^2 y z^m \left( \zeta_1^{m-1} \zeta_2^{m-2} \cdots \zeta_{m-1} \right) \\
    = & \, \fkc_{0,2m} s^3 w^4 z^{2m} \left( \zeta_1^{2(m-1)} \zeta_2^{2(m-2)} \cdots \zeta_{m-1} \right) + \fkc_{1,m} s^2 w^3 x z^m \left( \zeta_1^{m-1} \zeta_2^{m-2} \cdots \zeta_{m-1} \right) \\
    & \, + \fkc_2 s w^2 x^2 + \fkc_3 w x^3 \left( \zeta_1 \zeta_2^2 \cdots \zeta_m^m \right) \,,
  \end{aligned}
\end{equation}
and the irreducible Cartan divisors are
\begin{equation}
  \begin{tabular}{c|c}
    Section & Equation in $Y_4$ \\ \hline
    $z$ & $-c_{2,0} s w^2 x^2 + s y \left( b_{1,0} w x + y \right) + x^2 \zeta_1 \left( b_{0,0} y - c_{3,0} w x \right)$ \\
    $\zeta_{1\leq i<m}$ & $-c_{2,0} x^2 + y \left( b_{1,0} x + y \right)$ \\
    $\zeta_m$ & $-c_{2,0} x^2 + b_{1,0} x y + y^2 - c_{1,m} x \zeta_{m-1} + b_{2,m} y \zeta_{m-1} - c_{0,2m} \zeta_{m-2}^2$
  \end{tabular}
\end{equation}
For odd $n$, the geometry reads
\begin{equation}
  \begin{aligned}
    & \, y^2 s + \fkb_0 x^2 y \left( \zeta_1 \zeta_2^2 \cdots \zeta_m^m \right) + \fkb_1 s w x y + \fkb_{2,m+1} s^2 w^2 y z^{m+1} \left( \zeta_1^{m} \zeta_2^{m-1} \cdots \zeta_{m} \right) \\
    = & \, \fkc_{0,2m+1} s^3 w^4 z^{2m+1} \left( \zeta_1^{2m-1} \zeta_2^{2m-3} \cdots \zeta_{m} \right) + \fkc_{1,m+1} s^2 w^3 x z^{m+1} \left( \zeta_1^{m} \zeta_2^{m-1} \cdots \zeta_{m} \right) \\
    & \, + \fkc_2 s w^2 x^2 + \fkc_3 w x^3 \left( \zeta_1 \zeta_2^2 \cdots \zeta_m^m \right) \,,
  \end{aligned}
\end{equation}
and the Cartan divisors are
\begin{equation}
  \begin{tabular}{c|c}
    Section & Equation in $Y_4$ \\ \hline
    $z$ & $-c_{2,0} s w^2 x^2 + s y \left( b_{1,0} w x + y \right) + x^2 \zeta_1 \left( b_{0,0} y - c_{3,0} w x \right)$ \\
    $\zeta_{1\leq i\leq m}$ & $-c_{2,0} x^2 + y \left( b_{1,0} x + y \right)$ \\
  \end{tabular}
\end{equation}
For even $n$, the Cartan matrix one obtains from the ordering $(z, \zeta_1, \dots, \zeta_n)$ reproduces the $C_n$-type Dynkin diagrams, as expected.
\newpage



\def\cprime{$'$}
\providecommand{\href}[2]{#2}\begingroup\raggedright\endgroup


\end{document}